\documentclass[usenatbib]{mn2e}

\usepackage[pdftex]{graphicx}
\usepackage{savesym}
\usepackage{amsmath}
\savesymbol{iint}
\usepackage{txfonts}
\restoresymbol{TXF}{iint}

\graphicspath{{../c/ps/}{pict/}}

\usepackage{subfigure}
\usepackage{epsfig}
\usepackage{enumerate}
\usepackage{url}
\def\simless{\mathbin{\lower 3pt\hbox
{$\rlap{\raise 5pt\hbox{$\char'074$}}\mathchar"7218$}}}   
\def\simmore{\mathbin{\lower 3pt\hbox
{$\rlap{\raise 5pt\hbox{$\char'076$}}\mathchar"7218$}}}   
\newcommand       \be          {\begin{eqnarray}}
\newcommand       \ee          {\end{eqnarray}}
\newcommand       \apj          {ApJ}
\newcommand       \apjl         {ApJL}
\newcommand       \aap          {A\&A}
\newcommand       \nat          {Nature}
\newcommand       \mnras        {MNRAS}

\newcommand       \aj      {AJ}

\newcommand       \araa      {ARA\&A}

\newcommand      \apjs {ApJ Supplements}
\newcommand      \na {New Astronomy}

\def\apss{Astrophysics \& Space Science}

\def\simlt{\mathrel{\hbox{\rlap{\hbox{\lower4pt\hbox{$\sim$}}}\hbox{$<$}}}}
\def\simgt{\mathrel{\hbox{\rlap{\hbox{\lower4pt\hbox{$\sim$}}}\hbox{$>$}}}}
\def\lesssim{\mathrel{\hbox{\rlap{\hbox{\lower4pt\hbox{$\sim$}}}\hbox{$<$}}}}
\def\gtrsim{\mathrel{\hbox{\rlap{\hbox{\lower4pt\hbox{$\sim$}}}\hbox{$>$}}}}

\newcommand{\eps}{\epsilon}
\newcommand{\sls}{\nonumber \\}
\newcommand{\de}{\delta}
\newcommand{\De}{\Delta}

\newcommand{\g}{\gamma}
\newcommand{\ratdeg}[3]{\left(\frac{#1}{#2}\right)^{#3}}
\newcommand{\e}[2]{#1\cdot 10^{#2}}
\newcommand{\et}[2]{#1\times 10^{#2}}
\newcommand{\tvf}{\tilde{v}_f}
\newcommand{\tvfe}{v_8}
\newcommand{\ba}[1]{\begin{array}{#1}}
\newcommand{\ea}{\end{array}}

\usepackage{ulem}
\usepackage{color}

\title[Radio synchrotron emission from shocks in novae]{Shocks in nova outflows. II. Synchrotron radio emission}

\author[]{Andrey Vlasov$^{1}$, Indrek~Vurm$^{1,2}$, Brian D.~Metzger$^{1}\thanks{E-mail: bmetzger@phys.columbia.edu}$\\
$^{1}$Columbia Astrophysics Laboratory, Columbia University, New York, NY, 10027, USA\\ 
$^{2}$Tartu Observatory, T$\tilde{o}$ravere, Tartumaa EE-61602, Estonia\\ }

\begin{document}
\date{Received / Accepted}

\maketitle

\label{firstpage}
\begin{abstract}
The discovery of GeV gamma-rays from classical novae indicates that shocks and relativistic particle acceleration are energetically key in these events.  Further evidence for shocks comes from thermal keV X-ray emission and an early peak in the radio light curve on a timescale of months with a brightness temperature which is too high to result from freely expanding photo-ionized gas.  Paper I developed a one dimensional model for the thermal emission from nova shocks.  This work concluded that the shock-powered radio peak cannot be thermal if line cooling operates in the post-shock gas at the rate determined by collisional ionization equilibrium.  Here we extend this calculation to include non-thermal synchrotron emission.  Applying our model to three classical novae, we constrain the amplification of the magnetic field $\epsilon_B$ and the efficiency $\epsilon_e$ of accelerating relativistic electrons of characteristic Lorentz factor $\gamma \sim 100$.  If the shocks are radiative (low velocity $v_{\rm sh} \lesssim 1000$ km s$^{-1}$) and cover a large solid angle of the nova outflow, as likely characterize those producing gamma-rays, then values of $\epsilon_e \sim 0.01-0.1$ are required to achieve the peak radio brightness for $\epsilon_B = 10^{-2}$.  Such high efficiencies exclude secondary pairs from pion decay as the source of the radio-emitting particles, instead favoring the direct acceleration of electrons at the shock.  If the radio-emitting shocks are instead adiabatic (high velocity), as likely characterize those responsible for the thermal X-rays, then much higher brightness temperatures are possible, allowing the radio-emitting shocks to cover a smaller outflow solid angle.    
\end{abstract}
\section{Introduction}

Classical and symbiotic novae are luminous transients, powered by runaway thermonuclear burning of a hydrogen-rich layer accreted from a binary companion (e.g.,~\citealt{Gallagher&Starrfield78,Starrfield+00,Yaron+05,Townsley&Bildsten05,Casanova+11}).  The resulting energy release causes the white dwarf atmosphere to inflate, ejecting $\sim 10^{-5}-10^{-4}\,\, M_{\odot}$ of CNO-enriched matter at hundreds to thousands of kilometers per second (e.g., \citealt{Seaquist+80,Shore12}).

\begin{figure}
\begin{center}
\includegraphics[width=1.0\linewidth]{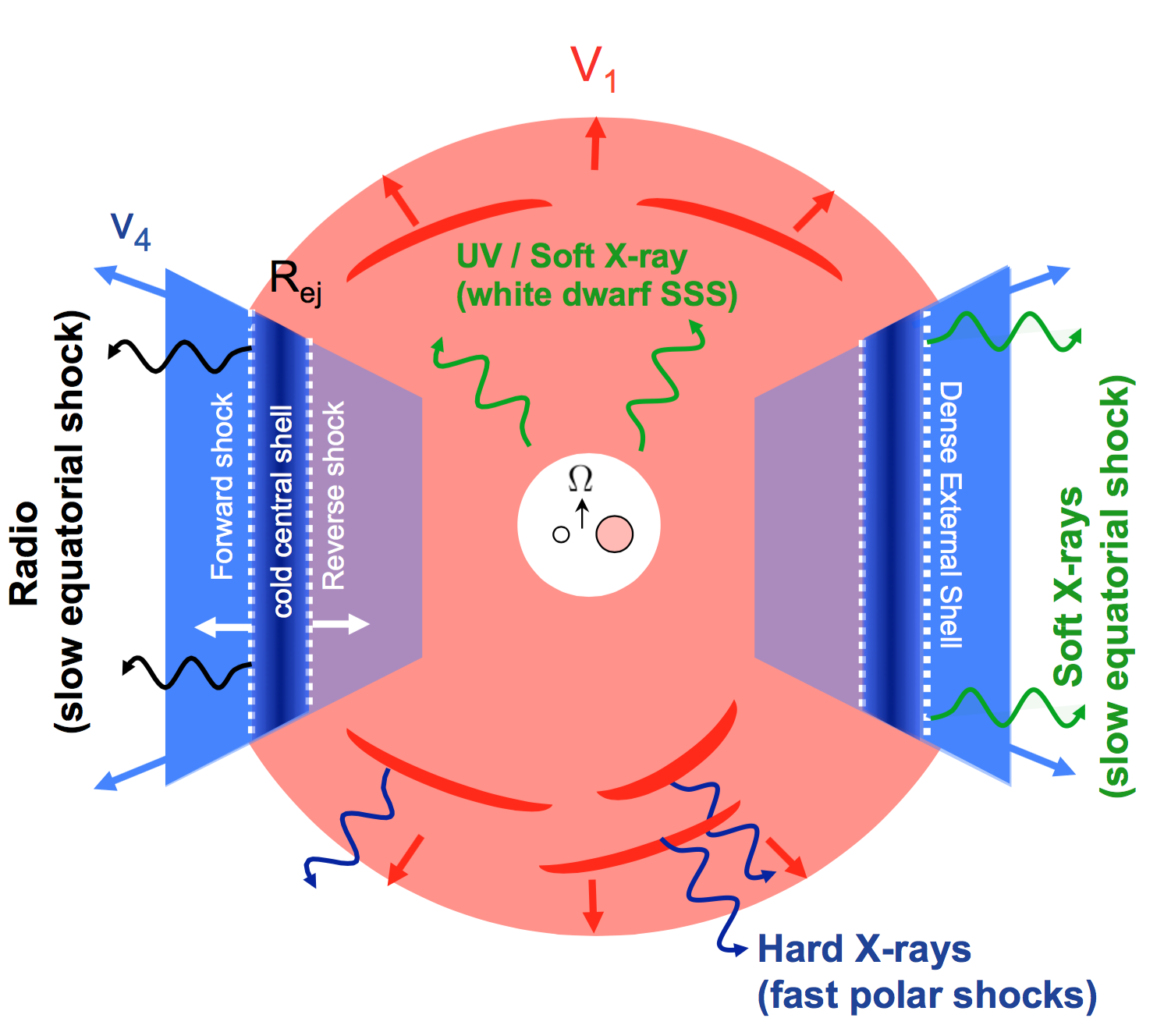}
\caption{Proposed scenario for the locations of X-ray and radio emitting shocks.  A slow outflow is ejected first, its geometry shaped into an equatorially-concentrated torus (blue). This is followed by a faster outflow or continuous wind with a higher velocity and more spherical geometry (red). The fast and slow components collide in the equatorial plane, producing powerful radiative shocks (Fig.~\ref{fig:shockzones}) which are responsible for the gamma-ray emission on timescales of weeks and the non-thermal radio emission on a timescale of months.  Adiabatic internal shocks within the fast, low density polar outflow power hard $\gg$ keV thermal X-rays and possibly radio emission at very early times.  Slower equatorial shocks produce the non-thermal radio peak on a timescale of months and softer X-rays ($kT \lesssim$ keV), which are challenging to detect due to their lower luminosity and confusion with supersoft X-rays from the white dwarf. The direction of the binary orbital angular velocity $\Omega$ is indicated by an arrow.}
\label{fig:cartoon}
\end{center}
\end{figure}

Radio and optical imaging \citep{Chomiuk+14a,Schaefer+14}, and optical spectroscopy (\citealt{Ribeiro+13,Shore+13}), suggests that a nova outburst proceeds in at least two stages (Figure \ref{fig:cartoon}).  The runaway is first accompanied by a low velocity outflow concentrated in the equatorial plane of the binary, perhaps influenced by the gravity of the companion star, as occurs in common envelope phases of stellar evolution (e.g., \citealt{Livio+90,Lloyd+97}).  Given the uncertain nature of the slow ejecta, we use the agnostic term `dense external shell' (DES; \citealt{Metzger+14}).

The outflowing DES is then followed by a more continuous wind (e.g., \citealt{Bath&Shaviv76}) with a higher velocity and a more spherical geometry.  A collision between this fast outflow and the slower DES results in strong internal shocks, which are most powerful near the equatorial plane where the density contrast is largest.  If the fast component expands relatively unimpeded along the polar direction, this creates a bipolar morphology (\citealt{Chomiuk+14a}; \citealt{Metzger+15}).  Such a scenario does not exclude fast shocks within the polar region, characterized by lower densities and hard X-ray emission (Fig.~\ref{fig:cartoon}).

Several lines of evidence support shocks being common features of nova outbursts.   Nova optical spectra exhibit complex absorption lines with multiple velocity components (e.g., \citealt{Friedjung&Duerbeck93,Williams+08,Williams&Mason10}).  In addition to broad P Cygni lines indicating high velocity $\gtrsim 1000$ km s$^{-1}$ matter, narrower absorption features ($\approx 500-900$ km s$^{-1}$) are observed near optical maximum and may be created within the DES.  A large fraction of novae show such narrow lines.  This indicates that the absorbing material has a high covering fraction (\citealt{Williams&Mason10}) and hence is likely to impact the faster P Cygni outflow.  

Many novae are accompanied by thermal X-ray emission of luminosity $L_{X} \sim 10^{33}-10^{35}$ erg s$^{-1}$ and temperatures $\gtrsim$ keV (e.g., \citealt{Lloyd+92,OBrien+94,Orio04,Sokoloski+06,Ness+07,Mukai+08,Schwarz+11,Krauss+11,Chomiuk+14a}; see \citealt{Osborne15} for a summary of {\it Swift} observations).  This emission is too hard to be thermal radiation from the white dwarf surface (\citealt{Wolf+13}), but is readily explained as free-free emisson  from $\gtrsim 1000$ km s$^{-1}$ shocks (e.g.~\citealt{Mukai&Ishida01}).  The X-ray emission is often delayed by weeks or longer after the optical maximum, perhaps due to absorption by the DES (\citealt{Metzger+14}). 

Although the presence of shocks in novae have been realized for some time, their energetic importance was only recently revealed by the {\it Fermi} LAT discovery of $\gtrsim 100$ MeV gamma-rays, coincident within days of the optical peak and last for weeks (\citealt{Ackermann+14}).  Gamma-rays were first detected in the symbiotic nova V407 Cyg 2010, in which the target material for the shocks could be understood as the dense wind of the companion red giant (\citealt{Abdo+10}; \citealt{Vaytet+11}; \citealt{Martin&Dubus13}).  However, six additional novae have now been detected by Fermi-LAT, at least four of which show no evidence for a giant companion and hence were likely ordinary classical novae with main sequence companions.  The DES is clearly present even in binary systems that are not embedded in the wind of an M giant or associated with recurrent novae, supporting the internal shock scenario.    


The high gamma-ray luminosities $L_{\gamma} \sim 10^{35}-10^{36}$ erg s$^{-1}$ require shocks with kinetic powers which are at least two orders of magnitude larger, i.e. $L_{\rm sh} \sim 10^{37}-10^{38}$ erg s$^{-1}$, approaching the bolometric output of the nova (\citealt{Metzger+15}).  Gamma-rays are produced by the decay of neutral pions created by collisions between relativistic protons and ambient protons in the ejecta (hadronic scenario), or by inverse Compton or bremsstrahlung emission from relativistic electrons (leptonic scenario).  Hadronic versus lepton emission scenarios cannot be distinguished based on the gamma-ray spectra alone \citep{Ackermann+14}, although \citet{Metzger+15} cite evidence in favor of a hadronic scenario.  


Additional evidence for shocks comes at radio wavelengths.  Novae produce thermal radio emission from the freely expanding photoionized ejecta of temperature $\sim 10^{4}$ K, which peaks as the ejecta becomes optically thin to free-free absorption roughly a year after the optical outburst (\citealt{Seaquist&Bode08}).  However, a growing sample of novae show an additional peak in the radio emission at earlier times ($\lesssim 100$ days; \citealt{Taylor+87,Krauss+11,Chomiuk+14a,Weston+15a}; Fig.~\ref{fig:TB}) with brightness temperatures $10^5-10^6$ K higher than that of photo-ionized gas.  This additional early radio peak requires sudden heating of the ejecta (e.g., \citealt{Lloyd+96,Metzger+14}) or non-thermal emission  (\citealt{Taylor+87,Weston+15a}), in either case implicating shocks. 


\citet{Metzger+14} (Paper I) developed a one-dimensional model for the forward-reverse shock structure in novae and its resulting {\it thermal} X-ray, optical, radio emission.  This initial work provided an acceptable fit to the radio light curves of the gamma-ray nova V1324 Sco, under the assumption that the dominant cooling behind the shock was provided by free-free emission.  However, for low velocity shocks $\lesssim 10^{3}$ km s$^{-1}$ line cooling of the CNO-enriched gas can greatly exceed free-free cooling.  For cooling rates determined by collisional ionization equilibrium (CIE), this additional cooling reduces the peak brightness temperature of thermal models to values $\lesssim 10^4$ K, which are too low to explain the early radio peak, unless line cooling is suppressed by non-LTE effects.  

In this work (Paper II) we extend the \citet{Metzger+14} model to include non-thermal synchrotron radio emission, which we demonstrate can explain the observed emission, even for low shock velocities. In addition to providing information on the structure of the ejecta and the nova outburst mechanism, synchrotron emission provides an alternative probe of relativistic particle acceleration in these events, complementary to that obtained from the gamma-ray band.  In leptonic scenario, relativistic electrons accelerated directly at the shock power the radio emission.  Radio-emitting $e^{\pm}$ pairs are also produced in hadronic scenarios by the decay of the charged pions. Radio observations can in principle help disentangle leptonic from hadronic models.

This paper is organized as follows.  We begin with an overview of shocks in novae ($\S\ref{sec:shocks}$), including the collision dynamics, observational evidence for shocks, the analytic condition for radio maximum, the radiative versus adiabatic nature of the shock, and thermal X-ray emission.  In $\S\ref{sec:synch}$ we describe key features of synchrotron emission, including leptonic and hadronic sources of non-thermal particles and their cooling.  In $\S\ref{sec:model}$ we provide a detailed description of our model for radio emission from the forward shock. In $\S\ref{sec:results}$ we describe our results, including analytic estimates for brightness temperature, and fits to the radio lightcurves of three novae: V1324 Sco, V1723 Aql, and V5589 Sgr.  In the discussion ($\S\ref{sec:discussion}$) we use the radio observations to constrain the acceleration efficiency of relativistic particles and magnetic field amplification in the shocks and their connection to gamma-ray emission.  We also discuss outstanding issues, including the unexpectedly monochromatic light curve of V1723 Aql and the role of adiabatic X-ray producing shocks.  In $\S\ref{sec:conclusions}$ we summarize our conclusions.

\begin{table*}
\caption{Commonly used variables and their definitions.}
\label{table:pars}
\begin{center}
\begin{tabular}{|c|l|}
 \hline
 Variable & Definition \\
 \hline
$v_1$ & Velocity of fast outflow (Fig.~\ref{fig:shockzones}) \\
$v_4$ & Velocity of DES \\
$n_1$ & Density of (unshocked) fast outflow  \\
$n_4$ & Density of (unshocked) DES \\
$T_4$ & Temperature of DES in the photo-ionized layer just ahead of the shock \\
$T_3$ & Temperature immediately behind the forward shock \\
$n_3$ & Density immediately behind forward shock \\
$v_{{\rm shock}}$ & Velocity of forward shock in the white dwarf frame \\
$v_{{\rm shell}}$ & Velocity of central shell in the white dwarf frame \\
$\tvf \equiv v_{{\rm shock}}-v_4 \equiv 10^8 \tvfe\, {\rm cm\,s}^{-1}$ & Velocity of the shock in the frame of the DES\\
$H=10^{14} H_{14}$ cm & Density scale height of DES \\
$T_b$ & Observed brightness temperature, corrected for pre-shock screening by the photo-ionized layer \\
$ \tau_{{\rm ff},4}$ & Free-free optical depth of unshocked DES \\
 $\alpha_{{\rm ff},4}$ &  Free-free absorption coefficient of unshocked DES \\   
 $t_{\rm cool}$ & Cooling time of gas in the post shock region \\
$\Delta_{\rm ion}$ & Thickness of ionized layer ahead of the shock (eq.~[\ref{eq:delta}]) \\
 $\eta_3\equiv t_{\rm cool}/t_{\rm fall}$ & Cooling efficiency: ratio of post-shock cooling timescale to shock expansion time down the density gradient of the DES (eq.~[\ref{eq:radiative}]) \\
 $n_{{\rm pk},\De}$ & Density of unshocked DES at time of radio peak ($\tau_{\rm ff,4}=1$) for case when $\De_{\rm ion}<H$ (eq.~[\ref{eq:npk}], upper line) \\
$n_{{\rm pk},H}$ &   Density of unshocked DES at time of radio peak ($\tau_{\rm ff,4}=1$) for case when $\De_{\rm ion}>H$ (eq.~[\ref{eq:npk}], lower line) \\
  $M_{{\rm ej}}$ & Total ejecta mass \\
  $R_{{\rm sh}}$ & Radius of cool central shell $\sim$ radius of shock \\
  $f_{\rm EUV}=0.1 f_{\rm EUV,-1}$ & Fraction of shock power placed into hydrogen-ionizing radiation \\

  $\eps_{{\rm p}}=0.1\eps_{{\rm p},-1}$ & Fraction of shock power placed into relativistic protons \\
  $\eps_{{\rm e}}$ & Fraction of shock power placed into relativistic electrons and positrons \\
 $\g_{{\rm pk}}$ & Lorentz factor of the electrons or positrons which determine the synchrotron emissivity at the radio peak \\
 $T_\nu$ & Brightness temperature of emission at generic location behind the shock \\
 $T_{\nu, sync}$ & Brightness temperature of synchrotron emission (eq. [\ref{eq:Tsyn}]) \\
 $\tau_\nu$ & Optical depth at arbitrary location behind the shock \\
$T_{\nu,\rm pk}^{\rm th}$ & Peak observed radio brightness temperature due to thermal emission \\
$T_{\nu,\rm pk}^{\rm nth}$ & Peak observed radio brightness temperature of nonthermal synchrotron emission \\
  $T_{\nu,H,pk}^{th}$ & Thermal contribution to the peak brightness temperature for adiabatic shocks (eq~[\ref{eq:Tthadpk}]) \\
  $T_{\nu,H,pk}^{nth}$ & Non-thermal contribution at the peak brightness temperature of adiabatic shocks (eq.~[\ref{eq:Tnthestpkad}]) \\
 \hline
\end{tabular}
\end{center}
\end{table*}

\section{Shocks in Novae}
 \label{sec:shocks} 

\begin{figure}
\begin{center}
\includegraphics[width=1.0\linewidth]{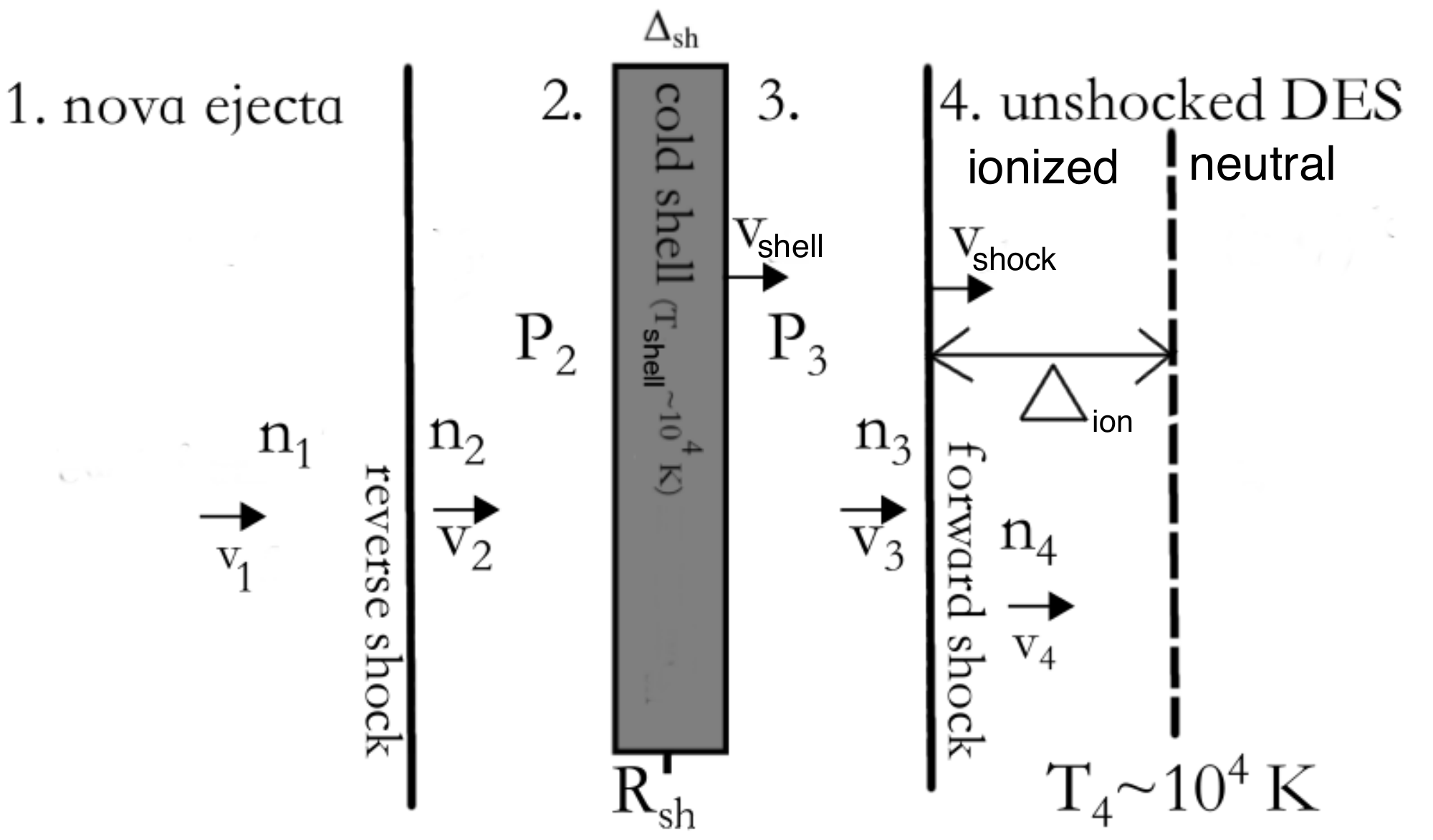}
\caption{Shock interaction between the fast nova outflow (Region 1) and the slower dense external shell [DES] (Region 4).  A forward shock is driven into the DES, while the reverse shock is driven back into the fast ejecta.  The shocked ejecta (Region 2) and shocked DES (Region 3) are separated by a cold central shell containing the swept up mass.  Observed radio emission originates from the forward shock, since emission from the reverse shock is absorbed by the cold central shell. The ionized layer of thickness $\Delta_{\rm ion}$ (eq. \ref{eq:delta}) in front of the forward shock is also shown.}
\label{fig:shockzones}
\end{center}
\end{figure}

Following \citet{Metzger+14}, we consider the collision between a fast outflow from the white dwarf of velocity $v_{1}$ and the DES of velocity $v_{4} < v_1$ and unshocked density $n_{4}$, as shown in Figure \ref{fig:shockzones}.  We assume that the DES of velocity $v_4$ is ejected at $t = 0$, corresponding to the time of the first optical detection.  The fast outflow of velocity $v_1$ is ejected after a delay of time $\De t$.  The fast outflow and DES collide at a radius and time given, respectively, by
\be
  t_{\rm 0}=\frac{v_{1}}{v_{1}-v_{4}}\De t; \,\,\,\,\,\, R_{\rm 0}=\frac{v_1v_4}{v_1-v_4}\De t
\label{eq:R0}
\ee
The mean density of the DES at the time of the collision is 
\be
\bar{n}_0 \approx \frac{M_{\rm DES}}{4\pi R_0^{3}f_{\Omega} m_p} \sim 10^{8}\left(\frac{M_{\rm DES}}{10^{-4}M_{\odot}}\right)\left(\frac{t_0}{60\,{\rm d}}\right)^{-3}\left(\frac{v_{\rm 4}}{10^{3}{\rm \,km\,s^{-1}}}\right)^{-3}\,{\rm cm^{-3}},
\label{eq:nDES}
\ee
where $f_{\Omega} \sim 0.5$ is the fraction of the total solid-angle subtended by the outflow and for purposes of an estimate we have assumed the thickness of the DES is $\sim R_0 \sim t_0 v_4$.  We define the ejecta number density as $n \equiv \rho/m_p$, where $\rho$ is the mass density.

This interaction drives a forward shock through the DES and a reverse shock back through the fast ejecta (see Fig.~\ref{fig:shockzones}).  We assume spherical symmetry and, for the time being, that the shocks are radiative ($\S\ref{sec:radshock}$).  For radiative shocks the post shock material is compressed and piles up in a central cold shell sandwiched by the ram pressure of the two shocks.  Neglecting non-thermal pressure, the shocked gas cools by a factor of $\sim 10^{3}$, its volume becoming negligible. Hence the shocks propagate outwards at the same velocity as the central shell, $v_{\rm shell} = v_{\rm shock} \equiv v_{\rm sh}$.  The velocity of the cold central shell is determined by equating the rate of momentum deposition from ahead and from behind, as described in \citet{Metzger+14}.
In what follows, we define the shock velocity in the upstream frame,
\be
  \tvf=v_{sh}-v_4= 10^8 \tvfe \,{\rm cm\,s^{-1}},
\label{eq:vshock}
\ee
normalized to a characteristic value of 1000 km s$^{-1}$.

Radio emission is assumed to originate from the forward shock because the reverse shock emission is highly attenuated by free-free absorption within the cold central shell.  The forward shock heats the gas to a temperature
\be
T_3 \simeq \frac{3\mu m_p \tvf^{2} }{16 k} \approx 1.7\times 10^{7} \tvfe^{2}\,{\rm K}
\label{eq:T3}
\ee
and compresses it to a density $n_{3} = 4n_{\rm 4}$, where $\mu$ is the mean molecular weight and $\tvfe \equiv \tvf/10^{8}$ cm s$^{-1}$.  We assume solar chemical composition with enhanced abundances as follows: [He/H]=0.08, [N/H]=1.7, [O/H]=1.3, [Ne/H]=1.9, [Mg/H]=0.7, [Fe/H]=0.7, typical of nova ejecta (e.g., \citealt{Schwarz+07}), which corresponds to $\mu = 0.76$.  However, our qualitative results are not sensitive to the precise abundances we have assumed.

Absent sources of external photo-ionization, gas well ahead of the forward shock is neutral due to the short timescale for radiative recombination.  The upstream is, however, exposed to ionizing UV and X-ray radiation from the shock, which penetrates gas ahead of the shock to a depth, $\Delta_{\rm ion}$.  The latter is set by the balance between photo-ionization and recombination, similar to an HII region (\citealt{Metzger+14}; Fig.~\ref{fig:shockzones}), and is approximately given by
\be
\Delta_{\rm ion} \approx 2\times 10^{14} f_{\rm EUV,-1}\left(\frac{n_{4}}{10^{7}\,{\rm cm^{-3}}}\right)^{-1}\tvfe^{3}\,{\rm cm},
\label{eq:delta}
\ee
where $f_{\rm EUV,-1} = f_{\rm EUV}/0.1$ is the fraction of the total shock power $L_{\rm sh} \propto n_4 v_{\rm sh}^{3}$ placed into ionizing radiation and absorbed by the neutral layer.  Although the neutral gas upstream of the shock effectively absorbs soft UV and X-ray photons, harder X-rays can escape from this region.  A minimum ionizing fraction of
\be f_{\rm EUV} \approx \frac{2 \, \mathrm{Ryd}}{kT_3} = 0.02 \, \tvfe^{-2}
\label{eq:feuvmin}
\ee 
is obtained in the limit that free-free emission is the sole source of ionization, where Ryd = 13.6 eV.  The value of $f_{\rm EUV}$ can in principle greatly exceed this minimum due to line emission and from the reprocessing of higher-energy photons to lower frequencies by the neutral gas ahead of the shock or in the central shell.\footnote{Ionizing X-rays from the white dwarf are likely blocked by the neutral central shell in the equatorial plane, although they may escape along the low density polar region (Fig.~\ref{fig:cartoon}).  Many novae are not detected as supersoft X-ray sources until after the early shock-powered radio emission has peaked (\citealt{Schwarz+11}).}   

\begin{figure}
\begin{center}
\includegraphics[width=1.0\linewidth]{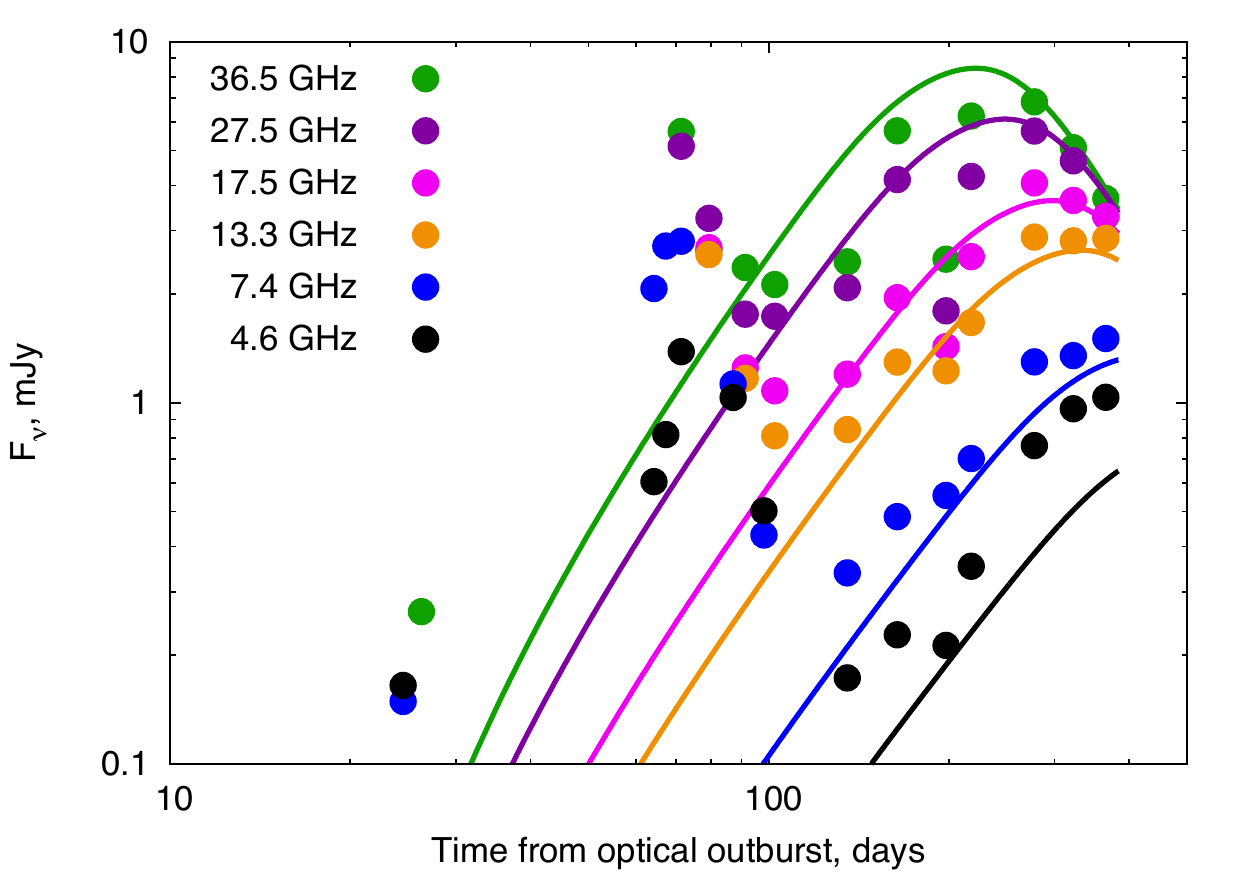}
\includegraphics[width=1.0\linewidth]{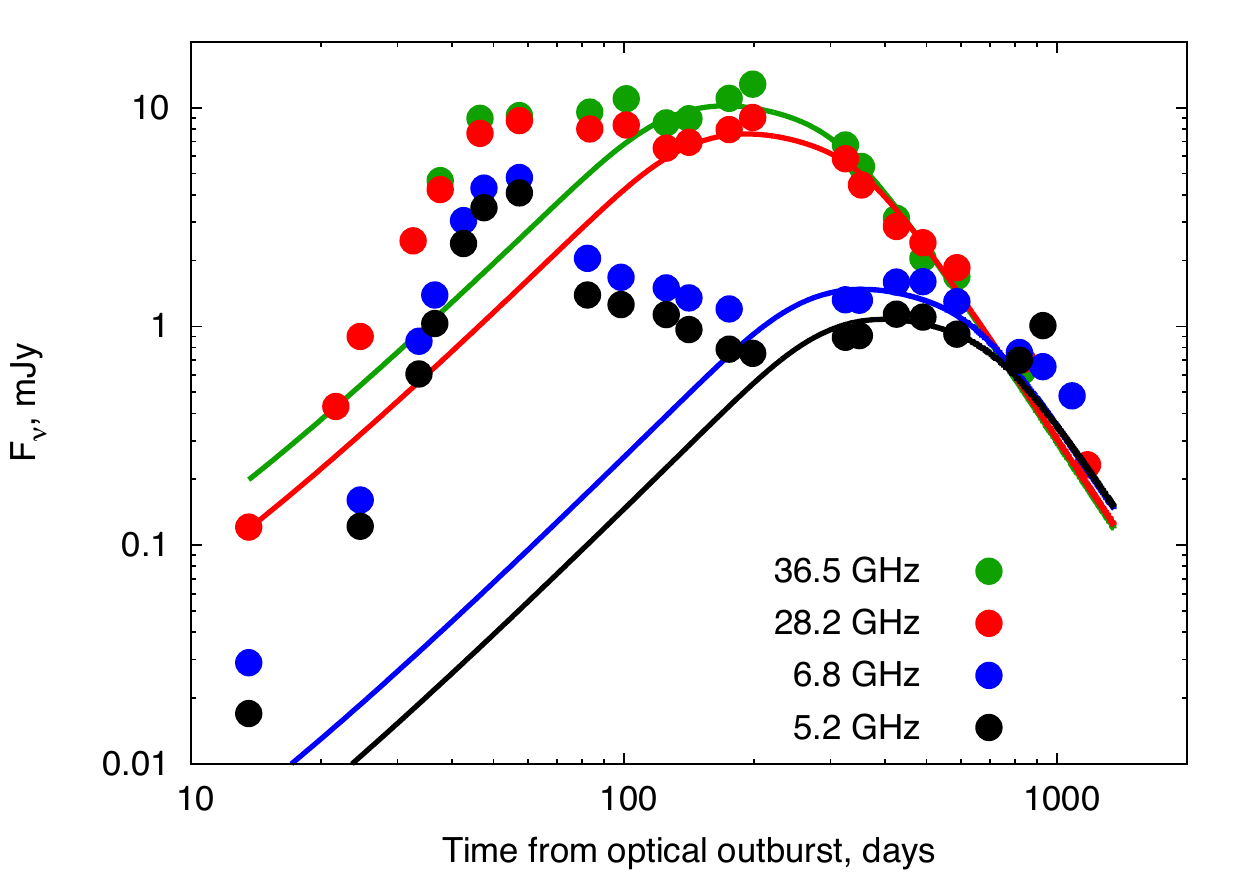}
\includegraphics[width=1.0\linewidth]{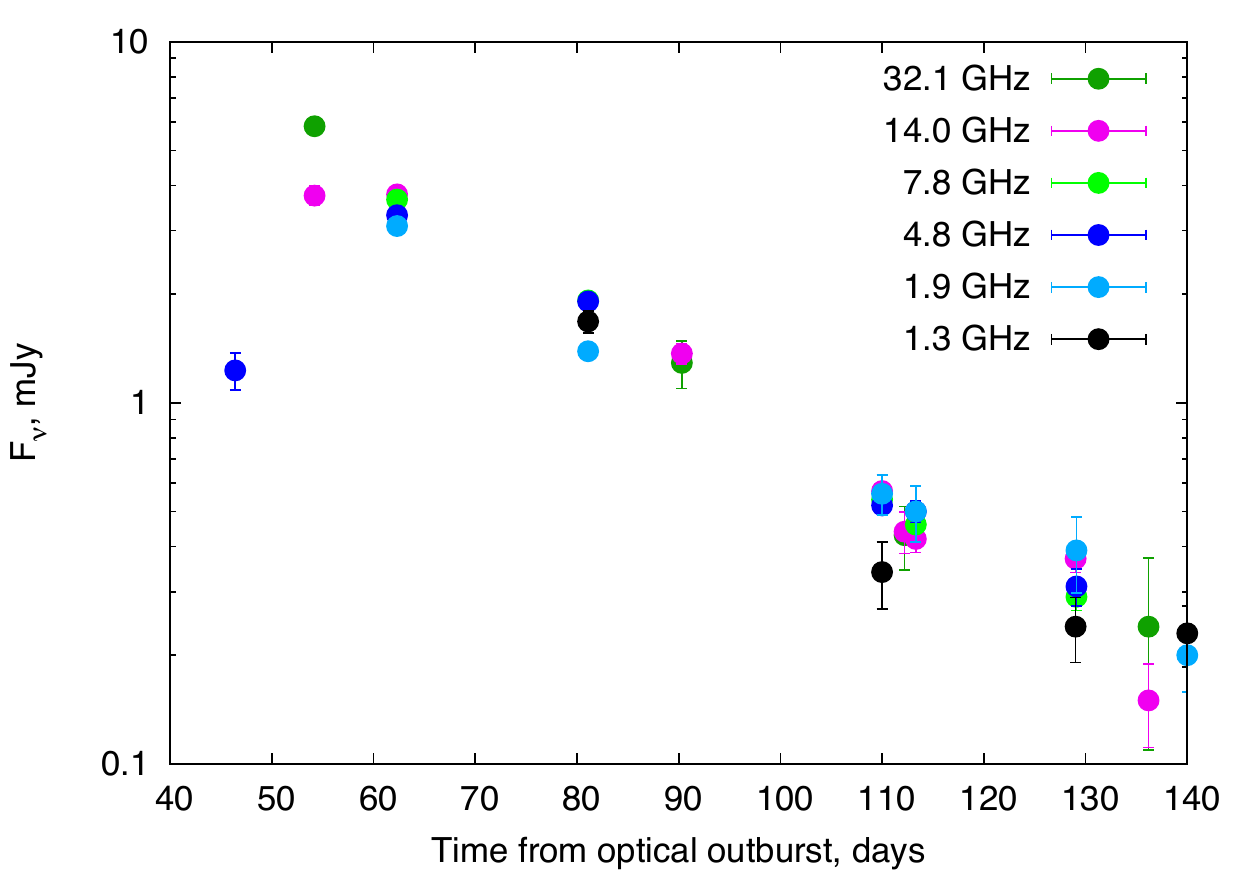}
\caption{Radio light curves of V1324 Sco (top panel; \citealt{Finzell+15}), V1723 Aql (middle panel; \citealt{Krauss+11}, \citealt{Weston+15a}), and V5589 Sgr (bottom panel; \citealt{Weston+15a}).  In V1723 Aql, only frequencies with full time coverage are shown.  In V1324 Sco and V1723 Aql, we show for comparison our best-fit thermal models (Table \ref{table:th}). In V1324 Sco and V1723 Aql, the maximum uncertainties in the data are 0.3 and 0.1 mJy, respectively, and hence the error bars are not discernible for most of the data points. In V5589 Sgr, we do not attempt to fit a thermal model, even though some of the emission after day 100 could in principle be thermal.}
\label{fig:fluxes}
\end{center}
\end{figure}

\begin{figure}
\begin{center}
\includegraphics[width=1.0\linewidth]{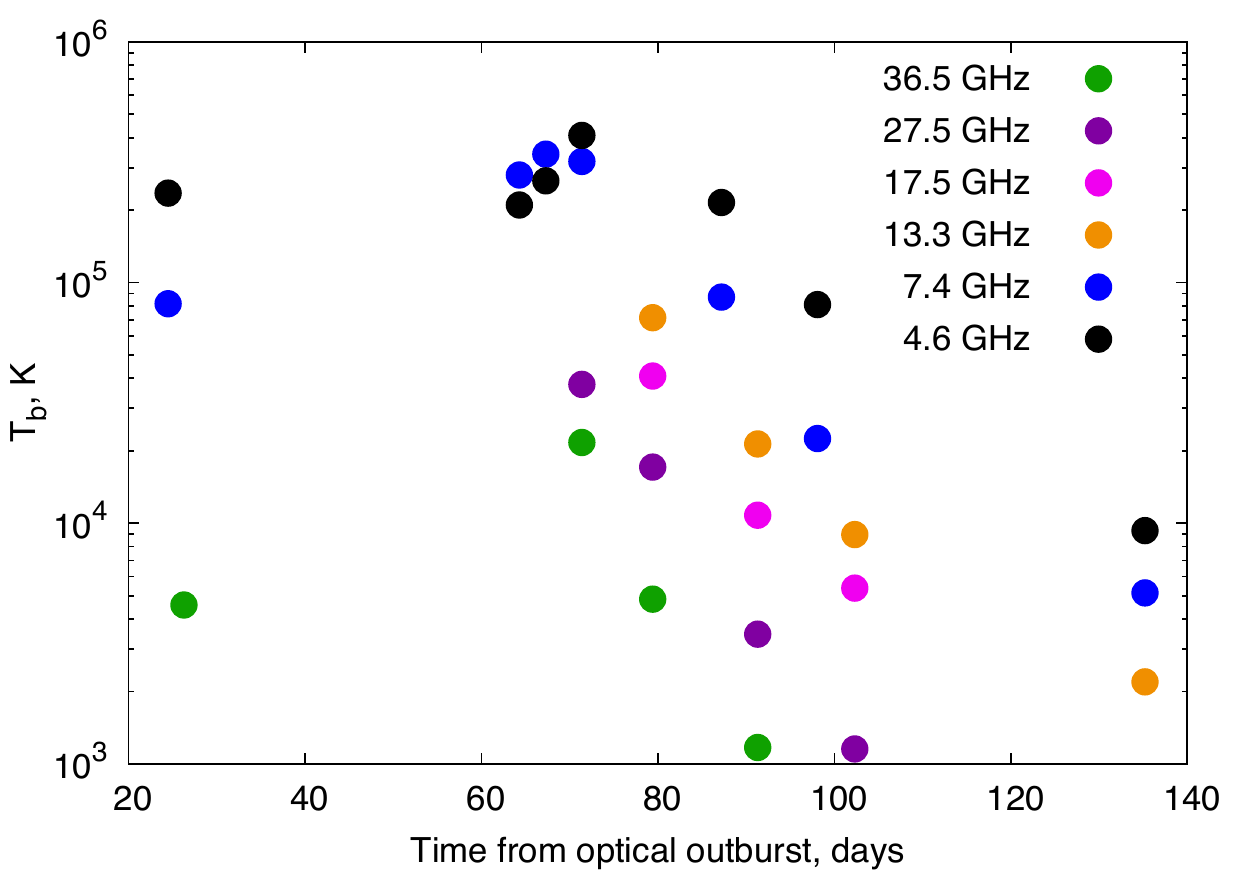}
\includegraphics[width=1.0\linewidth]{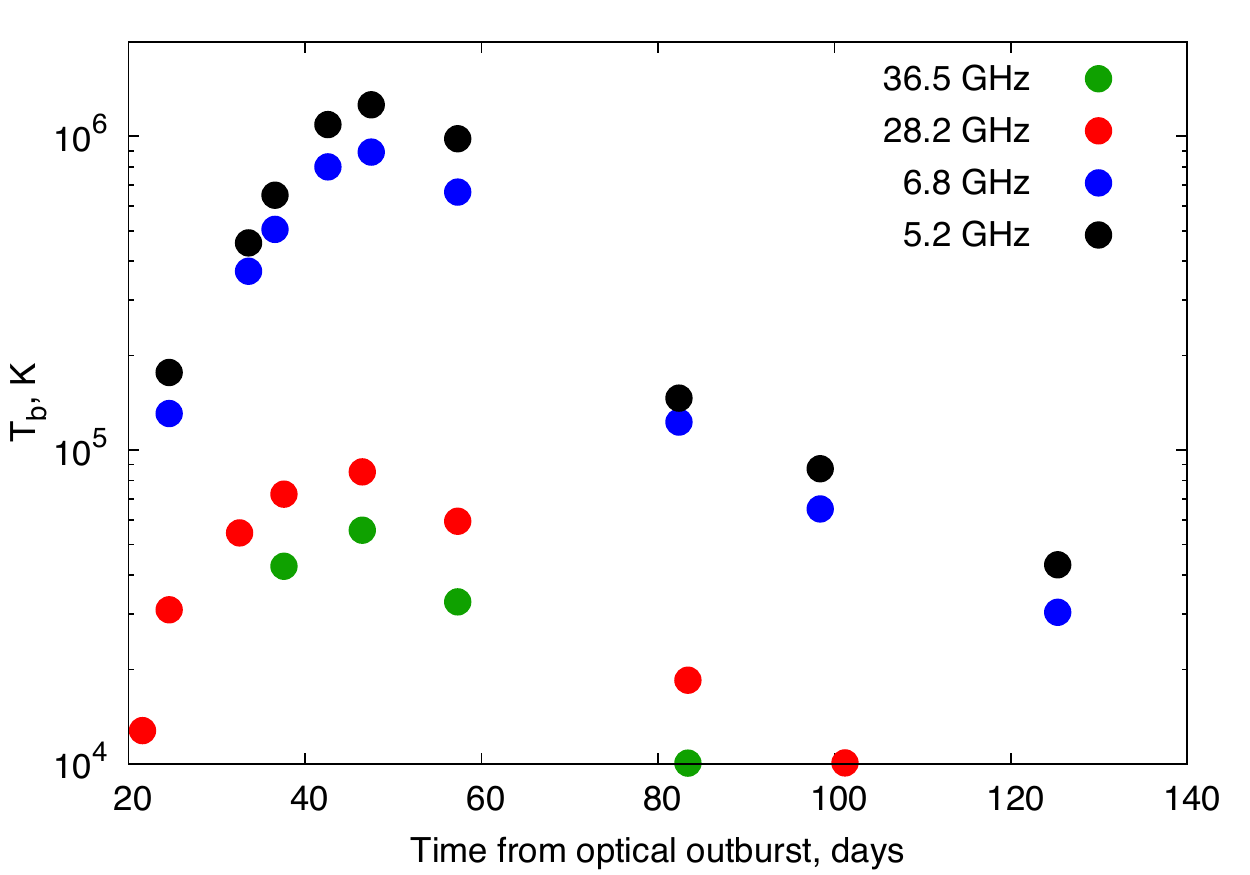}
\caption{Brightness temperature of the early component of the radio emission for V1324 Sco  (top panel) and V1723 Aquila (bottom panel), calculated by subtracting the thermal emission component off the raw fluxes (Fig.~\ref{fig:fluxes}) and assuming an emitting radius for the non-thermal emission corresponding to the fastest velocity of the ejecta inferred from the thermal fits.  In V1723 Aql, only frequencies with full time coverage are shown.  In V1324 Sco and V1723 Aql, the maximum uncertainties in the data are 0.3 and 0.1 mJy, respectively, and hence the error bars are not discernible for most of the data points. 
}
\label{fig:TB}
\end{center}
\end{figure}

\subsection{Evidence for Non-Thermal Radio Emission}
\label{sec:observations}
Figure \ref{fig:fluxes} shows the radio light curves of three novae with early-time shock signatures: V1324 Sco (\citealt{Finzell+15}), V1723 Aql (\citealt{Krauss+11}; \citealt{Weston+15a}), and V5589 Sgr (\citealt{Weston+15b}).  Thermal free-free emission from the photo-ionized nova ejecta peaks roughly a year after the outburst (\citealt{Seaquist&Bode08}).  In V1324 Sco and V1723 Aql, the light curve also shows a second, earlier peak on a characteristic timescale of a few months.  This early peak is also present in V5589 Sgr, although the late thermal peak in this event is only apparent at the highest radio frequencies (\citealt{Weston+15b}).

Thermal emission from an expanding sphere is initially characterized by an optically-thick, Rayleigh-Jeans spectrum $F_{\nu}\propto \nu^2$, with the total flux increasing with the surface area $\propto R^{2} \propto t^2$.  Then, starting at high frequencies, the ejecta begins to become optically thin and the radio lightcurve decays as the radio photosphere recedes back through the ejecta.  The shape of the radio lightcurve near its maximum depends on the density profile of the ejecta, which we take to be that of homologous expansion, $n \propto t^{-1}r^{-2}$.  Finally, after the entire ejecta becomes optically thin, the spectrum approaches that of optically thin free-free emission, $F_\nu\propto \nu^{-0.1}t^{-3}$.

 In order to isolate the non-thermal, shock-powered contribution to the radio emission, we first remove the thermal emission from the photo-ionized ejecta, using a model for the latter as outlined in \citet{BodeEvans08}.  We assume that shocks make no contribution to the emission at late times, $t\gtrsim 100-120$ days.  Our best-fit parameters for the ejecta mass of $M_{\rm ej} \approx 2-3\times 10^{-4}M_{\odot}$ and maximum ejecta velocities of $v_{\rm max} \approx 1300-1600$ km s$^{-1}$ are compiled in Table \ref{table:th}.  Our values are consistent with those found by \citet{Weston+15a} for V1723 Aql.  After subtracting the thermal component from the raw fluxes (Fig.~\ref{fig:fluxes}), the remainder should in principle contain only the shock-powered emission. 

Our model for the late thermal emission is admittedly simplified, as it assumes a homologous spherically symmetric and isothermal outflow.  However, we are primarily interested in modeling the shape of late thermal emission near the time of early radio peak, at around 80-100 days.  At this time the ejecta is still optically-thick and thus our only essential assumption is that the outer ejecta is isothermal and expanding ballistically. 

\begin{table}
\caption{Best-fit parameters for the thermal radio emission models.  The fits are based on observations at $t\gtrsim 100-120$ days in order to exclude contributions from the early radio peak.}
\label{table:th}
\begin{center}
\begin{tabular}{|c|c|c|c|c|c|c|} 
 \hline
 Nova & D & $v_{\rm min}^{(a)}$ & $v_{\rm max}^{(b)}$  & T$^{(c)}$ & $M_{\rm ej}^{(d)}$ & $\Delta t^{(e)}$  \\
 \hline
          &   kpc & km s$^{-1}$   &  km s$^{-1}$  &   K   &  $M_\odot$ & days \\
 \hline         
V1324 Sco & 7.8  & $550$  & 1300 &  $1.2\cdot 10^4$ & $\e{2.7}{-4}$ & 15 \\ 
\hline
V1723 Aq & 6.1 & $270$ & 1600 & $\e{1.0}{4}$  & $\e{2.2}{-4}$ & -2.2 \\
 \hline
\end{tabular}
\end{center}
$^{(a)}$Minimum velocity.  $^{(b)}$Maximum velocity.  $^{(c)}$Ejecta temperature.  $^{(d)}$Ejecta mass.  $^{(e)}$Time delay between outflow ejection and optical outburst.
\end{table}

\begin{table*}
\caption{Multiwavelength properties of novae}
\label{table:data}
\begin{tabular}{|c|c|c|c|c|c|c|c|c|} 
 \hline
 Name & D$^{\dagger}$ & $L_{\rm 10 GHz,\,pk}^{nth}$ & $L_{\rm 10 GHz,\,pk}^{th}$  & $L_{\rm X,\,pk}$ & $\frac{L_{\rm 10 GHz, pk}}{L_{\rm X,pk}}$ & $kT_{\rm pk}^{\ddagger}$  & $L_{\g,\,pk}$ & Orientation$^{\amalg}$  \\
 \hline
          &   {\rm (kpc)} & $({\rm erg\,s^{-1}})$   &  $({\rm erg\, s^{-1}})$  &   $({\rm erg \,s^{-1}})$  & -  & ({\rm keV}) & $({\rm erg\, s^{-1}})$ & -  \\
 \hline         
V1324 Sco & $6.5^{(l)}-8(7.8)$  & $\et{3.4}{30}{}^{(d)}$  & $\et{1.4}{30}{}^{(d)}$ &  $<10^{34}{}^{(h)}$ & $>\et{3}{-4}$  & - & $\et{2.7}{36}{}^{(m)}$ & ? \\
\hline
V1723 Aq & $5.3-6.1(6.1){}^{(e)}$ & $\et{2.5}{30}{}^{(d)}$ & $\et{6.3}{29}{}^{(d)}$ & $\et{1.2}{34}{}^{(f)}$ & $\et{2.1}{-4}$  & $1.8-3{}^{(f)}$  & - & I \\
\hline
V5589 Sgr & $3.2-4.6(4.0)^{(b)}$ & $\et{9}{29}{}^{(b)}$ & $<\et{3.6}{28}{}^{(b)}$ & $1.2\times 10^{34}{}^{(b)}$ & $\et{7.8}{-5}$ & $0.14-32.7^{(b)}$ & - & ? \\
\hline
V959 Mon & $1.0-1.8(1.4)^{(a)}$ & $<\et{1.1}{29}{}^{(c)}$   & $\et{1.8}{29}{}^{(c)}$ & $ \et{2.4}{33}{}^{(g)}$ & $<\et{5}{-5}$ & 3.2${}^{(g)}$  & $5.3\cdot 10^{34}{}^{(m)}$ & E \\
 \hline
V339 Del & $3.9-5.1(4.5)^{(i)}$ & $<\et{2.3}{28}{}^{(k)}$ & $\et{9.2}{29}{}^{(k)}$ & $\et{1.9}{33}{}^{(j)}$ & $<\et{1.2}{-4}$ & $>0.8^{(j)}$ & $\et{2.9}{35}{}^{(m)}$ & F
\end{tabular}\\
$^{\dagger}$Estimated uncertainty range of distance, followed in paranthesis by the fiducial value adopted in calculating luminosities; $^{\ddagger}$Temperature of X-ray emission; $^{\amalg}$Approximate inclination of binary (E = edge on; F = face on; I = intermediate; $?$ = unknown)  ${}^{(a)}$ \citealt{Linford+15} ${}^{(b)}$ \citealt{Weston+15b} ${}^{(c)}$ \citealt{Chomiuk+14b} ${}^{(d)}$ \citealt{Weston+13} ${}^{(e)}$ \citealt{Weston+15a} ${}^{(f)}$ \citealt{Krauss+11}, correct to unabsorbed value ${}^{(g)}$ \citealt{Nelson+12a} ${}^{(h)}$ for $kT \sim 2-25$ keV, \citealt{Page+12}, ${}^{(i)}$ \citealt{Munari+15} ${}^{(j)}$ \citealt{Page&Beardmore13} ${}^{(k)}$ Justin Linford, private communication ${}^{(l)}$ \citealt{Finzell+15} ${}^{(m)}$ \citealt{Ackermann+14} 
\end{table*}

Figure \ref{fig:TB} shows the brightness temperature $T_b$ of the early shock-powered radio emission, which we have calculated assuming that the shock-powered emission covers the entire solid angle of the outflow and originates from a radius equal to that of the fastest ejecta inferred from the thermal fit, $R = v_{\rm max}t$.  The brightness temperature peaks at values $\sim 10^{5}-10^{6}$ K which greatly exceed the temperature $\sim 10^4$ K of the photo-ionized ejecta inferred at late times (\citealt{Weston+15a}). 


The outer DES through which the forward shock is propagating at time $t$, can be modeled as an exponential density profile,
\be
 n_4=n_0 \exp[-\tvf (t-t_0)/H], \label{eq:rhopro}
\ee
where $n_{0}$ is a fiducial density at the time $t_0$ of the collision (eq.~[\ref{eq:R0}]) and $H$ defines the density scaleheight.  The light curves of V1324 Sco and V5589 Sgr exhibit a rapid post-maximum decline on a timescale of $t_{\rm fall} \sim$ weeks much shorter than the timescale of the collision $\sim t_0$ (Fig.~\ref{fig:fluxes}). This suggests that the forward shock is propagating down a steep radial gradient,
\be
H \sim \tvf t_{\rm fall} \approx 6\times 10^{13} \tvfe \left(\frac{t_{\rm fall}}{\rm week}\right)\,{\rm cm},
\label{eq:scaleheight}
\ee
which is smaller than the collision radius $R_0 \sim 10^{15}$ cm (eq.~[\ref{eq:R0}]).


\subsection{Condition for Radio Maximum \label{sec:radiopk}}

The detection of gamma-rays within days of the optical maximum (\citealt{Ackermann+14}) demonstrates that shocks are present even early in the nova eruption.  However, at such early times, the density of the shocked matter is highest, and radio emission from the shocks is absorbed by the photo-ionized gas ahead of the shock. 

The observed brightness temperature, $T_{b}$, is related to the unscreened value just ahead of the shock, $T_{\nu}|_{\rm shock}$, according to (\citealt{Metzger+14})
\be
  T_{b}=T_{\nu}|_{\rm shock}e^{-\tau_{\rm ff,4}}+(1-e^{-\tau_{\rm ff,4}})T_{4}, \label{eq:screening}
\ee
where $\tau_{\rm ff,4}=\alpha_{\rm ff,4}\Delta_{\rm ion}$ is the free-free optical depth of the ionized layer ahead of the shock of thickness $\Delta_{\rm ion}$ (eq.~[\ref{eq:delta}]) and
\be
\alpha_{\rm ff,4} \approx \alpha_0 \, \nu^{-2}T_{4}^{-3/2}n_{4}^{2}
\label{eq:alphaff}
\ee
is the free-free absorption coefficient 
 (\citealt{Rybicki&Lightman79}), where $T_4 \approx 2\times 10^{4}$ K is the temperature in the photo-ionized layer.  For conditions in photo-ionized layer, $\alpha_0=0.1\,\, {\rm cm}^{5}\, {\rm K}^{3/2}\,{\rm s}^{-2}$.

Although the {\it unscreened} flux $T_{\nu}$ is initially large, the emission reaching the observer is suppressed by a factor of $e^{-\tau_{\rm ff,4}} \ll 1$.  Shock-powered radio emission peaks at frequency $\nu$ once the shock propagates down the density gradient until (\citealt{Metzger+14})
\be
 \tau_{\rm ff,4}(\nu) = \alpha_{\rm ff,4}(\nu) \, \Delta_{\rm ion} \approx 1.
\label{eq:radiopeak}
\ee

Combining equations (\ref{eq:delta}) and (\ref{eq:radiopeak}), radio emission peaks once the upstream density decreases to a value of
\be
 n_{\rm pk} ={\rm max} \left\{\begin{array}{ll}
 n_{\rm pk,\Delta}\equiv 1 \times 10^{6} \, \nu_{10}^{2} \, f_{\rm EUV,-1}^{-1} \, \tvfe^{-3}\,{\rm cm^{-3}} \\
 n_{\rm pk,H}\equiv 5 \times 10^{6} \, \nu_{10} \, H_{14}^{-1/2}\,{\rm cm^{-3}}  \\
\end{array}  \right.
\label{eq:npk}
\ee
where $\nu =10\, \nu_{10}$ GHz, $H = H_{14}10^{14}$ cm, and the second line accounts for cases when the entire scaleheight is ionized ($\De_{\rm ion} > H$).  

The dependency $n_{\rm pk}\propto \nu^q$, where $q = 1-2$, illustrates that the time delay ($\De t_{\rm pk}$) between the maximum flux at different frequencies $\nu_{2}$ and $\nu_{1}$ is typically comparable to the rise time,
\be
 {\De t_{\rm pk}}=q\frac{H}{\tvf}\log\left(\frac{\nu_2}{\nu_1}\right) \label{eq:dtnu},
\ee
as was obtained by combining equations (\ref{eq:rhopro}) and (\ref{eq:npk}).

In summary, for characteristic parameters $\tvfe \sim 0.5-2$, $\nu \sim 3-30$ GHz, the radio light curve peaks when the shock reaches external densities of $n_{\rm pk} \sim 10^{6}-10^{8}$ cm$^{-3}$.  These are generally lower than the mean density of the shell at the time of the collision, $\bar{n}_0$ (eq.~[\ref{eq:nDES}]), and those required to produce the observed $\gamma-$ray emission \citep{Metzger+15}.  We now consider properties of the forward shock at the radio maximum, defined by $n_4 = n_{\rm pk}$.    

\subsection{Is the Forward Shock Radiative or Adiabatic?} 
\label{sec:radshock}

\begin{figure}
\begin{center}
\includegraphics[width=1.0\linewidth]{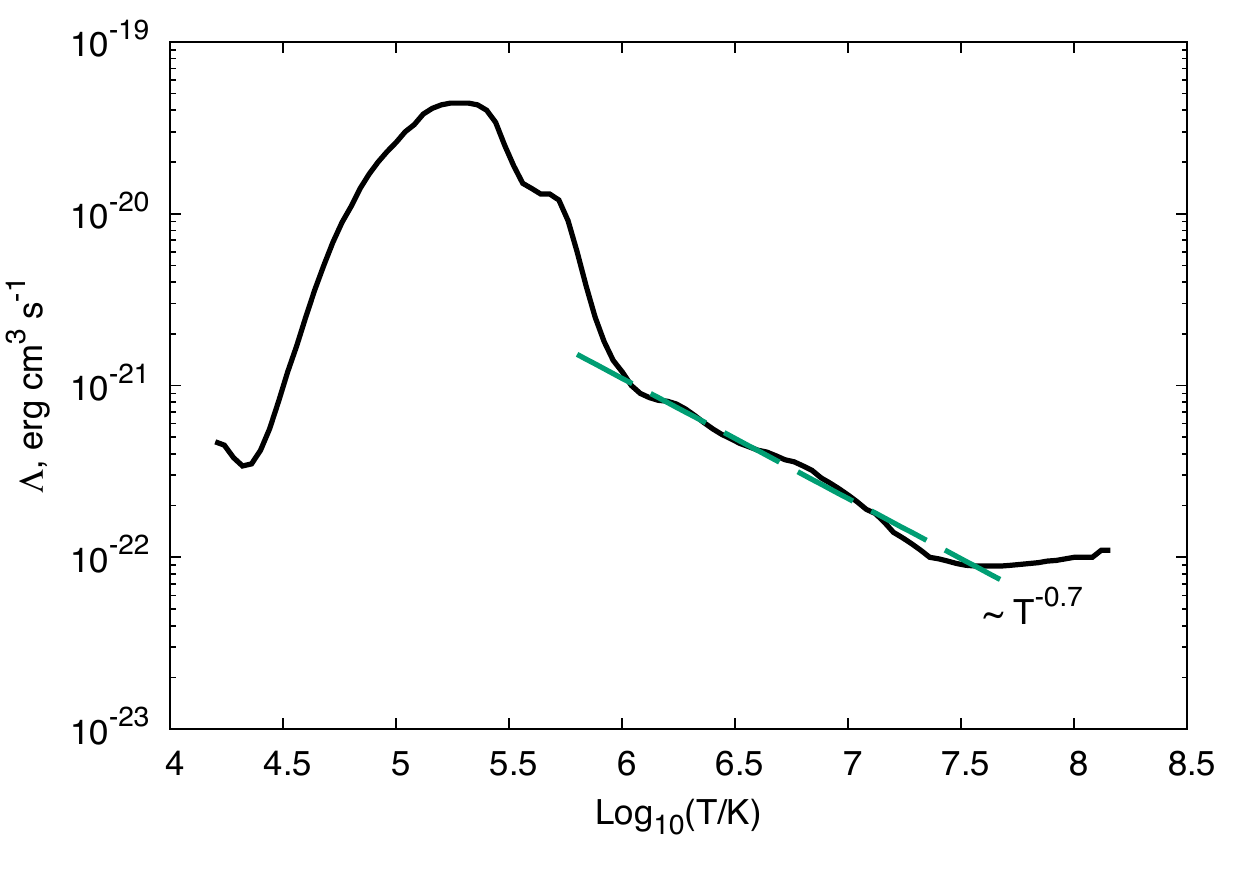}
\caption{Cooling function $\Lambda(T)$ from \citet{Schure+09}, adopted for our fiducial ejecta composition enriched in CNO elements. A green line shows the power-law $\Lambda \propto T^{-0.7}$ (eq.~[\ref{eq:lambda}]).}
\label{fig:lambda}
\end{center}
\end{figure}

Whether the shock is radiative or adiabatic near the radio peak depends on the cooling timescale of the post-shock gas, 
\be
t_{\rm cool} = \frac{3kT_{\rm 3}}{2\mu \Lambda n_{\rm 3, pk}},
\label{eq:tcool}
\ee 
where $n_{\rm 3, pk} = 4n_{\rm pk}$ (eq.~[\ref{eq:npk}]) and 
\be
\Lambda = \Lambda_{\rm lines} + \Lambda_{\rm ff}
\ee
is the cooling function.  The latter receives contributions from emission lines, $\Lambda_{\rm lines}$, and from free-free emission 
\be
 \Lambda_{\rm ff} \approx 3\times 10^{-27}(T/{\rm K})^{1/2}\,\,\, {\rm erg\,\, cm}^{3}\,\, {\rm s}^{-1}. \label{eq:lambdaff}
\ee
Figure \ref{fig:lambda} shows tabulated cooling function from \citet{Schure+09} for our assumed ejecta abundances and power-law approximation to the line cooling rate,
\be
 \Lambda_{\rm lines} \approx
  10^{-22}\Lambda_{c,-22}\left(\frac{T}{1.7\times 10^7{\rm K}}\right)^\de {\rm erg\,cm^{3}\,s^{-1}},
 \label{eq:lambda} 
\ee
with $\Lambda_{c,-22}=1.5, \de=-0.7$.  The cooling rate is larger than for solar metallicity gas due to the enhancements of CNO elements.
 
The ratio of the cooling timescale (eq.~[\ref{eq:tcool}]) at the time of radio maximum to the characteristic light curve decay timescale, $t_{\rm fall}=H/\tvf$ (eq.~[\ref{eq:scaleheight}]) is given by
\be
\eta_3 &\equiv& \left.\frac{t_{\rm cool}}{t_{\rm fall}}\right|_{n_{4} = n_{\rm pk}} = \frac{3kT_{3}\tvf }{8 \mu \Lambda n_{\rm pk}H}
\approx
\text{min}\left\{\begin{array}{ll}
4.8 \,
\tvfe^{7.4}H_{14}^{-1}f_{\rm EUV,-1}\nu_{10}^{-2}\\
1.3 \,
\tvfe^{4.4}H_{14}^{-1/2}\nu_{10}^{-1}\\
\end{array}  \right.
\label{eq:radiative}
\ee
where the final equality uses equations (\ref{eq:T3}), (\ref{eq:npk}).  Radiative shocks ($t_{\rm cool} \ll t_{\rm fall}$) and adiabatic shocks ($t_{\rm cool} \gg t_{\rm fall}$) are divided sharply at $\tvf \approx 1000$ km s$^{-1}$, due to the sensitive dependence of $t_{\rm cool}/t_{\rm fall}$ on the shock velocity.\footnote{Note that $\eta$ also equals the ratio of the post-shock cooling length 
\be
L_{\rm cool}=\tvf t_{\rm cool} = \eta_{3}H
 \label{eq:Lcool}
\ee to the scale-height, such that the condition for radiative shocks ($\eta_3 \ll 1$) can also be written as $L_{\rm cool} \ll H$.}

As we will discuss, in a radiative shock, only the immediate post-shock material contributes to the thermal or non-thermal emission, allowing for faster light curve evolution.  The rapid post-maximum decline of V1324 Sco and V5589 Sgr ($t_{\rm fall} \ll t$) is consistent with radiative shocks in these systems (\citealt{Metzger+14}). 



\subsection{Thermal X-ray Emission} \label{sec:xraysintro}
\begin{figure}
\begin{center}
\includegraphics[width=1.0\linewidth]{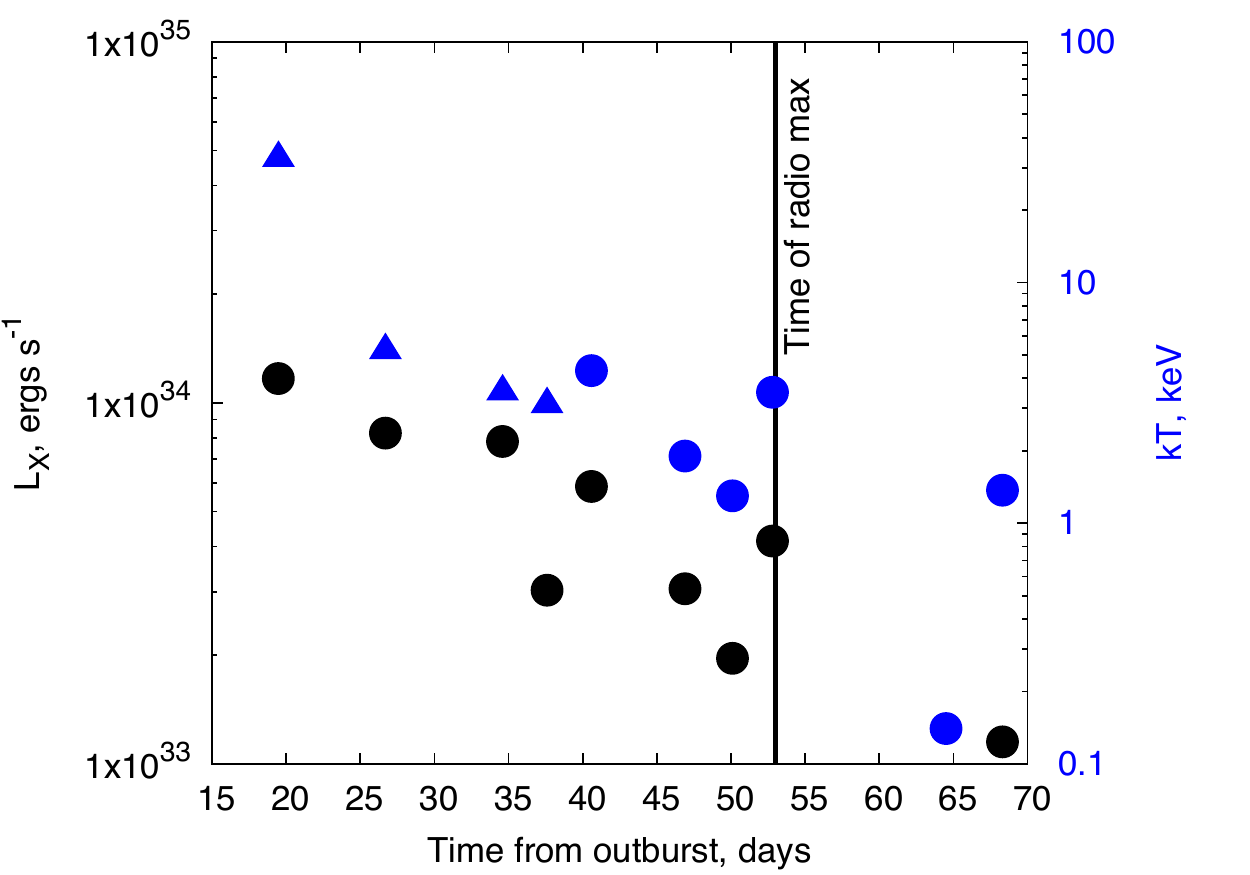}
\caption{X-ray luminosities (black circles; left axis) and temperature $kT$ (blue circles; right axis) of V5589 Sgr as measured by {\it Swift} XRT.  Blue triangles indicate temperature lower limits.}
\label{fig:Lxdata}
\end{center}
\end{figure}

Thermal X-rays are diagnostic of the shock velocity and hence of their radiative nature.  Figure \ref{fig:Lxdata} shows that the X-ray light curve of  V5589 Sgr peaks at a luminosity of $L_X \sim 10^{34}$ erg s$^{-1}$ a few weeks after the optical outburst (\citealt{Weston+15b}).  The temperature is very high $kT \gtrsim 30$ keV initially, decreasing to $\sim 1$ keV by the radio peak around day 50.  Large velocities $v_8 \gg 1$ are required to produce the high X-ray temperatures at the earliest times, clearly implicating adiabatic shocks.  However, the X-ray and radio emission may not originate from the same shocks, while even in a given band different shocks may dominate the emission at different times.  

The X-ray luminosity of the forward shock is given by (\citealt{Metzger+14}; their eq.~[38])
\begin{eqnarray}
L_X&\approx&\frac{4\pi f_\Omega R_{\rm sh}^2\tvf}{1+5\eta_{3}/2} \left(\frac{3}{2}\frac{n_4}{\mu}kT_3\right)\left(\frac{\Lambda_{\rm ff}}{\Lambda}\right)
 \label{eq:Lx}
\end{eqnarray}
where $f_{\Omega}$ is the fraction of the solid angle of the outflow covered by shocks.  The factor $(1+5\eta_3/2)^{-1}$ accounts for the radiative efficiency of the shock and the factor $\Lambda_{\rm ff}/\Lambda$ accounts for the fraction of the shock power emitted as free-free emission in the X-ray band (line emission occurs primarily in the UV).

X-ray emission can be attenuated by bound-free absorption.  For our adopted composition we find that the bound-free opacity is reasonably approximated as $\kappa \approx 2000(E_X/{\rm keV})^{-2}$ cm$^{2}$  g$^{-1}$ across the range of X-ray energies $E_X$ of interest \citep{Verner+96}.  The X-ray optical depth at the energies corresponding to the peak of the free-free emission ($E_{X} = kT_{3} \approx 1.4 \tvfe^{2}$ keV) is thus given by
\begin{eqnarray}
\tau_{X} \approx 3.3\left(\frac{E_{X}}{\rm keV}\right)^{-2}\left(\frac{n_{\rm 4}}{10^{7}\,{\rm cm^{-3}}}\right)H_{14} \approx 1.8v_{8}^{-4}\left(\frac{n_{\rm 4}}{10^{7}\,{\rm cm^{-3}}}\right)H_{14},  
\label{eq:tauX}
\end{eqnarray}
Analogous to the radio light curve, the X-ray luminosity from the shock propagating down a density gradient peaks at $\tau_X \sim 1$.  This occurs once the density of the pre-shock gas decreases to a value of
\be
n_{\rm X} \approx 6\times 10^{6}H_{14}^{-1}v_{8}^{4}\,{\rm cm^{-3}}.
\label{eq:nX}
\ee

\begin{figure}
\begin{center}
\includegraphics[width=1.0\linewidth]{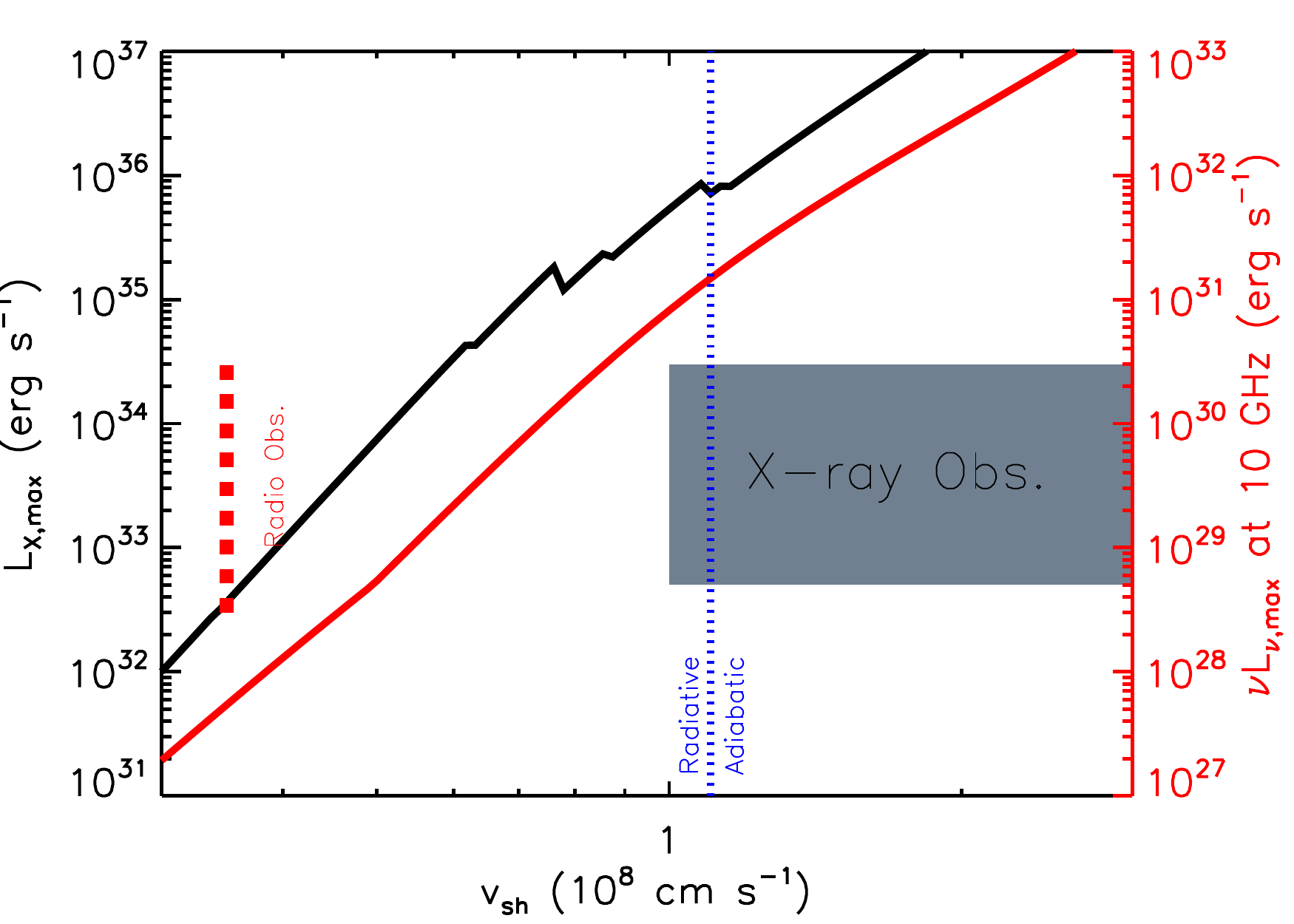}
\caption{Thermal X-ray luminosity at the time of X-ray peak (black line - left axis; $n = n_x$) and 10 GHz radio luminosity $\nu L_{\nu}$ at the time of non-thermal radio peak (red line - right axis; $n = n_{\rm pk}$) as a function of shock velocity.  Both luminosities are maximum allowed values, because they are calculated assuming $\tau_X = 1$ and $\tau_{\rm \nu = 10 \,GHz} = 1$, respectively.  We have assumed fiducial parameters for the density scale-height $H = 10^{14}$ cm and shock microphysical parameters $\epsilon_e = \epsilon_B = 0.01$, and a common radius $R_{\rm sh} = 10^{15}$ cm and covering fraction $f_{\Omega} = 1$ of the shock.  Also shown is the division between radiative and adiabatic shocks (vertical dashed blue line) and the range of observed radio luminosities (dashed red line) and X-ray luminosities and shock velocities corresponding to the observed X-ray temperature (grey region; Table \ref{table:data}, Fig.~\ref{fig:Lxdata}).   }
\label{fig:LradLX}
\end{center}
\end{figure}

Figure \ref{fig:LradLX} shows the maximum X-ray luminosity as a function of the shock velocity, obtained by combining $n_X$ with equation (\ref{eq:Lx}) for $f_{\Omega} = 1$ and assuming a shock radius of $R_{\rm sh} \sim 10^{15}$ cm.  Although this maximum value is calculated assuming that the density of the shocked matter equals or exceeds $n_X$, this assumption is unlikely to be valid for the highest shock velocities because the density required to achieve $\tau_X \gtrsim 1$ becomes unphysically high.

For high velocity shocks ($v_8 \gtrsim 1$), the predicted X-ray luminosity is several orders of magnitude greater than the observed range in classical novae, $L_{X} \sim 10^{33}-10^{35}$ erg s$^{-1}$ (\citealt{Mukai&Ishida01, Mukai+08, Osborne15}; see Table \ref{table:data}).  This indicates that the observed X-ray producing shocks either (a) cover a small fraction of the outflow solid angle $f_{\Omega} \sim 0.01-0.1$, or (b) are produced in regions of the outflow with much lower densities, $n \ll n_X$, than the mean values $\bar{n} \sim 10^{7}-10^{9}$ cm$^{-3}$ (eq.~[\ref{eq:nDES}]).  


The second condition is challenging to satisfy at early times when the densities are probably highest, while both conditions are at odds with the high covering fractions $f_{\Omega} \sim 1$ and high shock densities needed to power the gamma-ray luminosities (\citealt{Metzger+15}).  We therefore postulate that the hard $\gg$ keV X-ray emission may not originate from the same shocks responsible for the gamma-ray and non-thermal radio peak, but instead from the low density polar region  (Fig.~\ref{fig:cartoon}). 

In such a scenario, the early radio peak could be powered either by the same fast polar shocks, or by lower velocity ($v_{8} \lesssim 1$) radiative shocks in the higher density equatorial region (Fig.~\ref{fig:cartoon}).  
In the latter case the radio-producing shocks still produce thermal X-rays; however, being comparable in brightness, yet much softer in energy, they may not be readily observable.  This would also be true if the X-ray luminosity of radiative shocks is suppressed due to the role of thin-shell instabilities (\citealt{Kee+14}).


\section{Synchrotron Radio Emission}
\label{sec:synch}

Amplification of the magnetic field is required to produce the observed synchrotron emission and is expected to result from instabilities driven by cosmic ray currents penetrating the upstream region (e.g.~\citealt{Bell04}; \citealt{Caprioli&Spitkovsky14b}).  The strength of the post shock magnetic field can be estimated as
\be 
 B_{\rm sh}= (24\pi P_3 \epsilon_B)^{1/2} \approx 0.22 n_{4,7}^{1/2} \tvfe \eps_{B,-2}^{1/2}
 \, {\rm G}, \label{eq:Bgrowth}
\ee
where $n_{4,7} \equiv n_{4}/10^{7}$ cm$^{-3}$, $\eps_B = 10^{-2}\eps_{B,-2}$ is the fraction of the post-shock energy density in the magnetic field.  Substituting $n_4=n_{\rm pk} \approx n_{\rm pk,\Delta}$ (eq.~[\ref{eq:npk}]) into equation (\ref{eq:Bgrowth}) gives the field strength when the radio emission reaches its peak,
\be
B_{\rm pk} =
8.2\times 10^{-2} \,
\eps_{B,-2}^{1/2}f_{\rm EUV,-1}^{-1/2}\tvfe^{-1/2}\nu_{10}\,{\rm G}.
\ee

An electron or positron of Lorentz factor $\gamma$ produces synchrotron emission mostly near a characteristic frequency
\be
\nu_{\rm syn}  = \frac{1}{2\pi}\frac{eB\gamma^{2}}{m_e c} \approx
92\,{\rm GHz}\,\,\eps_{B,-2}^{1/2}\tvfe^{-1/2}f_{\rm EUV,-1}^{-1/2}\nu_{10}
\left(\frac{\gamma}{{200}}\right)^{2}.
\ee
There thus exists a special electron Lorentz factor,
\be
\gamma_{\rm pk} \approx
210 \,\, \eps_{B,-2}^{-1/4}\tvfe^{1/4}f_{\rm EUV,-1}^{1/4}, \label{eq:gammapk}
\ee
which determines the peak emission ($\nu_{\rm syn} = \nu$) and, remarkably, is independent of the observing frequency.\footnote{The `Razin' effect suppresses synchrotron radiation below a critical frequency $\nu_{R} = \nu_{p}\gamma$ (\citealt{Rybicki&Lightman79}), where $\nu_p = (n_{\rm pk}e^{2}/\pi m_{\rm e} )^{1/2}$.  For $\gamma = \gamma_{\rm pk}$ we thus have $\nu/\nu_{\rm R}=  4.5 \, \eps_{B,-2}^{1/4}\tvfe^{5/4}f_{\rm EUV,-1}^{1/4}$.  Since for parameters of interest we have $\nu \gtrsim \nu_{R}$, modifications from the Razin effect should be weak and are hereafter neglected.}

\subsection{Leptonic Emission}

The relativistic leptons responsible for synchrotron radiation originate either from the direct acceleration of electrons via diffusive shock acceleration (e.g., \citealt{Blandford&Ostriker78}), or from electron-positron pairs produced by the decay of charged pions from inelastic proton-proton collisions.  Both theory and simulations predict the spectrum of accelerated electrons $(dN/dE)dE \propto E^{-p}dE$, with $p \gtrsim 2$ (when $E\gg m_ec^2$) in the case of strong shocks (Figure \ref{fig:piplspem}, top panel).  Two key parameters are the fractions of the shock kinetic power placed into non-thermal electrons and ions, $\eps_{\rm e}$ and $\eps_{\rm p}$, respectively.  

Particle-in-cell (PIC) and hybrid kinetic simulations (e.g., \citealt{Wolff&Tautz15}, \citealt{Caprioli&Spitkovsky14a}, \citealt{Caprioli&Spitkovsky14b}, \citealt{Caprioli&Spitkovsky14c}, \citealt{Kato14}, \citealt{Park+14}) indicate that the values of $\eps_{\rm e}$ and $\eps_{\rm p}$ depend on the Mach number of the shock, and the strength and the inclination angle $\theta$ of the upstream magnetic field with respect to the shock normal.  Shocks which propagate nearly perpendicular to the direction of the magnetic field ($\theta \approx 90^{\circ}$) do not efficiently accelerate protons or electrons  (\citealt{Riquelme&Spitkovsky11}), while those propagating nearly parallel to the magnetic field ($\theta \approx 0^{\circ}$) accelerate both (\citealt{Kato14}; \citealt{Park+14}). 

What is the geometry of the magnetic field in the ejecta?  On timescales of months, the ejecta has expanded several orders of magnitude from its initial size at the base of the outflow (\citealt{Metzger+15}).  This expansion both dilutes the magnetic field strength via flux freezing and stretches the field geometry to be perpendicular to the radial direction in which the shocks are likely propagating.  Based on the above discussion, such a geometry would appear to disfavor hadronic scenarios.  However, leptonic scenarios are also strained because the values of $\eps_e \gtrsim 10^{-2}$ required to explain the $\gamma$-ray emission in V1324 Sco \citep{Metzger+15} greatly exceed the electron acceleration efficiencies seen in current numerical simulations (\citealt{Kato14}, \citealt{Park+14}).


Due to global asymmetries, or inhomogeneities in the shocked gas caused by radiative instabilities (\citealt{Metzger+15}), the shocks may not propagate perpendicular to the magnetic field everywhere, allowing hadronic acceleration to operate across a fraction of the shock surface.  This possibility motivates considering alternative, hadronic sources for the radio-emitting leptons. 


\subsection{Hadronic Emission}

 Relativistic protons accelerated near the shock collide with thermal protons in the upstream or downstream regions, producing neutral ($\pi^{0}$) and charged ($\pi^-$, $\pi^+$) pions.  The former decay directly into $\gamma-$rays, while $\pi^{\pm}$ decay into neutrinos and muons, which in turn decay into relativistic $e^{\pm}$ pairs and neutrinos.  In addition to electrons accelerated directly at the shock, these secondary pairs may contribute to the observed radio emission.

Figure \ref{fig:piplspem} shows the $e^{\pm}$ spectrum from pion decay, calculated for a flat input proton energy spectrum ($p = 2$).  Radio emission near the time of peak flux is produced by electrons of energy $\g_{\rm pk}m_e c^{2} \approx 100$ MeV, where $\g_{\rm pk}\approx 200$ (eq.~[\ref{eq:gammapk}]).  By coincidence, this is close to the pion rest energy $m_{\pi} c^{2} \approx 140$~MeV and hence to the peak of the $e^{\pm}$ distribution produce by pion decay.  Thus, secondary pairs can contribute significantly to the radio emission {\it if they are deposited in regions where their radiation is observable}.  

Although pion-producing collisions occur both ahead of the shock and in the post-shock cooling layer, pairs produced in the cold central shell cannot contribute to the radio emission due to the high free-free optical depth in this region (see Fig.~\ref{fig:taut}). Only $e^{\pm}$ pairs produced in the first cooling length behind the shock contribute to the observable radio emission, but these contain only a small fraction of the shock power,
\be 
\eps_{\rm e} \sim
 \eps_{\rm p} \, \frac{t_{\rm cool}}{t_{\pi}} \, f_\pi=
  4\times 10^{-4} \,
  (\eps_{\rm p}/0.1) (f_{\pi}/0.1) \tvfe^{3.4},
 \label{eq:eehadron}
\ee
where $f_{\pi}$ is the fraction of proton energy per inelastic collision placed into pions, $t_{\pi} = (n_3 \sigma_{\pi}c)^{-1}$ is the timescale for pion production, and $\sigma_{\pi } \approx 4\times 10^{-26}$ cm$^{2}$ is the characteristic inelastic p-p cross section (\citealt{Kamae+06}).

Pairs are also produced by p-p collisions {\it upstream} of the shock, which then radiate after being advected into the magnetized downstream.  However, proton acceleration via DSA is confined to the narrow photo-ionized layer ahead of the shock (\citealt{Metzger+16}), the narrow radial thickness of which, $\Delta_{\rm ion}$ (eq.~[\ref{eq:delta}]), limits the radial extent of the pion production region ahead of the shock.\footnote{Very high energy protons of energy $\sim E_{\rm max} \gtrsim 10$ GeV leak out of the DSA cycle, escaping into the neutral upstream with a nearly mono-energetic distribution with a comparable energy flux to the power-law spectrum ultimately advected downstream (e.g.~\citealt{Caprioli&Spitkovsky14a}).  However, the upstream-stream protons inject pairs at multi-GeV energies $\gg \gamma_{\rm pk} m_e c^{2}$ (eq.~[\ref{eq:gammapk}]) which radiate most of their synchrotron emission at higher frequencies than that responsible for the bulk of the radio emission.}

\section{Radio Emission Model}
\label{sec:model}

\subsection{Pressure Evolution of the Post-Shock Gas \label{sec:pev}}
Gas compresses behind the shock due to radiative cooling at approximately constant pressure (neglecting the effect of thermal instabilities; \citealt{Chevalier&Imamura82}).  The total pressure includes both the thermal gas and relativistic ions (`cosmic rays'),
\be
 P_{\rm tot}=P_{\rm th}+P_{\rm CR}=P_{\rm th,0}\left(\frac{n}{n_3}\right)\left(\frac{T}{T_3}\right)+P_{\rm CR,0}\ratdeg{n}{n_3}{4/3},
\label{eq:Ptot}
\ee


The energy fraction which goes into relativistic ions is twice the pressure ratio between relativistic ions and total pressure $\eps_{\rm p} = 2P_{\rm CR,0}/(P_{\rm th,0}+P_{\rm CR,0})$. Our numerical calculations use $\eps_p=0.2$, although our results are not sensitive to this assumption because the observed emission is dominated by that originating within the first cooling length behind the shock, i.e. before compression becomes important.

\subsection{Electron and Positron Spectra}
 \label{sec:relel}

The electron and $e^{\pm}$ energy spectra, $N(E)$, in both leptonic and hadronic scenarios are normalized such that a fraction $\eps_e$ of shock power is placed into relativistic electrons.  In the leptonic scenario, we assume that 
\be
N(E) {= C_E} E^{-p},\,\,\,E_{\rm min} < E < E_{\rm max} \label{eq:CEdef}
\ee
where $E = \gamma m_e c^{2}$, $E_{\rm min} = 2 m_e c^{2} \approx 10^{6}$ eV and $E_{\rm max} = 10^{12}$ eV (\citealt{Metzger+16}).  The normalization constant is given by
\be
 C_{E}=
{
  \frac{9 m_p n_4 \tvf^2 \eps_e}{8}\ratdeg{n}{n_3}{\frac{p+2}{3}} \times
  \begin{cases}
 \dfrac{(p-2) E_{\rm min}^{p-2}}{1 - (E_{\rm max}/E_{\rm min})^{-p+2}}, 	& p\ne 2 \\
 [\ln(E_{\rm max}/E_{\rm min})]^{-1}, 					& p= 2,
 \end{cases}
}
\label{eq:CE}
\ee
where the term $\propto (n/n_3)^{(p+2)/3}$ accounts for adiabatic heating as gas compresses downstream.  
As shown in Appendix $\ref{sec:timescales}$, most sources of cooling of relativistic electrons in the post shock thermal cooling layer are negligible, with the possible exception of Coulomb losses, which we include in a few select cases as described below.

 In the hadronic scenario, we model the spectrum of $e^{\pm}$ pairs from pion decay following \citet{Kamae+06}, assuming an input power-law spectrum of relativistic protons $N_p(E) \propto E^{-p}$ with the same energy distribution in the leptonic case.

\subsection{Radiative Transfer Equation}
\label{sec:TBevo}

The radio emission downstream of the shock is governed by the radiative transfer equation,
which can be conveniently written in terms of the brightness temperature $T_{\nu}$ (\citealt{Rybicki&Lightman79}),
\be
\frac{dT_{\nu}}{d\tau_{\nu}} = T - T_{\nu} + T_{\nu, \rm syn},
\label{eq:TBevo}
\ee
where $T$ is the gas temperature,
\be
 T_{\nu, \rm syn} = \frac{j_{\nu, \rm syn} c^{2}}{2\alpha_{\rm ff} k\nu^{2}}.
\label{eq:Tsyn_1}
\ee
is the synchrotron brightness temperature, and
$\tau_{\nu} = \int\alpha_{\rm ff}dz$ is the free-free optical depth.  Here $z$ is the depth behind the shock and $\alpha_{\rm ff}$ is the free-free absorption coefficient (eq.~[\ref{eq:alphaff}]), where the relevant pre-factor in is now $\alpha_0=0.28 \,\, {\rm cm}^{5}\, {\rm K}^{3/2}\,{\rm s}^{-2}$ for the higher temperatures appropriate to the post-shock gas.  We neglect synchrotron self-absorption, which is only relevant for relativistic brightness temperatures of $T_{\nu} \gtrsim 10^{11}$~K, much higher than those observed in novae.

It is useful to change variables from free-free optical depth $\tau_{\nu}$ to temperature, according to (\citealt{Metzger+14})
\be
 d\tau_\nu=\alpha_{\rm ff}\frac{dT}{dT/dz}=\frac{5k\alpha_0 n_4 \tilde v_{f}}{2\nu^2 T^{3/2}\Lambda(T)} dT \label{eq:dtau}
\ee
using the temperature gradient $dT/dz = 2n^{2}\Lambda(T)/({5k}n_{4}\tvf)$ behind the shock set by cooling.  As both the free-free absorption coefficient and the cooling rate scale as $\propto n^{2}$, this allows for a one-to-one (up to a coefficient) mapping $\tau_{\nu}(T)$ given $\Lambda(T)$.  This mapping is notably independent of the details of how the gas compresses behind the shock.

\begin{figure}
\begin{center}
\includegraphics[width=1.0\linewidth]{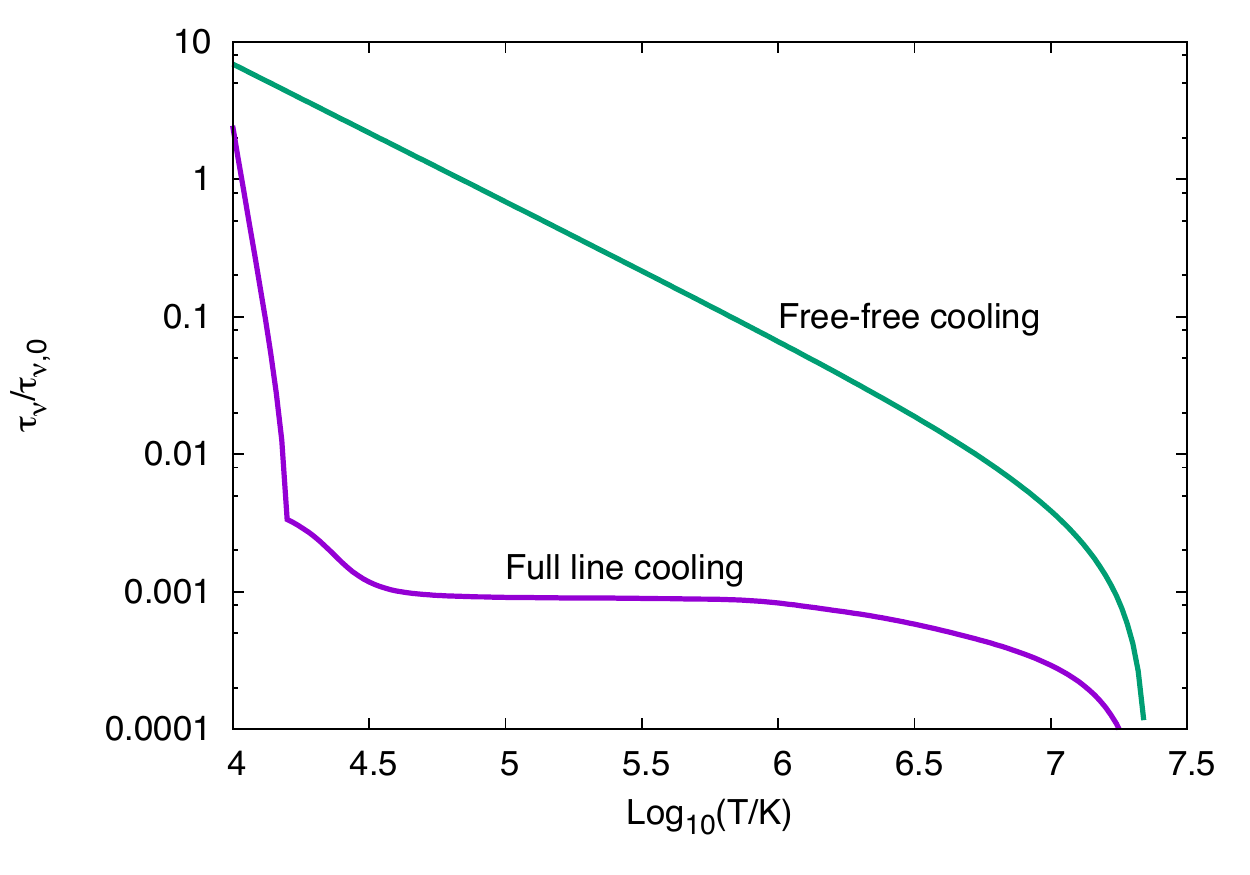}
\caption{Free-free optical depth $\tau_\nu$ as a function of gas temperature $T$ behind the shock. The optical depth is normalized to a fiducial value $\tau_{0,\nu}\equiv (n_{4}/10^{7}{\rm cm^{-3}}) \tvfe/\nu_{10}^2$.  Different lines correspond to cases calculated for a cooling function including full line cooling (purple) and one including just free-free cooling (green).}
\label{fig:taut}
\end{center}
\end{figure}

Figure \ref{fig:taut} shows the evolution of the optical depth $\tau_{\nu}/\tau_{\nu,0}$ as a function of temperature $T$ behind the shock, where $\tau_{\nu,0}=n_{4,7}\tvfe/\nu_{10}^2$ and $\tau_{\nu}=0$ is defined at the forward shock surface ($T = T_{3}$; eq.~[\ref{eq:T3}]).   If only free-free cooling is included, then the radio photosphere ($\tau_\nu = 1$) can occur at higher temperatures of $\gtrsim 10^{5}$ K, depending on $\tau_{0,\nu}$.  However, if full line cooling is included \citep{Schure+09}, then the photosphere temperature is much lower, $\approx 10^{4}$ K.  The observed brightness temperatures $\gtrsim 10^{5}$ K of nova shocks (Fig.~\ref{fig:TB}) can therefore only be thermal in origin if line cooling is highly suppressed from its standard CIE value (\citealt{Metzger+14}).  When line cooling is present, additional non-thermal synchrotron emission above the photosphere is needed to reproduce the observations.
 
The synchrotron emissivity is given by an integral over the electron distribution function,
\be
 j_{\nu, \rm syn}=\frac{1}{4\pi}\int_{E_{min}}^{E_{max}} P(\nu,E) \, {N}(E)dE,
\ee
where $P(\nu,E)$ is the emissivity for a single electron.  For analytical estimates it is convenient to use the so-called delta-function approximation,
\be
P(\nu,E) \approx \frac{4}{3} c\sigma_{\rm T} \, U_B \gamma^2 \delta\left( \nu - \gamma^2\nu_B\right),
\ee
where $\nu_B=eB/(2\pi m_{\rm e} c)$.  This yields reasonably accurate results for broad and smooth electron distributions.  Nevertheless, our numerical calculations use the exact expression for synchrotron emissivity.
For a power-law distribution of electrons,
\be
j_{\nu, \rm syn} =
\frac{1}{9}
\tilde{C}_E h\alpha_{\rm f} \nu_B \left(\frac{\nu}{\nu_B}\right)^{-\frac{p-1}{2}},
\label{eq:jnusyn}
\ee
where $\alpha_{\rm f}$ is the fine structure constant and
$\tilde{C}_E = C_E (m_{\rm e} c^2)^{-p+1}$ is a rescaled normalization constant of the electron distribution.

\begin{figure}
\subfigure{
\includegraphics[width=0.5\textwidth]{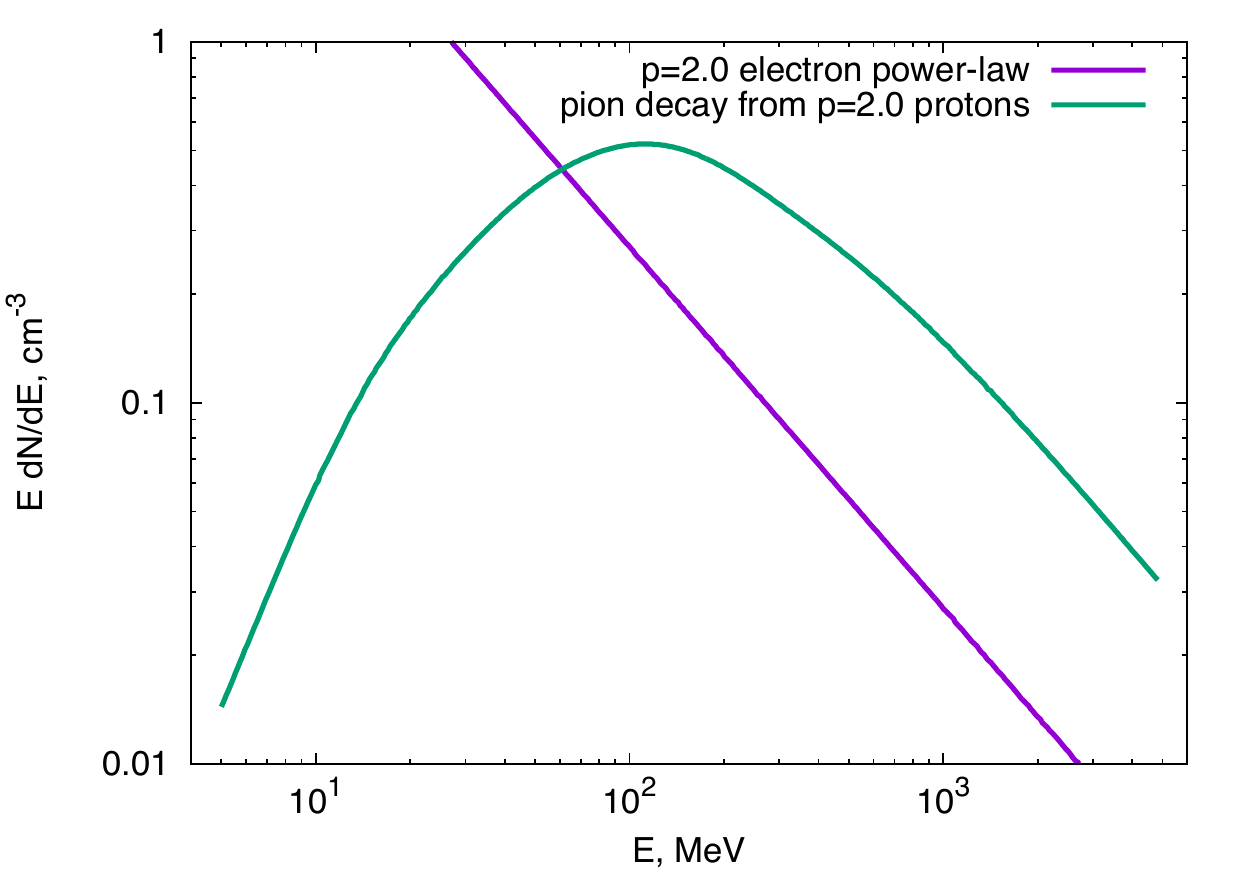}}
\subfigure{
\includegraphics[width=0.5\textwidth]{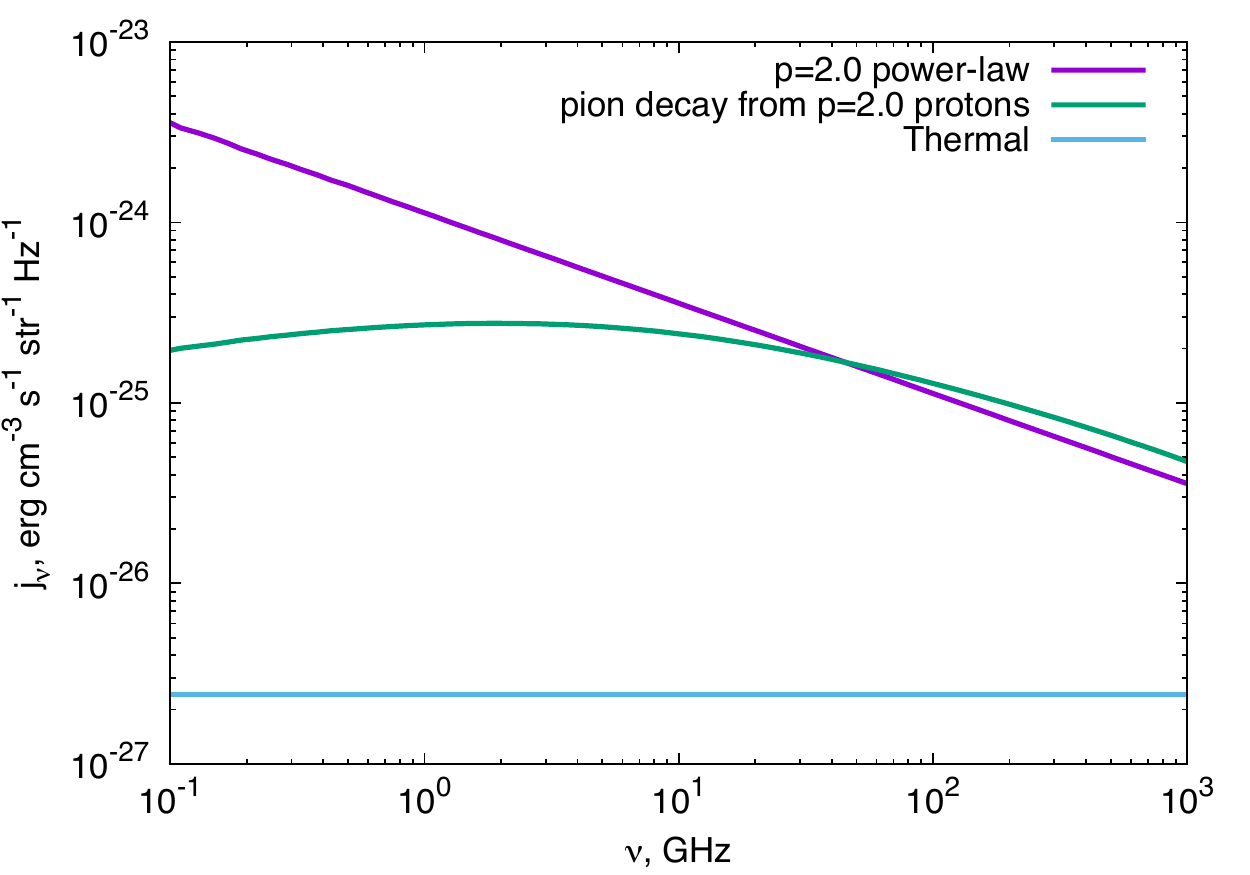}}
\caption{{\it Top Panel:}  Energy spectrum of relativistic leptons, calculated for direct shock acceleration of electrons (electron index $p = 2$) and for secondary pairs from pion decay (proton index $p=2$). {\it Bottom Panel:} Synchrotron emissivity as a function of frequency for the particle spectra in the top panel.  The thermal free-free emissivity is shown for comparison.  Calculations were performed adopting the best-fit shock parameters for V1324 Sco (Table \ref{table:sync}) of $\tvfe=0.63, \eps_e=0.08, f_{\rm EUV}=0.05$ at the $10$ GHz maximum ($t=71.5$ days, $n_4=\e{4.5}{6}$ g cm$^{-3}$).}
\label{fig:piplspem}
\end{figure}

We assume that the magnetic field does not decay as the plasma cools and compresses downstream of the shock, its strength increasing as $B = B_{\rm sh} (n/n_3)^{\Gamma}$, where $\Gamma = 2/3$ for the flux freezing of a tangled (statistically isotropic) field geometry.  Our results are not sensitive to this choice because, for any physical value of $\Gamma$, the integrated emission is dominated by the first cooling length behind the shock ($\S\ref{sec:analytic}$).  Using $C_E \propto n^{(p+2)/3}$ appropriate for adiabatic compression, the emissivity scales as $
 j_{\nu,\rm syn}\propto {n^{(2p+3)/3}}$.  Figure \ref{fig:piplspem} compares the frequency dependence of the emissivity for electrons from direct shock acceleration and pion decay.  


\section{Results} \label{sec:results} 

\subsection{{Analytical estimates}}
\label{sec:analytic}

The formal solution of equation (\ref{eq:TBevo}) for the brightness temperature at the forward shock surface ($T = T_3$) is given by
\be
 T_{\nu}=
\int_{\infty}^{0}
 (T_{\nu,\rm syn}+T)e^{-\tau_\nu}d\tau_{\nu} \label{eq:TBsol},
\ee
where, using equations (\ref{eq:alphaff}), , (\ref{eq:CE}), (\ref{eq:Tsyn_1}), and (\ref{eq:jnusyn}), 
\begin{eqnarray}
{T_{\nu, \rm syn} = 2 \times 10^{9} \, n_{4,7}^{-\frac{3-p}{4}}\nu_{10}^{\frac{1-p}{2}}
\tvfe^{\frac{11+p}{2}}\eps_{B,-2}^{\frac{p+1}{4}}\eps_{e,-2}\left(\frac{T}{T_3}\right)^{3/2}\ratdeg{n}{n_3}{\frac{2p-3}{3}} \, \mathrm{K},
 }
\label{eq:Tsyn} 
\end{eqnarray}
and the prefactor varies between $1.4-2\times 10^9$ for $p = 2-2.5$ due to the complex dependence of $C_E$ (eq. [\ref {eq:CE}]).

\subsubsection{Radiative Shocks}
\label{sec:radiative}
For low velocity, radiative shocks ($v_8 \lesssim 1$) line cooling dominates free-free cooling behind the shock, such that the radio photosphere $\tau_\nu = 1$ is reached at gas temperatures $T \lesssim 10^4$ K (Fig.~\ref{fig:taut}).  Because the latter is significantly lower than the observed brightness temperatures (Fig.~\ref{fig:TB}), non-thermal radiation with $T_{\nu,\rm syn} \gg T$ must dominate the emission.  This justifies neglecting $T$ compared to $T_{\nu,\rm syn}$ in equation (\ref{eq:TBsol}).  As justified below, a further simplification is allowed because $\int T_{\nu,\rm syn}d\tau_{\nu}$ is dominated by the first cooling length behind the shock, where $T \gtrsim 10^{6}$ K and $\tau_{\nu}\ll 1$, allowing us to approximate  $e^{-\tau_{\nu}}\approx 1$ in equation (\ref{eq:TBsol}).  Setting $dT=T$ in equation (\ref{eq:dtau}), the optical depth accumulated over a cooling length centered around temperature $T$ is given by
\be
 \De\tau_\nu=
 \frac{5 \alpha_0 k n_4 \tilde v_{f}}{2\nu^2 T^{1/2}\Lambda(T)} =
2.3\times 10^{-3} \,
 n_{4,7}\tvfe^{-2\de}\nu_{10}^{-2}
\Lambda_{c,-22}^{-1}
 \left(\frac{T}{T_3}\right)^{-\de-\frac12}
 \label{eq:Dtau},
\ee
where we have used the power-law approximation $\Lambda \propto T^{\delta}$ for the line cooling function in equation (\ref{eq:lambda}) assuming $v_{8} \lesssim 1$. 
From equations (\ref{eq:Tsyn}) and (\ref{eq:Dtau}), the brightness temperature accumulated over a cooling length centered around temperature $T$ is given by
\begin{eqnarray}
T_{\nu}^{\rm nth} \approx T_{\nu,\rm syn} \Delta \tau_{\nu} \approx \nonumber \\
\approx 5\times 10^{6} \,
n_{4,7}^{\frac{1+p}{4}}\nu_{10}^{-\frac{p+3}{2}} \tvfe^{\frac{11+p-4\de}{2}} \eps_{B,-2}^{\frac{p+1}{4}} \eps_{e,-2} \Lambda_{c,-22}^{-1} \left(\frac{T}{T_3}\right)^{1-\de} \ratdeg{n}{n_3}{\frac{2p-3}{3}} \,\mathrm{K} \sls
\underset{n/n_3=\left(T/T_3\right)^{-1}}{\propto} 
\left(\frac{T}{T_3}\right)^{\frac{6-2p-3\de}{3}}  \label{eq:Test}.
\end{eqnarray}
Because $T_{\nu}^{\rm nth}$ is an increasing function of temperature for $p<(6-3\de)/2\approx 4$, this illustrates that most of the flux is accumulated over the first cooling length behind the shock.

The {\it thermal} contribution to the brightness temperature at post-shock temperature $T$ is likewise approximately
\be
 T_{\nu}^{\rm th} \approx \De\tau_{\nu} T=
 3.5\times 10^4 \,
 n_{4,7}\tvfe^{2-2\de}\nu_{10}^{-2} \Lambda_{c,-22}^{-1} \, \left(\frac{T}{T_3}\right)^{\frac{1}{2}-\de} \, \mathrm{K}. \label{eq:Tthest}
\ee
If free-free emission dominates the cooling ($\de = 0.5$), then the emission receives equal contributions from each decade in temperature behind the shock (\citealt{Metzger+14}).  For line cooling ($\de=-0.7$) the first cooling length again provides the dominant contribution.  

\begin{figure}
\begin{center}
\includegraphics[width=1.0\linewidth]{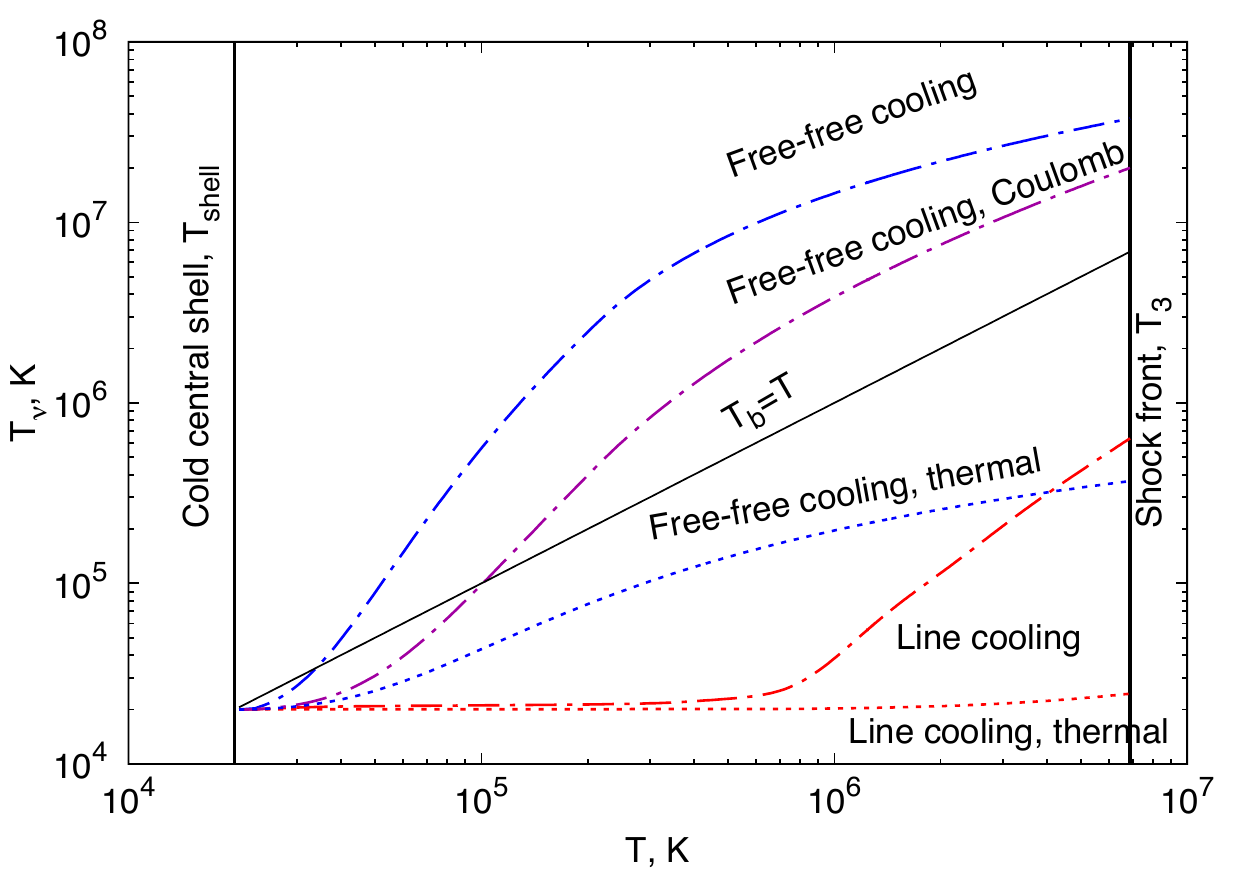}
\caption{10 GHz brightness temperature in the post-shock region as a function of gas temperature $T < T_3$ behind the shock ($T = T_3$).  Different lines correspond to different assumptions about (1) whether the emission accounts only for thermal free-free emission [dashed lines] or also includes non-thermal synchrotron emission [dash-dotted lines], and (2) whether the assumed cooling function is just free-free emission [blue, purple] or also includes emission lines [red].  A purple line shows the non-thermal case with free-free cooling, including the effect of Coulomb losses on the emitting relativistic leptons. 
All calculations were performed for a pre-shock density of $n_4=\e{4.5}{6}$ g cm$^{-3}$ corresponding to the 10 GHz peak time, adopting the best-fit shock parameters for V1324 Sco (Table \ref{table:sync}) of $\tvfe=0.63, \epsilon_e=0.08, f_{\rm EUV}=0.05$.}
\label{fig:tbacc}
\end{center}
\end{figure}

Figure \ref{fig:tbacc} shows how the 10 GHz brightness temperature grows as a function of temperature behind the shock for different assumptions about the cooling and emission of the post-shock gas.  At high optical depths, the radiation and gas possess the same temperature ($T_{\nu} = T$) because synchrotron emission is not present in the cold central shell.  However, differences in $T_{\nu}$ become pronounced as the shock front is approached at $T = T_3$.  When line cooling is included, only the first cooling length behind the shock significantly contributes to the final brightness temperature at the shock surface. 

Coulomb losses of relativistic electrons only noticeably impacts the non-thermal emission well downstream of the shock, in regions where the gas temperature is $\lesssim 10^5$ K (eq.~[\ref{eq:tcoul}]).  Coulomb losses therefore significantly impact the observed radiation temperature only if free-free cooling dominates, for which contributions to the emission from the post-shock gas $T_\nu \propto T^{1/6}$ (eq.~[\ref{eq:Test}]) vary only weakly with gas temperature.  Even in this case, however, the observed temperature reduced only by a factor of $\sim 2$ compared to an otherwise identical case neglecting Coulomb losses.  Coulomb corrections are completely negligible in our fiducial models when line cooling is included because the bulk of the emission comes from immediately behind the shock, where the temperature is high and the density is low. 

Thus far, we have focused on the emission temperature at the forward shock (the `unscreened' temperature, $T_{\nu}|_{\rm shock}$ in eq.~[\ref{eq:screening}]), and we have not yet accounted for free-free attenuation by the ionized layer ahead of the shock.  The observed brightness temperature $T_b$ only equals that of the shock, $T_{\nu}|_{\rm shock}$, after the upstream density has become sufficiently low ($n_4 \lesssim n_{\rm pk}$; eq.~[\ref{eq:screening}]).  The measured peak brightness temperature, $T_{\nu, \rm pk}$, is thus obtained by substituting $n_4 = n_{\rm pk}$ (eq.~[\ref{eq:npk}]) into equations (\ref{eq:Test}) and (\ref{eq:Tthest}) for $T=T_3, n=n_3$.  Multiplying the resulting expression by a factor of $1/e$ (since at the radio peak $\tau_\nu = 1$) yields
\begin{eqnarray}
T^{\rm th}_{\nu, \rm pk}  \approx \text{max}\left\{
\begin{array}{lr} 
1800 \,
\tvfe^{-2\de - 1}f_{\rm EUV,-1}^{-1} \Lambda_{c,-22}^{-1} \, \mathrm{K}\\ 
6900 \,
\tvfe^{2-2\de} \nu_{10}^{-1} H_{14}^{-1/2}\Lambda_{c,-22}^{-1}  \, \mathrm{K}\\
\end{array}
\right. 
 \label{eq:Tthestpk} 
\end{eqnarray}
in the thermal case and 
\begin{eqnarray}
T^{\rm nth}_{\nu, \rm pk}  \approx \text{max}\left\{
\begin{array}{lr} \et{3.9}{5} \, 
f_{\rm EUV,-1}^{-\frac{p+1}{4}} \nu_{10}^{-1} \tvfe^{\frac{19-p-8\de}{4}} \eps_{B,-2}^{\frac{p+1}{4}} \eps_{e,-2} \Lambda_{c,-22}^{-1}  \, \mathrm{K} \\
{1.0\times10^6} \,
\nu_{10}^{-\frac{p+5}{4}} \tvfe^{\frac{11+p-4\de}{2}} \eps_{B,-2}^{\frac{p+1}{4}} \eps_{e,-2} \Lambda_{c,-22}^{-1} H_{14}^{-\frac{p+1}{8}}  \, \mathrm{K} \\
\end{array}
\right.
 \label{eq:Tnthestpk} 
\end{eqnarray}
in the non-thermal case.

\begin{figure}
\begin{center}
\includegraphics[width=1.0\linewidth]{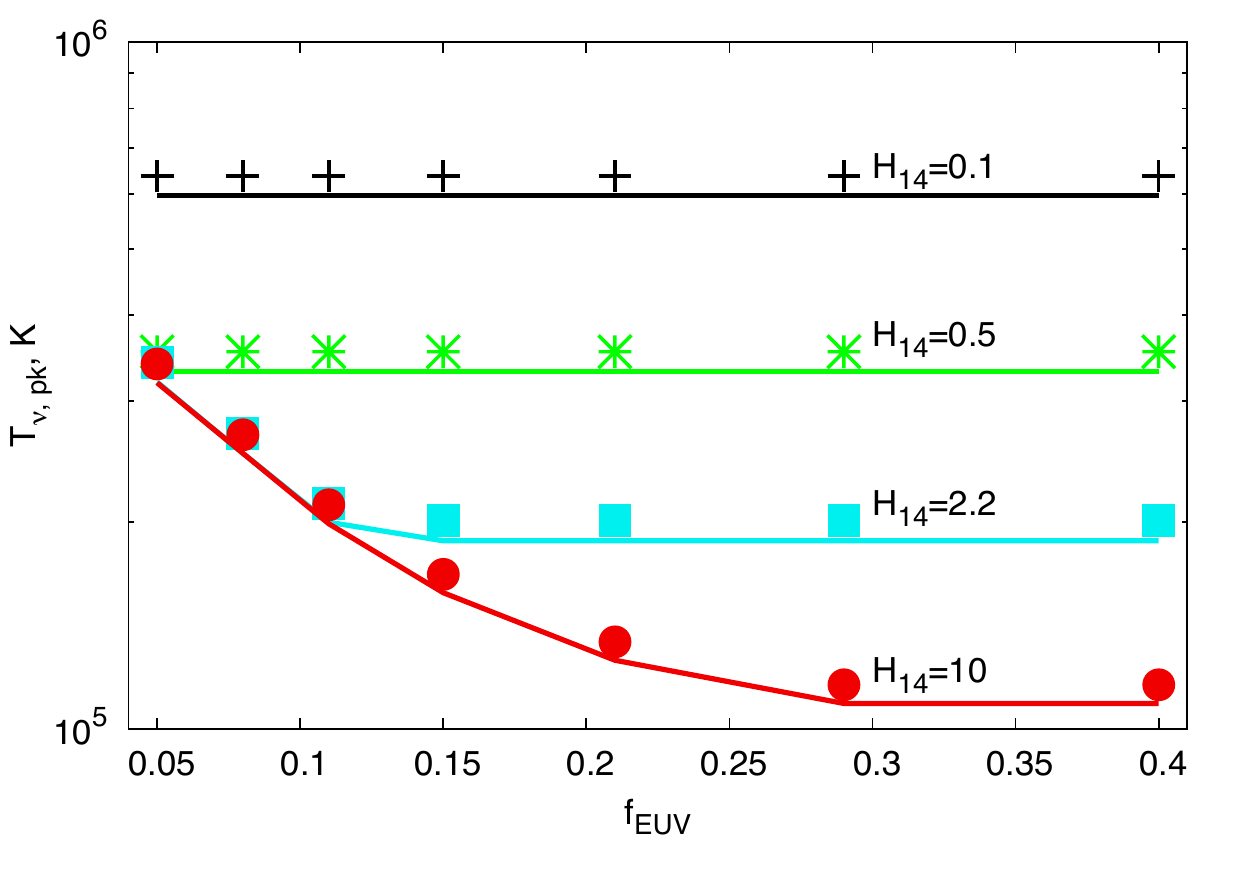}
\includegraphics[width=1.0\linewidth]{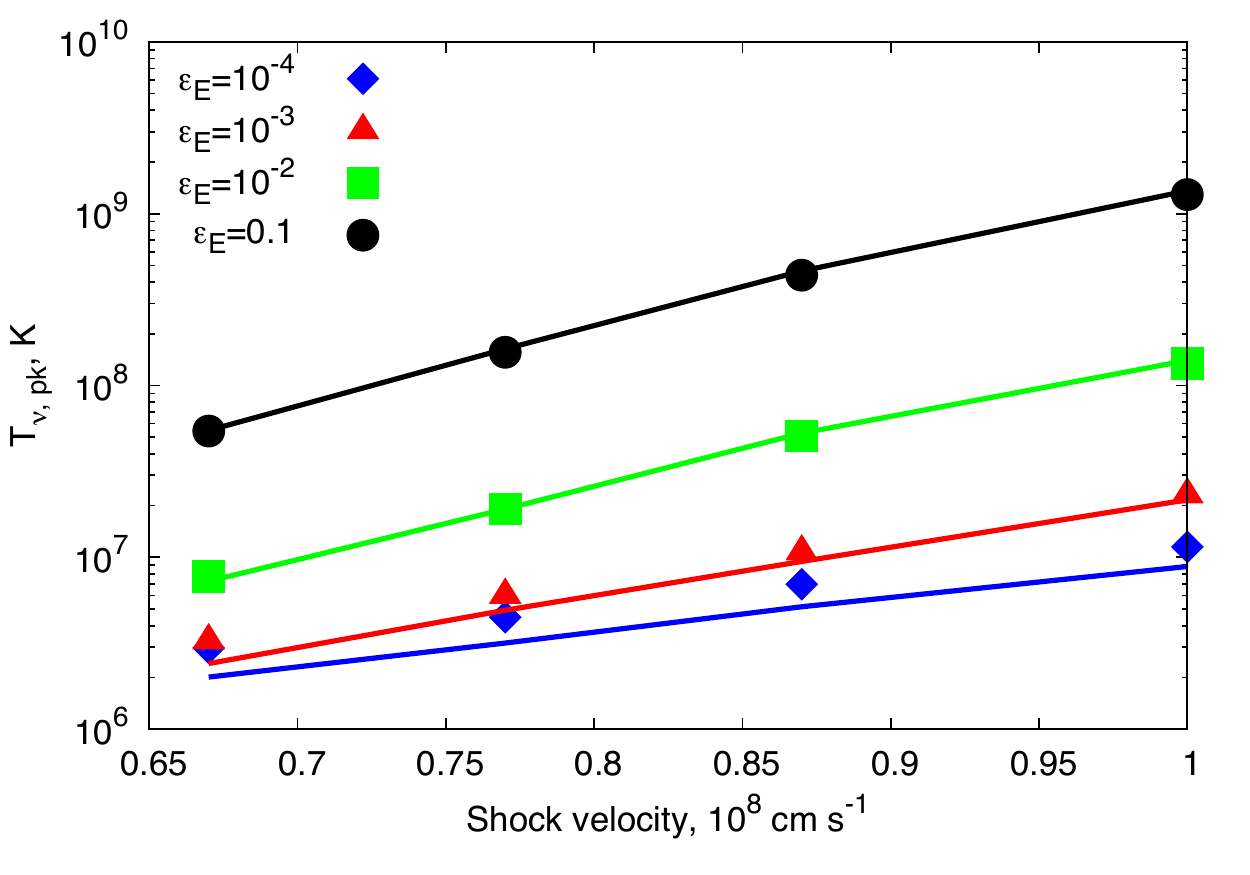}
\caption{Peak brightness temperature of non-thermal emission as function of shock parameters.  The top panel shows $T_{\nu\rm, pk}^{\rm nth}$ as a function of the fraction of shock energy used for ionization $f_{\rm EUV}$ and density scaleheight $H$.  The bottom panel shows $T_{\nu\rm, pk}^{\rm nth}$ as a function of the shock velocity, $\tvfe$, and fraction of the shock energy placed into relativistic electrons, $\epsilon_e$.  Results of the full calculation are shown as symbols, while the analytic estimates from equation (\ref{eq:Test}) are shown as lines. The values of parameters not varied in these figures are taken as best-fit values for V1324 Sco (Table \ref{table:sync}).}
\label{fig:tbfuvH}
\end{center}
\end{figure}

\begin{figure}
\begin{center}
\includegraphics[width=1.0\linewidth]{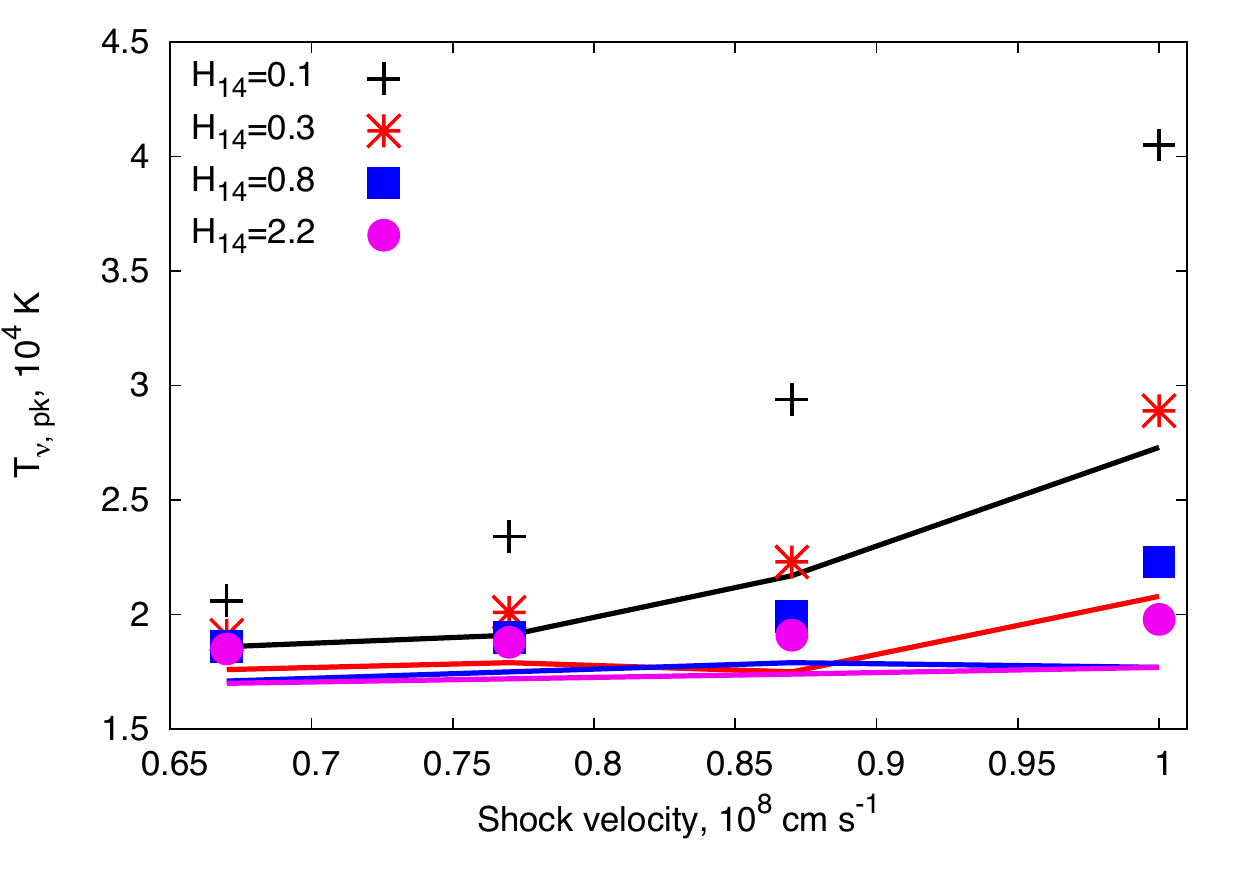}
\caption{Same as Figure \ref{fig:tbfuvH} but showing the observed thermal brightness temperature as a function of the shock velocity $\tvfe$ and scale-height $H_{14}$.  Results of our full calculation are shown as symbols, while the analytic estimates from equation (\ref{eq:Tthest}) are shown as lines.}
\label{fig:tbparsth}
\end{center}
\end{figure}

For $p = 2$, and for our fiducial power-law fit to the line cooling function ($\Lambda_{c,-22} = 1.5, \delta = -0.7$), equation (\ref{eq:Tnthestpk}) becomes
\begin{eqnarray}
T^{\rm nth}_{\nu, \rm pk}  \approx \text{max}\left\{
\begin{array}{lr}
\et{2.6}{5} \, 
f_{\rm EUV,-1}^{-3/4} \nu_{10}^{-1} \tvfe^{5.7} \eps_{B,-2}^{3/4} \eps_{e,-2}   \, \mathrm{K},
 \\
 6.9\times 10^5 \,
\nu_{10}^{-7/4} \tvfe^{7.9} \eps_{B,-2}^{3/4} \eps_{e,-2}  H_{14}^{-3/8}  \, \mathrm{K}.
\\
\end{array}
\right.
 \label{eq:Tnthestpk2} 
\end{eqnarray}
Figures \ref{fig:tbfuvH} and \ref{fig:tbparsth} compare our analytic expressions for the peak observed brightness temperature (eqs.~[\ref{eq:Tthestpk}, \ref{eq:Tnthestpk}]) and the peak value resulting from a direct integration of equation (\ref{eq:TBsol}).  

The top panel of Figure \ref{fig:tbfuvH} shows how $T_{\nu, \rm pk}^{\rm nth}$ depends on the density scale-height $H$ and the ionized fraction of the preshock layer, $f_{\rm EUV}$.  When the value of $H$ is small, or $f_{\rm EUV}$ is sufficiently high, then the entire DES scaleheight is ionized ($H = \Delta_{\rm ion}$) and the peak brightness temperature is independent of $f_{\rm EUV}$.  On the other hand, when $\Delta_{\rm ion} < H$, then the brightness temperature is independent of $H$ but decreases with increasing ionizing radiation as $T_{\rm \nu,pk}^{\rm nth}\propto f_{\rm EUV}^{-(p+1)/4}\underset{p=2}=f_{\rm EUV}^{-0.75}$ (eq.~[\ref{eq:Tnthestpk}]).  The bottom panel of Figure \ref{fig:tbfuvH} shows how the peak temperature depends on the shock velocity, $\tvf$, and the electron acceleration efficiency, $\eps_e$.  

Figure \ref{fig:tbparsth} compares the brightness temperature calculated in a purely thermal model ($\epsilon_e = 0$) to our analytic estimate of $T_{\nu, \rm pk}^{\rm th}$ (eq.~[\ref{eq:Tthestpk}]).  For a broad range of parameters, the peak temperatures fails to exceed $10^{5}$ K, making thermal models of radiative shocks challenging to reconcile with observations (see Fig.\ref{fig:TB}). 
\subsubsection{Adiabatic shocks}

For higher velocity $\tvfe \gtrsim 10^{8}$ cm s$^{-1}$, adiabatic shocks, the brightness temperature is again estimated by assuming that the first scaleheight $H$ behind the shock dominates the emission, i.e. neglecting ongoing emission from matter shocked many dynamical times earlier. However, unlike with radiative shocks, this assumption cannot be rigorously justified without a radiation hydrodynamical simulation, an undertaking beyond the scope of this paper.

The peak brightness temperature in the adiabatic case is estimated using the expressions for radiative shocks from $\S\ref{sec:radiative}$, but replacing the first cooling length behind the shock $t_{\rm cool}\tvf$ with $H$.  For purely thermal emission, the brightness temperature is
\be
  T_{\nu, H}^{th}=1.1\times 10^5 n_{4,7}^2 H_{14} \nu_{10}^{-2} \tvfe^{-1} {\rm K}
\label{eq:THth}
\ee
Substituting $n_{4} = n_{\rm pk}$ into equation (\ref{eq:THth}) gives a peak thermal temperature of
\be
 T_{\nu,H,pk}^{th}=\et{1.2}{4}\tvfe^{-1}{\rm K} \label{eq:Tthadpk},
\ee
where we have assumed that the scaleheight is fully ionized ($n_{\rm pk} = n_{\rm H,pk}$), which is generally well satisfied for high velocity shocks.

For adiabatic shocks, the peak thermal brightness temperature is a decreasing function of the shock velocity due to the $\alpha_{\rm ff}\propto T^{-3/2}$ dependence of free-free absorption.  Comparing equations (\ref{eq:Tnthestpk2}) and (\ref{eq:Tthadpk}), a maximum thermal brightness temperature of $\approx 10^{4}$ K is thus obtained for the intermediate shock velocity $\tvfe \sim 1$ separating radiative from adiabatic shocks.  The fact that this falls well short of observed peak brightness temperatures (Fig.~\ref{fig:TB}) again disfavors thermal models for the early radio peak..

For non-thermal emission from adiabatic shocks, the brightness temperature and its peak value are given, respectively, by
\be
 T_{\nu,H}^{\rm nth} = 1.2\times 10^7 \, n_{4,7}^{\frac{5+p}{4}}\nu_{10}^{-\frac{p+3}{2}}  \tvfe^{\frac{p+5}{2}}\eps_{B,-2}^{\frac{p+1}{4}}\eps_{e,-2} H_{14} \,{\rm K} \label{eq:Tnthestad}  \\
 T_{\nu,H,pk}^{\rm nth}= 1.4\times 10^6 \, \nu_{10}^{-\frac{p+1}{4}} \tvfe^{\frac{5+p}{2}}\eps_{B,-2}^{\frac{p+1}{4}}\eps_{e,-2} H_{14}^{\frac{3-p}{8}} \label{eq:Tnthestpkad}.
\ee

\begin{figure}
\subfigure{
\includegraphics[width=0.5\textwidth]{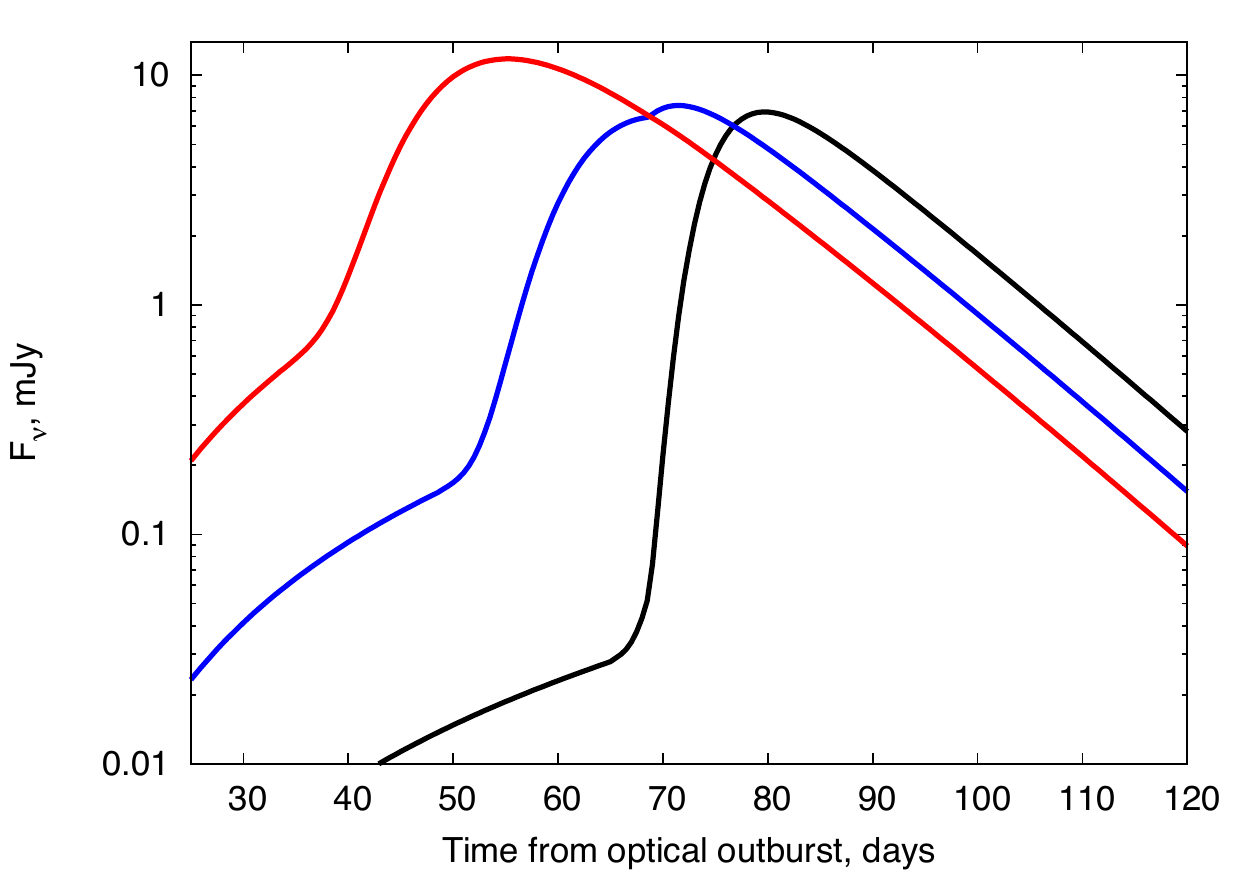}}
\subfigure{
\includegraphics[width=0.5\textwidth]{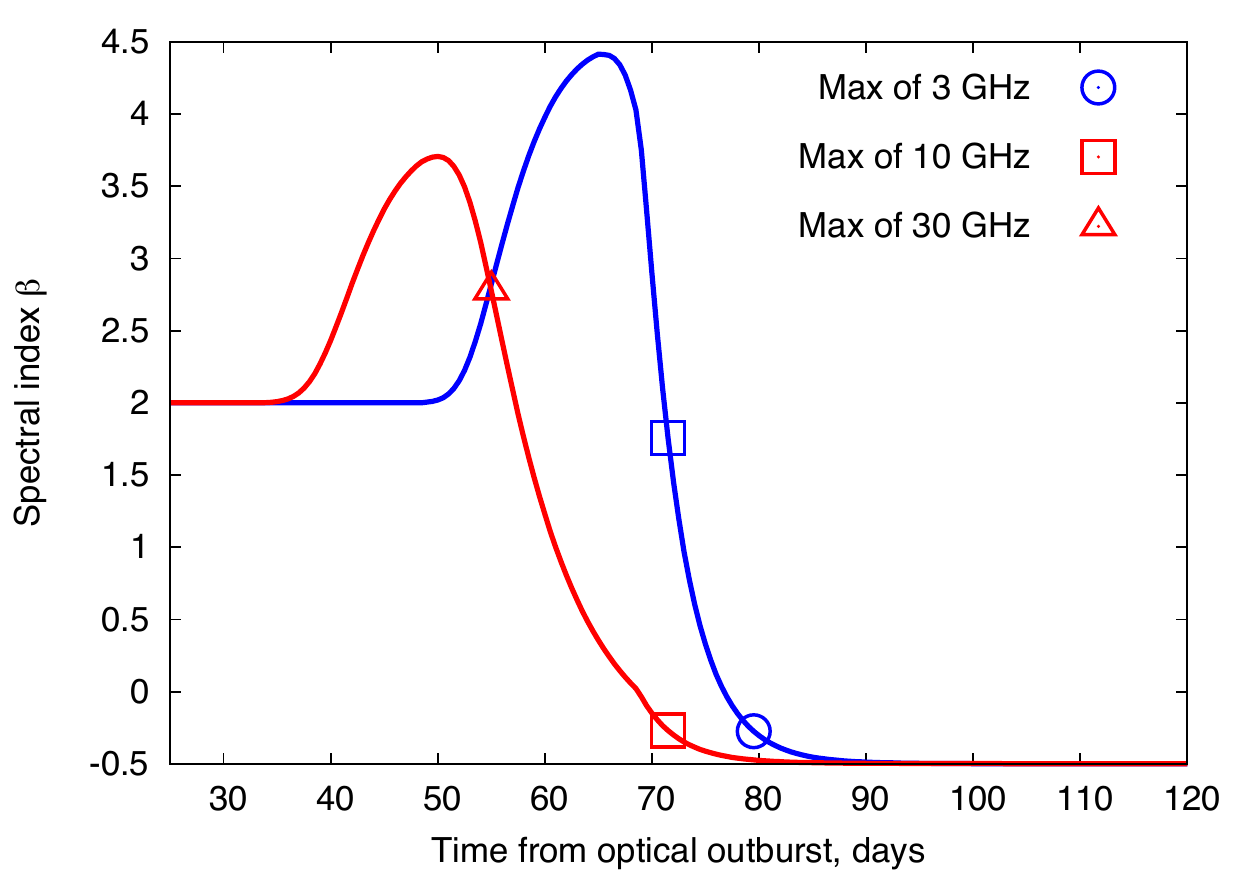}}
\caption{{\it Top:} Example model radio lightcurves at radio frequencies of 3 GHz (black), 10 GHz (blue), and 30 GHz (red), calculated for our best-fit parameters for V1324 Sco (Table \ref{table:sync}). The lightcurve does not include thermal emission from the cold central shell or other sources of photo-ionized ejecta.  {\it Bottom:} Measured spectral index $\beta$ between the 3 and 10 GHz bands (red) and the 10 and 30 GHz bands (blue). The time of light curve maxima are marked with symbols for 3 GHz (circle), 10 GHz (square), and 30 GHz (triangle).  The plateau at $\beta=2$ at early times corresponds to when the photo-ionized layer ahead of the shock is still optically thick.} 
\label{fig:tbs}
\end{figure}

\subsection{Radio lightcurves and spectra}

The shape of the radio light curve and spectrum are driven by the effects of free-free opacity, which are independent of the emission mechanism and hence apply both to thermal (\citealt{Metzger+14}) and non-thermal models.  Figure \ref{fig:tbs} shows an example model light curve (upper panel) and corresponding evolution of the spectral index across two representative frequency ranges (bottom panel).  At high frequencies the light curve peaks earlier and reaches a larger maximum flux, due to the frequency dependence of free-free absorption, $\alpha_{\rm ff} \propto \nu^{-2}$, and the monotonically declining shock power.  When such behaviour is not observed, this indicates that our one zone model is inadequate or that free-free opacity does not determine the light curve maximum ($\S\ref{sec:aql}$).

The spectral index $\beta$ initially rises to exceed that of optically-thick isothermal gas ($\beta > 2$) because the higher frequency emission peaks first. Then, once the lower frequency emission peaks, the shock becomes optically thin and hence the spectral index approaches the spectral index of optically thin synchrotron emission.  Importantly, a flat spectral index {\it does not itself provide conclusive evidence for non-thermal synchrotron emission}, even though the optically thin spectral indices are different for synchrotron and thermal bremsstrahlung emission.  The emission mechanism is instead more accurately distinguished from thermal emission based on the higher peak brightness temperatures which can be achieved by non-thermal models.  Also, at late times the radio emission should eventually come to be dominated by thermal emission of the photo-ionized ejecta and hence the spectral index will again rise to $\beta=2$.  This feature is not captured by Figure \ref{fig:tbs} because we have not included thermal emission from the cool central shell or other sources of photo-ionized ejecta.

\subsection{Fits to Individual Novae \label{sec:fits}}
\begin{figure}
\subfigure{
\includegraphics[width=0.5\textwidth]{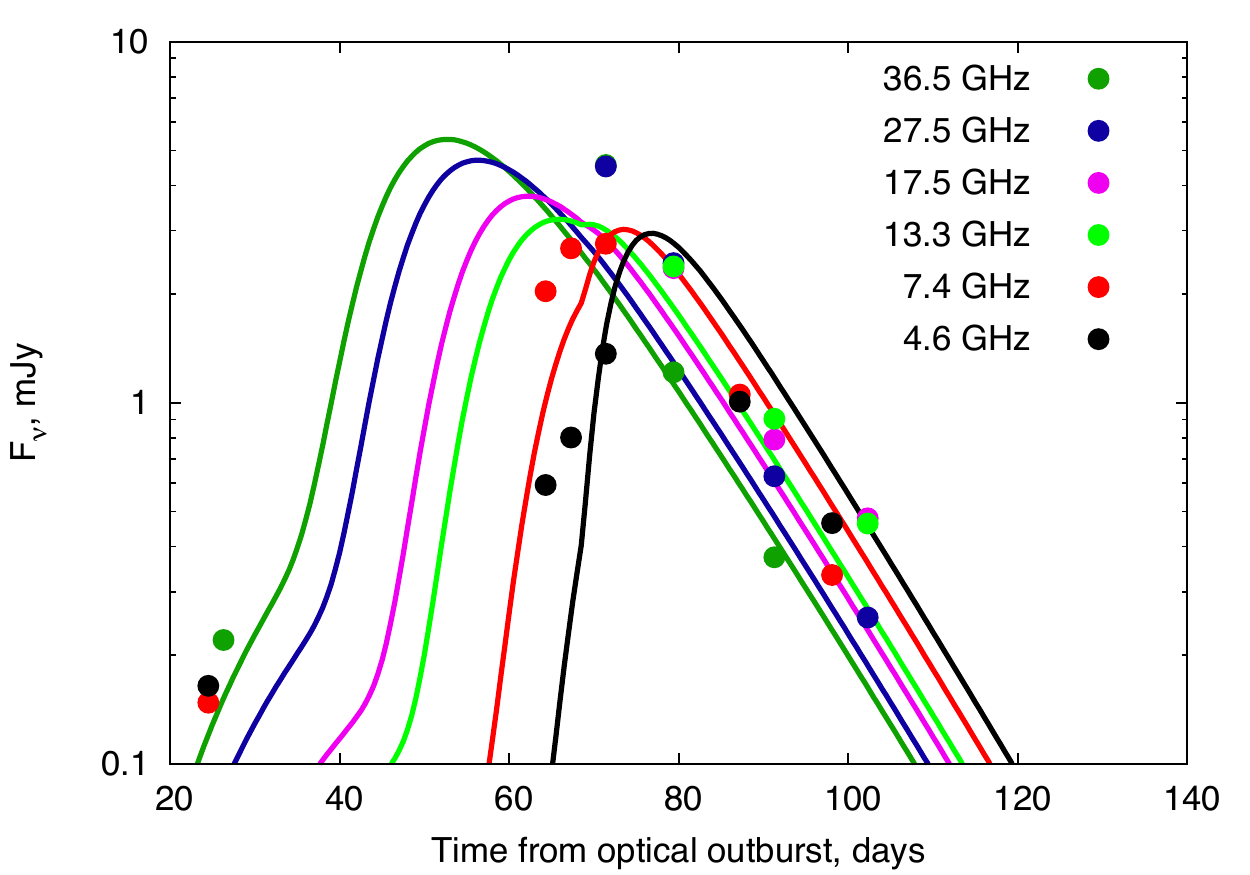}}
\subfigure{
\includegraphics[width=0.5\textwidth]{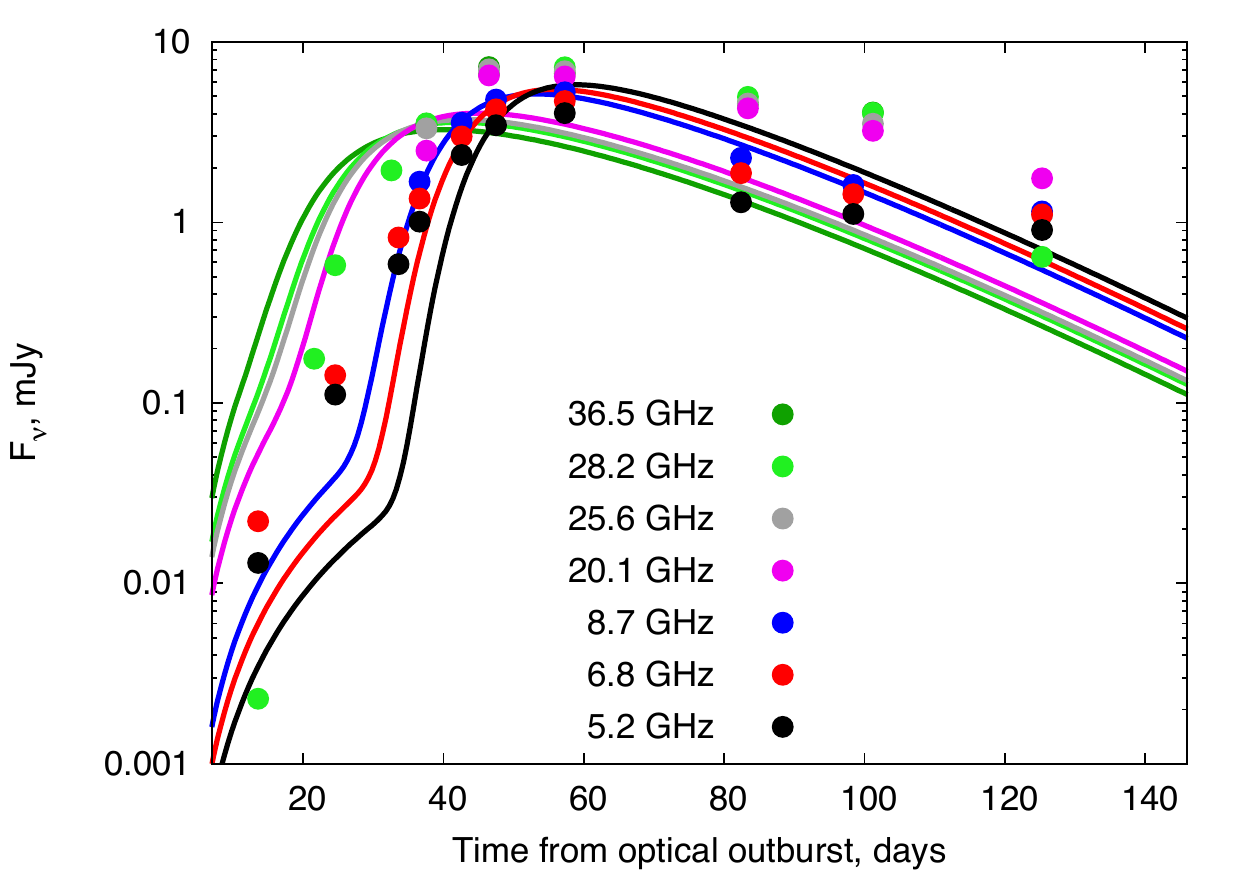}}
\subfigure{
\includegraphics[width=0.5\textwidth]{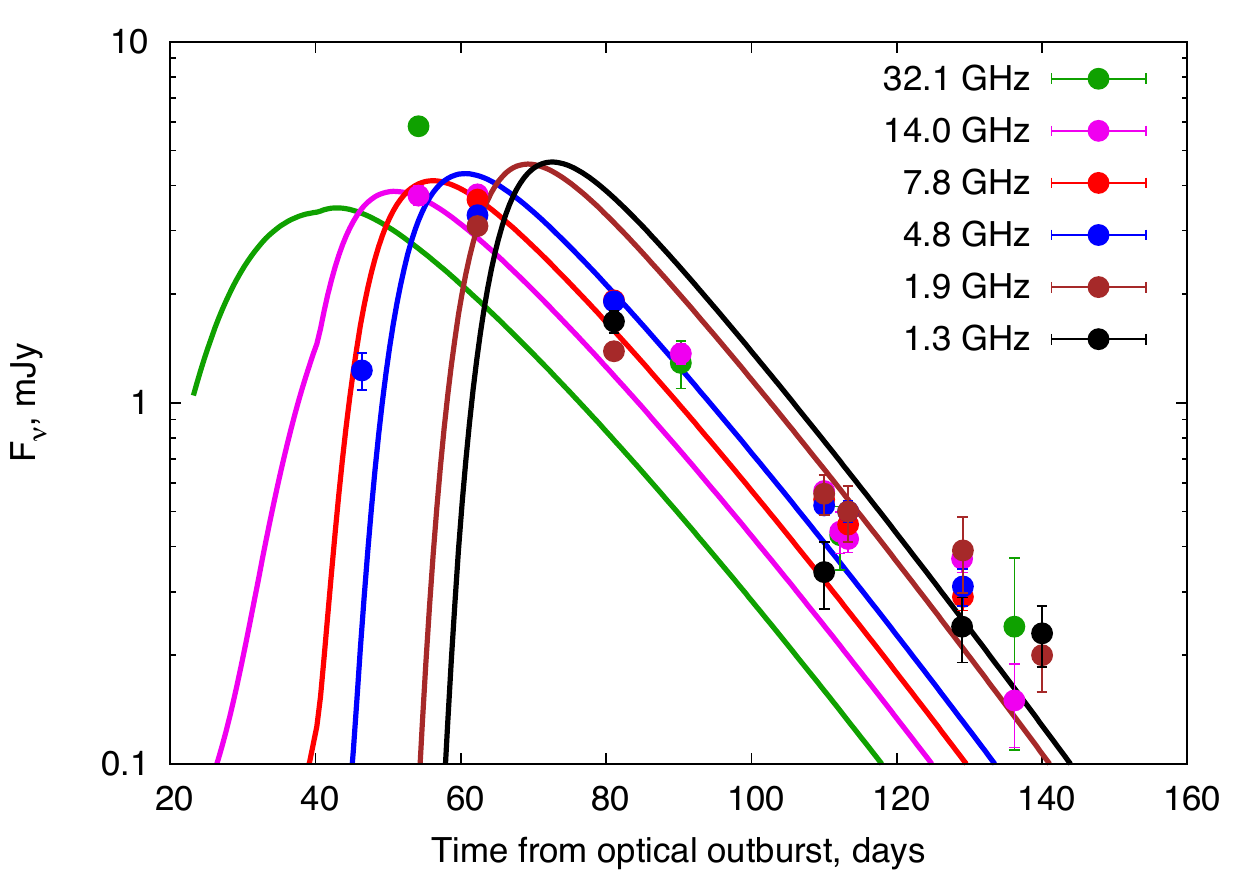}}
\caption{Synchrotron shock models fit to V1324 Sco (top panel), V1723 Aql (middle) and V5589 Sgr (bottom).  Parameters of the fits are provided in Table \ref{table:sync}.  In all cases the late-time thermal emission has been subtracted prior to the fit. Uncertainties are cited in the caption of Fig. \ref{fig:fluxes}.} 
\label{fig:bestfit}
\end{figure}

\begin{table*}
\caption{Best-fit parameters of synchrotron model.}
\label{table:sync}
\begin{center}
\begin{tabular}{cccccccccc} 
 \hline
 Nova & $v_1^{(a)}$ & $v_4^{(b)}$ & $n_{4}/n_{1}^{(c)}$ & $\tvf^{(d)}$  & $\De t^{(e)}$  & $H$$^{(f)}$ & $n_0$$^{(g)}$ & $\eps_e$$^{(h)}$ & $f_{\rm EUV}$$^{(i)}$ \\
 \hline
& (km s$^{-1}$)  & (km s$^{-1}$)   & -  & (km s$^{-1}$)  & (days) & (cm) &  (cm$^{-3}$) & - & -  \\
 \hline         
V1324 Sco & 1700  & 670  & 0.43 & 630 & 12 & $\et{3.7}{13}$ & $\et{7.8}{9}$ & 0.081 & 0.051  \\
\hline
V1723 Aql & 2300 & 770 & 0.66 & 830 & $3.2$ & $\et{9.5}{13}$  & $\et{6.7}{7}$ & $0.015$ & 0.31 \\
 \hline
 V5589 Sgr & 2200 & 190 & 2.3 & 810 & 5 & $\et{6.8}{13}$ & $\et{4.1}{8}$ & $0.014$ & 0.07 \\
 \hline
\end{tabular}
\end{center}
$^{(a)}$velocity of fast outflow, $^{(b)}$velocity of slow outflow (DES), $^{(c)}$ratio of densities of DES and fast outflow, $^{(d)}$velocity of the shock in the frame of the upstream gas, calculated from $v_1, v_4, n_4/n_1$ using equation (\ref{eq:vshock}), $^{(e)}$time delay between launching fast and slow outflows, $^{(f)}$scale-height of slow outflow, $^{(g)}$normalization of density profile of slow outflow (eq.~[\ref{eq:rhopro}]), $^{(h)}$fraction of shock power placed into power-law relativistic electrons/positrons, $^{(i)}$fraction of shock power placed into hydrogen-ionizing radiation (eq.~[\ref{eq:delta}])
\end{table*}


We fit our model to the radio light curves of three novae with early-time coverage, V1324 Sco, V1723 Aql and V5589 Sgr (Figs.~\ref{fig:fluxes}, \ref{fig:TB}).  We employ a $\chi$-squared minimization technique across 9 free parameters: $v_1, v_4, n_{4}/n_{1}, \Delta t, H, n_{0}, p, \eps_e, f_{EUV}$.  We assume a shock covering fraction of $f_{\Omega} = 1$ and take $\epsilon_B = 0.01$, as the latter is degenerate with $\epsilon_e$.  Table \ref{table:sync} provides the best-fit parameters of each novae.  Although the electron power-law index is a free parameter, in practice it always converges to a best-fit value of $p \simeq 2$.


Our best fit to V1324 Sco is shown in the upper panel of Figure \ref{fig:bestfit}.  The best-fit shock velocity of 640 km s$^{-1}$ is within the range of radiative shocks, consistent with the rapid post-maximum decline. For the parameters of our best-fit model, $n_{pk,\De}/n_{pk, H}=0.96\nu_{10}$, implying that for frequencies above(below) 10 GHz the thickness of ionization layer is less than(greater than) the scale height at the time of peak flux.  The transition between these two regimes can be seen as a small break in the 7.4 GHz light curve around day 68. Substituting our best-fit parameters into equation (\ref{eq:R0}), we find that the collision occurred 20 days after the start of the outflow, corresponding to 7 days after the onset of the gamma-ray emission \citep{Ackermann+14}.  This discrepancy is not necessarily worrisome, as the colliding `shells' may be ejected over days or longer.

V1723 Aql was more challenging to fit (Fig.\ref{fig:bestfit}, middle panel), mainly because different frequencies peak at nearly the same time.  This is contrary to the the expectation that high frequencies will peak first if free-free absorption indeed controls the light curve rise, as our model assumes (see $\S$\ref{sec:aql} for alternative interpretations).  Although we cannot fit V1723 Aql in detail, we can nevertheless reproduce the magnitudes of the peak fluxes for reasonable parameters.  Our best-fit model for V5589 Sgr is shown in the bottom panel of Figure \ref{fig:bestfit}, although the data is more sparse than for the other two events.  


\section{Discussion}
\label{sec:discussion}
\subsection{Efficiency of Relativistic Particle Acceleration}
 \label{sec:concconst}
The non-thermal radio emission from novae probes relativistic particle acceleration and magnetic field amplification at non-relativistic shocks, as parameterized through $\epsilon_e$ and $\epsilon_B$.  Considering all three novae, we infer electron acceleration efficiencies in the range $\epsilon_e \sim 0.01-0.08$ for $\eps_B = 0.01$ (Table \ref{table:sync}).  However, the peak radio luminosity is largely degenerate in $\eps_e,\eps_B$ and $\tvf$.  By combining the peak 10 GHz luminosity of V1324 Sco, V1723 Aql, and V5589 Sgr with analytic estimates of the peak brightness temperature (eq.~[\ref{eq:Tnthestpk}]) and assuming a shock radius equal to that of our best fit model ($R_{\rm sh} = v_{4}t_{\rm peak}$; Table \ref{table:sync}), we obtain the following limits,
\be
 f_\Omega \tvfe^{7.9} \eps_e \eps_B^{3/4} \approx 7\times 10^{-5}-1.2\times 10^{-4}, \label{eq:eefovfl} 
\ee 
where we have again assumed radiative shocks $(v_{8} \lesssim 1)$.

Now consider some implications of these constraints.  Making the very conservative assumption that $\epsilon_e + \epsilon_B \lesssim 1$, i.e. $\eps_e\eps_B^{3/4}\lesssim 0.5$, we find
\be
 \tvfe \gtrsim 0.4 f_\Omega^{-0.13} 
 \Rightarrow kT\gtrsim 0.2 f_\Omega^{-0.25}\,\,{\rm keV} \label{eq:lowerkTconst}.
\ee
This lower limit on the shock temperature ensures, for example, that thermal free-free emission from the shocks will fall within the spectral window of {\it Swift} and {\it Chandra}.  The radio emitting electrons should  produce a measurable X-ray signature, even if its too weak to detect or is overpowered by faster adiabatic shocks.

Next, using our best fit parameters for the shock velocities $\tvfe$ of each nova (Table \ref{table:sync}) and assuming an error on this quantity of 10 per cent, each event separately provides a constraint
\be
 \et{1.2}{-3} \le f_\Omega\eps_e \eps_B^{3/4} \le \et{5.4}{-3}\quad {\rm V1324\,\, Sco} \label{eq:firsteeconstr} \\
 \et{2.3}{-4} \le f_\Omega\eps_e \eps_B^{3/4} \le 10^{-3} \quad {\rm V1723\,\, Aql} \\
 \et{2}{-4} \le f_\Omega\eps_e \eps_B^{3/4} \le \et{9}{-4}\quad {\rm V5589\,\, Sgr} \label{eq:lasteeconstr}
\ee
For physical values of $\eps_B < 0.1$ (\citealt{Caprioli&Spitkovsky14b}), we thus require $\epsilon_e \gtrsim 10^{-3}-10^{-2}$.  

Such high acceleration efficiencies {\it disfavor hadronic scenarios for the radio-emitting leptons}, which are estimated to produce $\eps_e\lesssim 10^{-4}$ (eq.~[\ref{eq:eehadron}]).  They are also in tension with PIC simulations of particle acceleration at non-relativistic shocks (\citealt{Kato14}; \citealt{Park+14}) which find $\epsilon_e \sim 10^{-4}$ when extrapolated to shock velocities $v_8 \lesssim 10^{3}$ km s$^{-1}$, modeling of supernova remnants (\citealt{Morlino&Caprioli12}; $\epsilon_e \sim 10^{-4}$), and galactic cosmic ray emission (\citealt{Strong+10}; $\epsilon_e \sim 10^{-3}$).  On the other hand, the inferred  cosmic ray efficiencies are dependent on the shock fraction of the accelerated electrons which escape the supernova remnant.  Modeling of unresolved younger radio supernovae typically find higher values of $\epsilon_e$ (\citealt{Chevalier82}; \citealt{Chandra+12}), consistent with our results.  Given these observational and theoretical uncertainties, we tentatively favor a leptonic source for the radio-emitting electrons.   

In the above, we have assumed that the shocks cover a large fraction of the outflow surface ($f_{\Omega} \sim 1$).  However, if instead we have $f_{\Omega} \ll 1$, then the required values of $\epsilon_e$ and $\epsilon_B$ would be even higher than their already strained values.  If the radio-emitting shocks are radiative, they must therefore possess a large covering fraction $f_{\Omega} \sim 1$.  

Lower values of $\epsilon_e$ and $\epsilon_B$ are allowed if the radio emission is instead dominated by high-velocity $\tvfe \gtrsim 1$, adiabatic shocks covering a large solid fraction $f_{\Omega} \sim 1$.  The sensitive dependence of the brightness temperature on the shock velocity implies that even a moderate increase in $\tvf$ can increase the radio flux by orders of magnitude for fixed $\epsilon_e, \epsilon_B$.  On the other hand, the adiabatic shocks responsible for the hard X-rays appear to require a {\it small} covering fraction $f_{\Omega} \ll 1$ so as not to overproduce the observed X-ray luminosities ($\S\ref{sec:xraysintro}$).  


In V1324 Sco, \citet{Metzger+15} place a lower limit of $\eps_{\rm nth} = \epsilon_p + \epsilon_e > 10^{-2}$ on the total fraction of the shock energy placed into non-thermal particles.  Assuming that the microphysical parameters of the radio and gamma-ray emitting shocks are identical, and that the gamma-rays are leptonic in origin ($\epsilon_{\rm nth} = \epsilon_e \sim 0.01-0.1$), then by combining this constraint with our constraints on V1324 Sco from equation (\ref{eq:firsteeconstr}), we find
\be
 \et{3}{-3} \le \eps_B \le \et{2}{-2}\,\,\, {\rm V1324\,\, Sco},
\ee
providing evidence for magnetic field amplification.  

\subsection{The DES is not a MS progenitor wind} \label{sec:wind}
Gamma-rays are observed not only in novae with red giant companions (symbiotic novae), but also in systems with main-sequence companions (classical novae). Could the DES required for shock-powered radio and gamma-ray emission be the companion stellar wind?  In this section we estimate the radio emission from the fast nova ejecta interacting with the (assumed stationary) stellar wind of the binary companion. The density profile of a spherically symmetric, steady-state wind is given by
\be
 n_{w}\simeq \frac{\dot{M}}{4\pi r^2 v_w m_p}=300\frac{\dot{M}_{-10}}{r_{14}^2v_{w,8}}\,\, {\rm cm}^{-3}, \label{eq:nw}
\ee
where 
$r=r_{14}10^{14}$ cm is radius, $v_w= v_{w,8}10^8$ cm s$^{-1}$ is the wind speed, and $\dot{M}=\dot{M}_{-10} 10^{-10} M_{\odot}$ yr$^{-1}$ is the mass-loss rate normalized to one specific for main-sequence stars.


The density and radius of the radio photosphere are given by equating (\ref{eq:nw}) with $n_{\rm pk,H}$ (eq.~[\ref{eq:npk}]) for a density scaleheight $H \sim r$,
\be
 n_{w,pk}=\et{1.4}{8} \nu_{10}^{4/3} v_{w,8}^{1/3} \dot{M}_{-10}^{-1/3} \\
 r_{w, pk}=\et{1.5}{11} \nu_{10}^-{2/3} v_{w,8}^{-2/3} \dot{M}_{-10}^{2/3}
\ee
Although $n_{w,pk}$ is several orders larger than in our fiducial DES models, the radius of the photosphere is 3 orders of magnitude smaller.  Under these conditions the shock is adiabatic and hence the peak brightness temperature is found by substituting $n_{w,pk}$ and $r_{w, pk}$ into equation (\ref{eq:Tnthestad}), which gives (for $p=2$)
\be
 T_{w,pk}=\et{1.7}{6} \dot{M}_{-10}^{1/12}\tvfe^{3.5}\eps_{B,-2}^{3/4}\eps_{e,-2}\nu_{10}^{-5/6}v_{w,8}^{-1/12}\,{\rm K}
\ee
Although the value of $T_{\rm w,pk} \sim 10^{6}$ K is comparable to those of observed radio emission, the much smaller radius of the photosphere $r_{\rm w,pk}$ compared to that in our fiducial DES models of $\sim 10^{14}-10^{15}$ cm would result in a peak radio flux $\propto r_{\rm w,pk}^{2}T_{\rm w,pk}$ which is approximately 8 orders of magnitude smaller than the observed values.  

Radio emission from the interaction of the nova ejecta with the main sequence progenitor wind is thus undetectable for physical values of $\dot{M}$, unless the mass loss rate is comparable to that of the nova eruption itself.

\subsection{Light Curve of V1723 Aquila }
\label{sec:aql}

Our model does not provide a satisfactory fit to the light curve of V1723 Aql because all radio frequencies peak nearly simultaneously (Fig.~\ref{fig:bestfit}), contrary to the expectation if the light curve peaks due to the decreasing free-free optical depth ($\S\ref{sec:fits}$).  Here we describe modifications to the standard picture that could potentially account for this behavior.

\subsubsection{Steep density gradient}


The V1723 Aql light curves are consistent with all frequencies peaking within a time interval $\Delta t_{\rm pk} \lesssim$ few days.  From equation (\ref{eq:dtnu}) this requires a scale height of thickness
\be
 H \lesssim \Delta t_{\rm pk}\tvf=2.6\times 10^{13} v_8\left(\frac{\Delta t_{\rm pk}}{\rm 3\,\, day}\right) \,\,{\rm cm},
\ee
which is much smaller than the shock radius of $R_{\rm sh} \sim 10^{14}-10^{15}$ cm.  Such a steep gradient should also result in a short rise and fall time, which is inconsistent with the relatively broad peak observed in V1723 Aql.  This contradiction could be alleviated if the shock reaches the critical density of peak emission $n = n_{\rm pk}$ at different times across different parts of the ejecta surface.  In this case, the total light curve, comprised of the emission from all locations, would be ``smeared out" in time while still maintaining the nearly frequency-independent peak time set by the steep gradient.

\begin{figure}
\subfigure{
\includegraphics[width=0.5\textwidth]{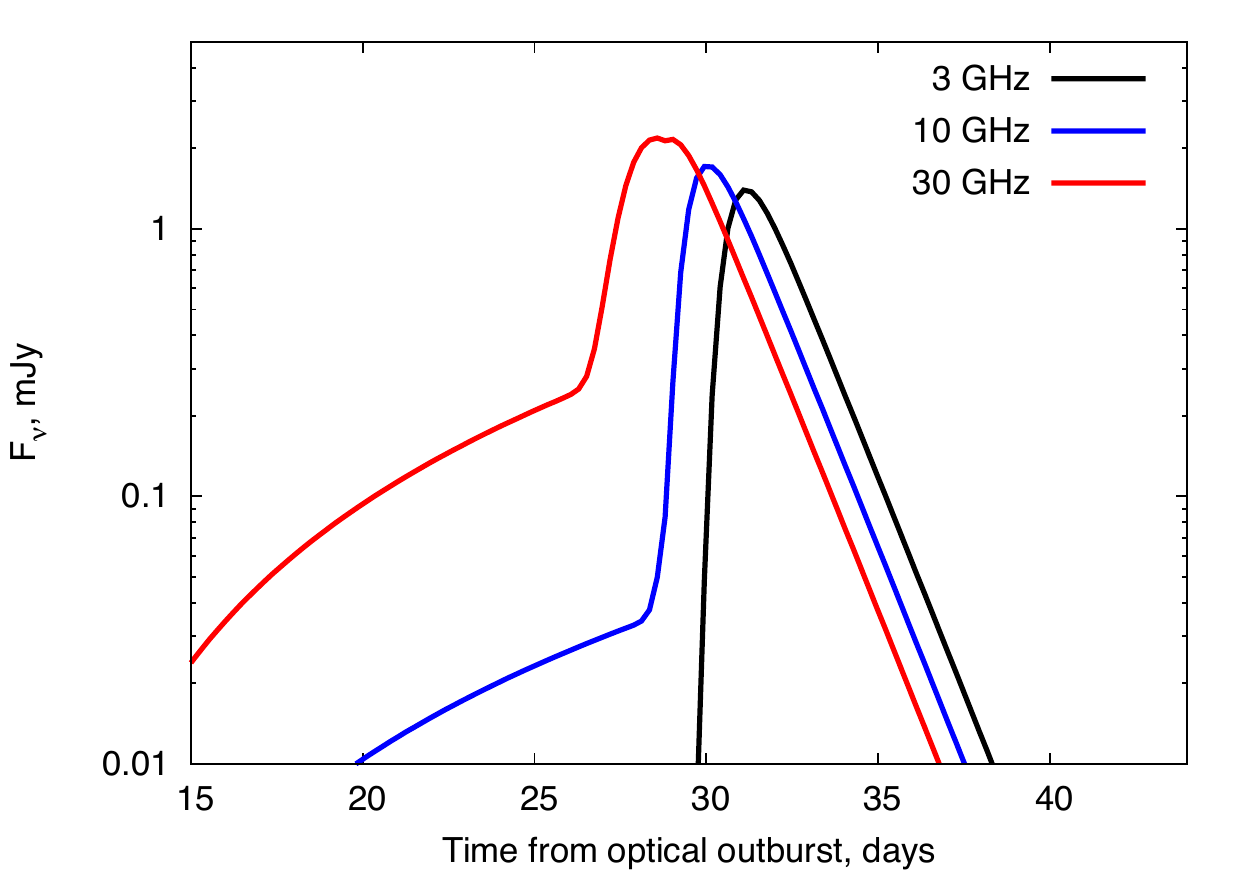}}
\subfigure{
\includegraphics[width=0.5\textwidth]{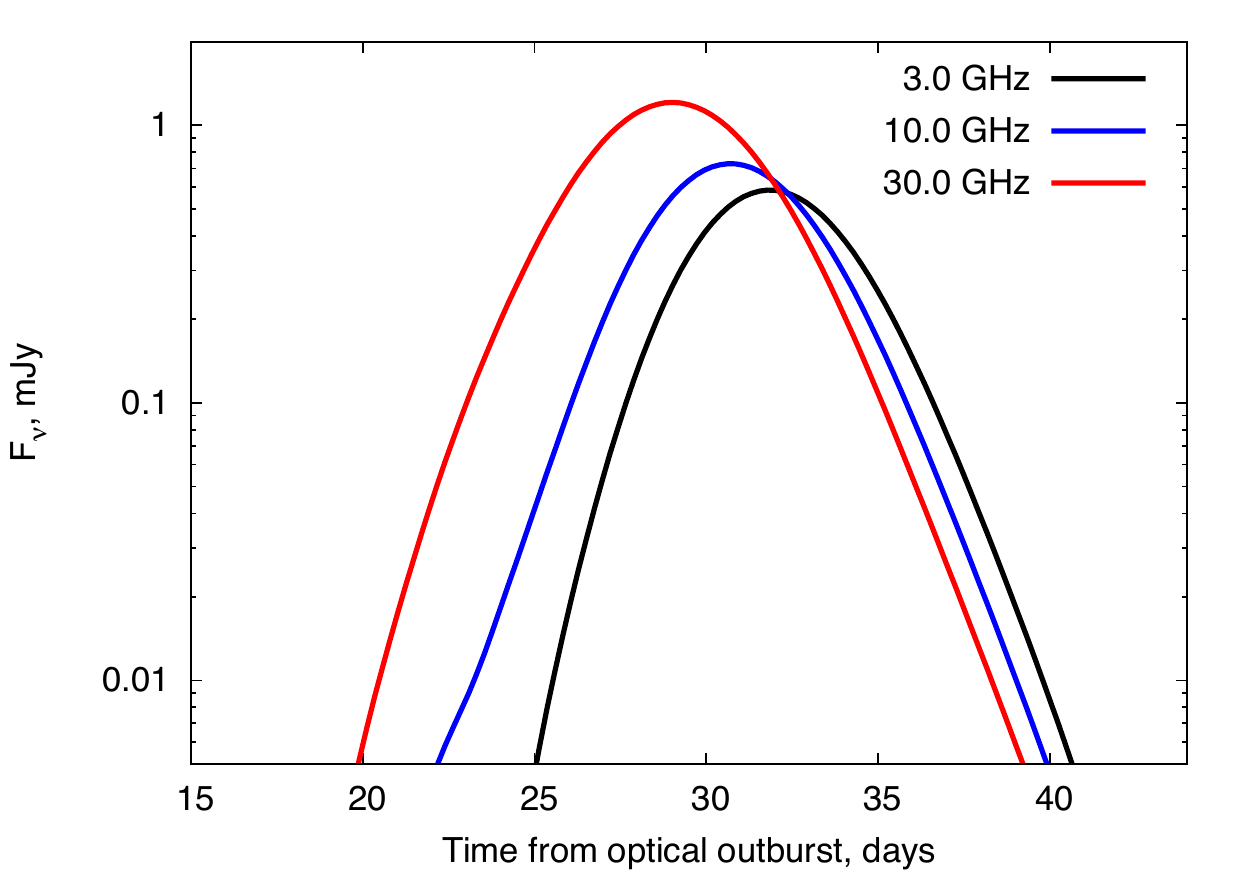}}
\subfigure{
\includegraphics[width=0.5\textwidth]{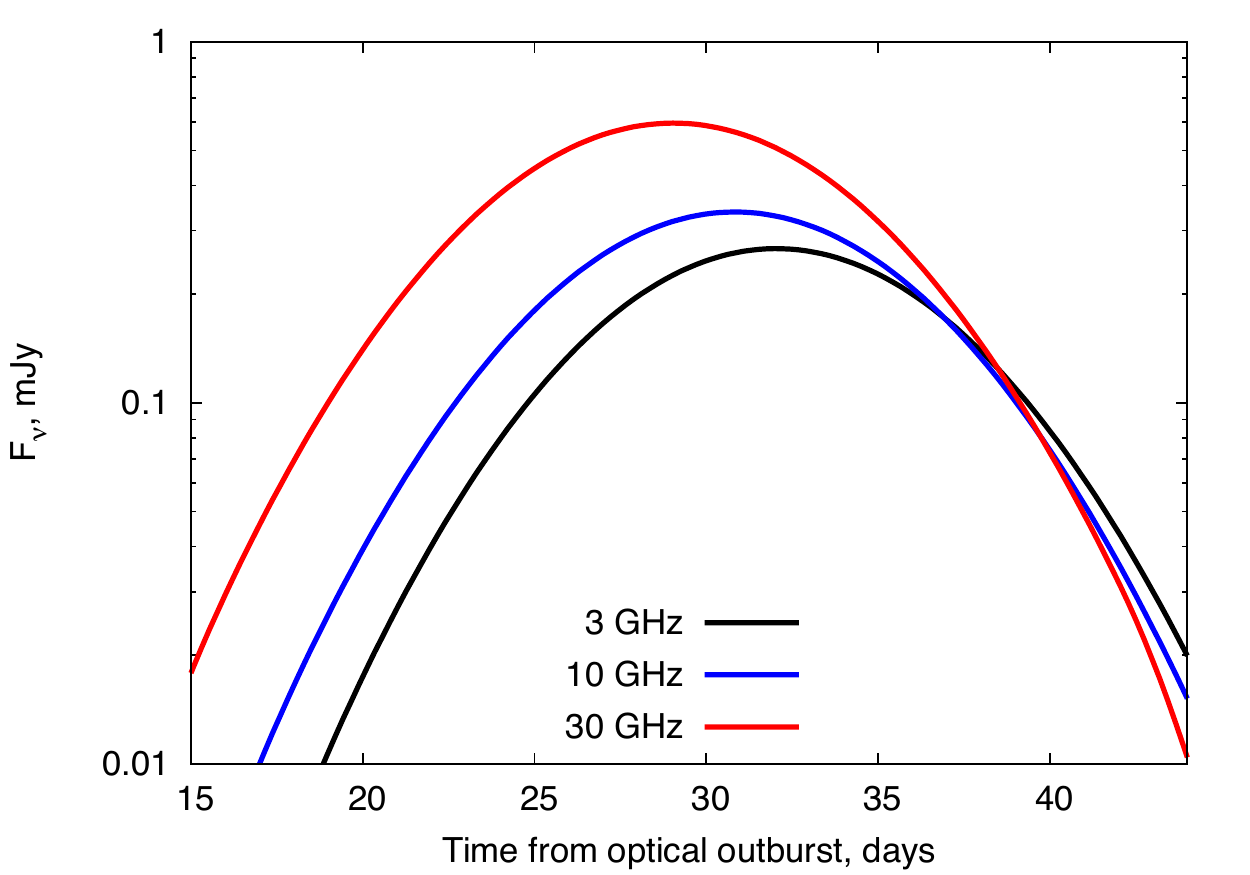}}
\caption{Example light curves (with late-time thermal emission subtracted) for a steep density gradient $H = 5\times 10^{12}$ cm, which has been convolved with a Gaussian of different widths $\sigma = 0$ (top panel), $\sigma$ = 2 days (middle panel), and $\sigma = 5$ days (bottom panel).  This illustrates how a shock propagating down a steep density profile, but reaching the DES surface at different times at different locations across the ejecta surface, can produce the appearance of an achromatic light curve.  }
\label{fig:smeared}
\end{figure}

Figure \ref{fig:smeared} shows the light curves of our single-zone model (for an assumed scale height of $H = 5\times 10^{12}$ cm) convolved with a Gaussian profile of different widths $\sigma$, in order to crudely mimic the effects of non-simultaneous shock emergence.  The top panel shows the unaltered radio lightcurve, i.e. assuming simultaneous shock emergence, which again makes clear that the high frequencies peak first ($\S\ref{sec:radiopk}$).  The middle and bottom panels show the same light curve, but smeared over time intervals of $\sigma=2$ days and $\sigma=5$ days, respectively.  
In this way, the difference between the times of maximum at different frequencies can be substantially smaller than the rise or fall time, as observed in V1723 Aql. 

\subsubsection{Collision With Optically Thin Shell}
\label{sec:thin}

Another possibility is that the radio-producing shock occurs in a dilute thin shell, which has a lower density than the mean density of the DES, i.e. $n_0 \ll n_{\rm pk} \sim 10^{7}$ cm$^{-1}$.  Absorption then plays no part in the lightcurve rise, which is instead controlled by the time required for the shock to cross the radial width of the shell.  

The free-free optical depth of a shell of radial thickness $D = 10^{14}D_{14}$ cm and temperature $T_{\rm sh}=2\cdot 10^4$ K is given by
\be
 \tau_{\rm sh}=3.5 n_{4,7}^2 D_{14} \nu_{10}^{-2}, 
\ee
 The density and radiative efficiency of the shell can be written as
\be
 n_{\rm 4}=1.7\times 10^{6} \tau_{\rm sh,-1}^{1/2} D_{14}^{-1/2} \nu_{10}\,\, {\rm cm^{-3}} \\
 \eta \equiv \frac{t_{\rm cool}}{t_{\rm sh}}=3.1 \tvfe^{4.4}\tau_{\rm sh,-1}^{-1/2}D_{14}^{-1/2}\nu_{10}^{-1},
\ee
where $\tau_{\rm sh}=0.1\tau_{\rm sh,-1}$ is normalized to a sufficiently low value so as not to influence the radio lightcurve significantly.  

Although the parameter space for an optically thin shell ($\tau_{\rm sh} < 0.1$) which is radiative ($\eta \lesssim 1$) is not large, this represents a viable explanation for the behavior of V1723 Aquila. 

\subsection{Radio Emission from Polar Adiabatic Shocks} 
\label{sec:fluxadiab}

\begin{figure}
\begin{center}
\includegraphics[width=1.0\linewidth]{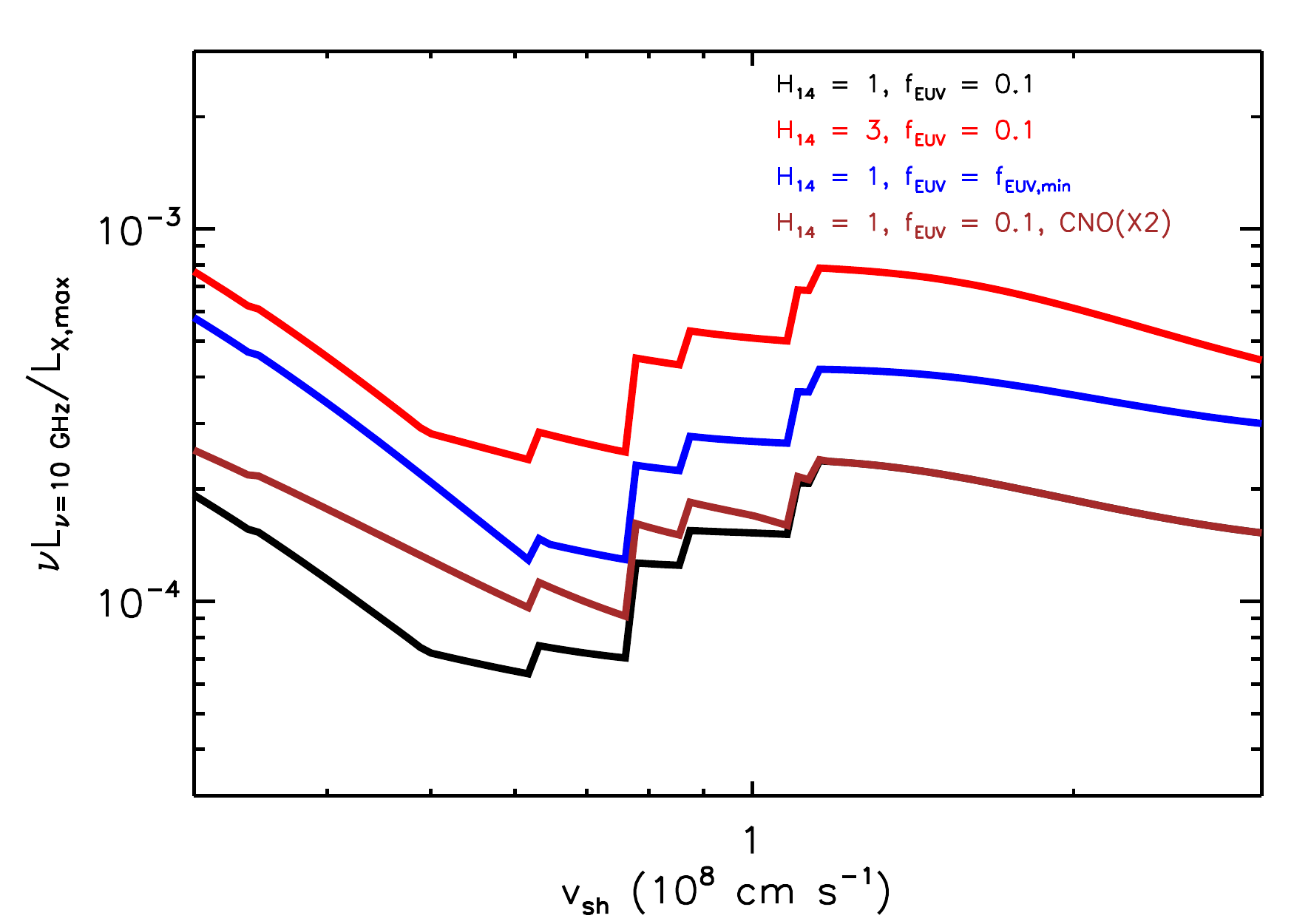}
\includegraphics[width=1.0\linewidth]{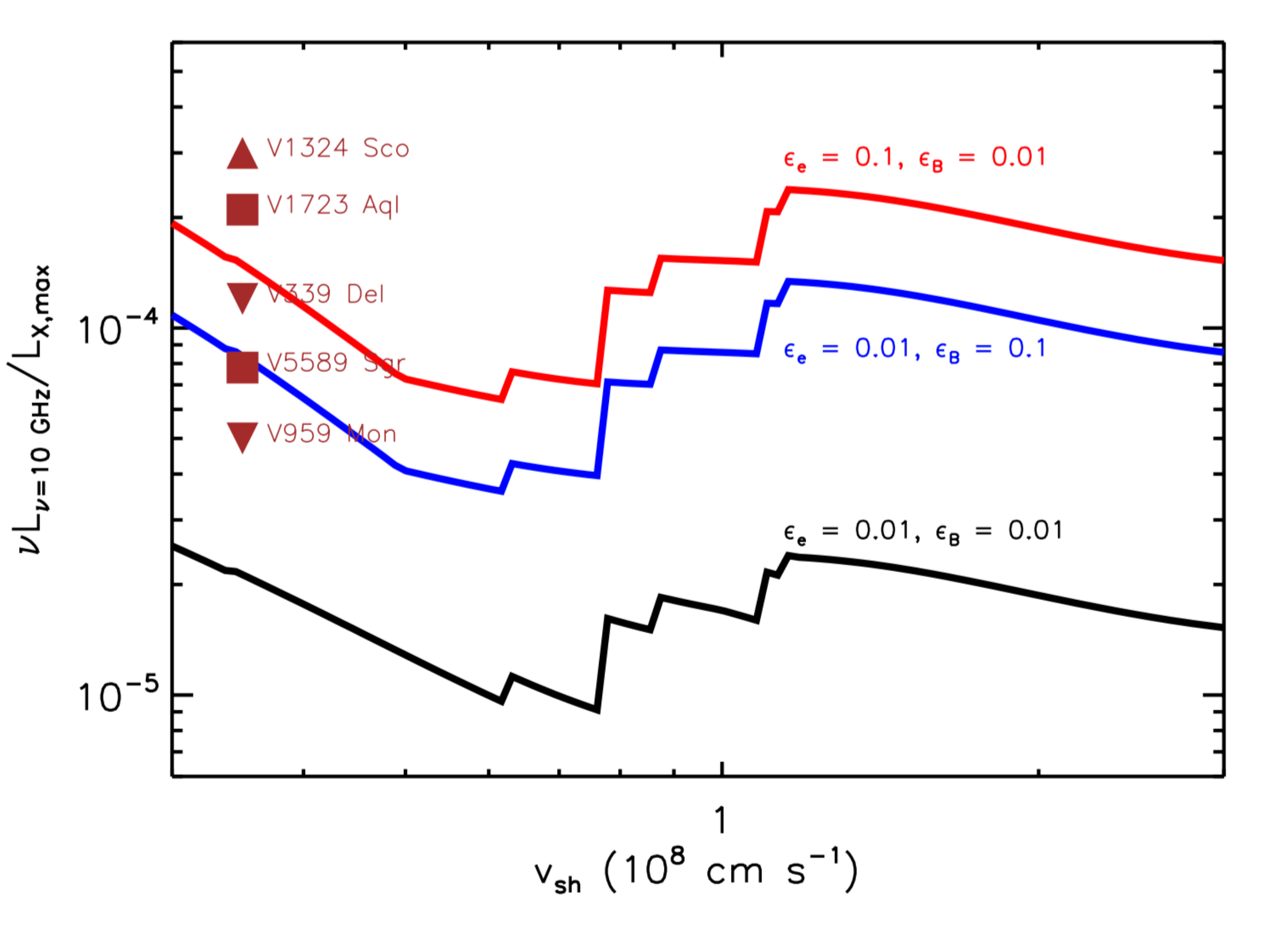}
\caption{Ratio of nonthermal peak 10 GHz radio luminosity (eq.~[\ref{eq:Tnthestpk2}]) to peak thermal X-ray luminosity (eq.~[\ref{eq:Lx}]) as a function of shock velocity $v_{\rm sh}$.  The X-ray luminosity is calculated for the preshock density $n_4 = n_{X}$ (eq.~[\ref{eq:nX}]), the latter calculated using the bound-free cross sections from \citet{Verner+96} for our assumed ejecta composition.  The top panel shows that the ratio of luminosities varies by only a factor of a few for realistic variations in the density scale-height $H$, the ionized fraction $f_{\rm EUV}$, or if the CNO mass fraction is doubled.  The bottom panel shows the more sensitive dependence of the luminosity ratio on the shock microphysical parameters $\epsilon_e$ and $\epsilon_B$, calculated for $H_{14} = 1, f_{\rm EUV} = 0.1$, and standard CNO abundances.  Shown for comparison are measurements (squares), or limits (triangles) on the luminosity ratio of the novae compiled in Table \ref{table:data}.   }
\label{fig:LRLXratio}
\end{center}
\end{figure}

Figure \ref{fig:LradLX} shows the maximum X-ray luminosity and the maximum 10 GHz radio luminosity as a function of the shock velocity, spanning the range from radiative ($v_8 \lesssim 1$) to adiabatic ($v_8 \gtrsim 1$) shocks.  Both radio and X-ray luminosities increase by many orders of magnitude across this range.  However, their {\it ratio} $L_{\rm R}/L_{\rm X}$ coincidentally varies only weakly with $v_8$, as shown in Figure \ref{fig:LRLXratio}.  Absolute luminosities depend on the covering fraction and radius of the shock, but this ratio is obviously independent of these uncertain quantities.  Reasonable variations in the density scaleheight $H$, ionization fraction $f_{\rm EUV}$, and the CNO abundances result in a factor $\lesssim 3$ variation in this ratio (Fig.~\ref{fig:LRLXratio}, top panel), which is instead most sensitive to the microphysical parameters, $\propto \epsilon_e \epsilon_B^{3/4}$ for $p = 2$ (Fig.~\ref{fig:LRLXratio}, bottom panel).  Any shock producing X-rays will therefore also produce radio emission of intensity comparable to the observed values.   Fig.~\ref{fig:LRLXratio} shows for comparison the measured value of $L_{\rm R}/L_{\rm X}$, or limits on its value, using the observed peak X-ray and non-thermal radio luminosities for five novae.

The $L_{\rm R}/L_{\rm X}$ ratio can deviate from its value shown in Fig.~\ref{fig:LRLXratio} if the density of the shocked shell is sufficiently low that the optical depth of the DES to radio or X-rays is $\lesssim 1$, i.e. if its central density obeys $\bar{n} \ll n_{\rm X}, n_{\rm pk}$.  The ratio of the density of maximum X-ray emission $n_{X}$ (eq.~[\ref{eq:nX}]) to that of maximum radio emission $n_{\rm pk}$ (eq.~[\ref{eq:npk}]) is given by
\be
\frac{n_{X}}{n_{\rm pk}} \approx 1.2 \nu_{10}^{-1}H_{14}^{-1/2}v_{8}^{4}
\ee
For high velocity $v_{8} \gg 1$ adiabatic shocks we can therefore have $n_{\rm pk} \lesssim \bar{n} \lesssim n_{X}$, in which case the X-ray emission is optically thin at peak and hence the ratio $L_{\rm R}/L_{\rm X}$ will be higher than its value in Figure \ref{fig:LRLXratio}.  Likewise, when $\bar{n} \lesssim n_{\rm pk}, n_{\rm X}$ (both X-rays and radio are optically thin at peak), we have that $L_{\rm R}/L_{\rm X} \propto \bar{n}^{-1/4}$ also increases with decreasing $\bar{n}$.  Thus, the $L_{\rm R}/L_{\rm X}$ ratio in Fig.~\ref{fig:LRLXratio} is a conservative minimum, achieved in the limit of a high density DES.

The early-time X-rays in some novae are too hard to originate from radiative shocks (Table \ref{table:data}, Figure \ref{fig:Lxdata}) and thus could instead originate from fast adiabatic shocks in the low density polar regions (Fig.~\ref{fig:cartoon}; $\S\ref{sec:xraysintro}$).  
If all the observed X-rays originate from fast polar shocks, and all radio emission from slower equatorial radiative shocks, then the measured ratio $L_{\rm R}/L_{\rm X} = L_{\rm R,rad}/L_{\rm X} + L_{\rm R,ad}/L_{\rm X}$, comprised of contributions from both radiative and adiabatic shocks, should exceed the ratio shown in Figure \ref{fig:LRLXratio}.  For V1324 Sco, V1723 Aql, V5589 Sgr this requires that $\epsilon_e \epsilon_B^{3/4} \approx 10^{-3}$, consistent with our best-fit values found earlier.  The non-detections of non-thermal radio emission in V339 Del and V959 Mon also do not contradict $\epsilon_e \epsilon_B^{3/4} \approx 10^{-3}$. We conclude that, within the uncertainties, the thermal X-ray and non-thermal radio emission in novae is consistent with originating from distinct shocks.  


\section{Conclusions}
\label{sec:conclusions}

We have explored non-thermal synchrotron radio emission from radiative shocks as a model for the early radio peaks in novae by means of a one-dimensional model.  Broadly speaking, we find that the measured brightness temperatures can be explained for physically reasonable parameters of the shock velocity and microphysical parameters $\epsilon_e$ and $\epsilon_B$ (Table \ref{table:sync}).  The presence of a detectable early radio peak requires the presence of DES of density $\gtrsim 10^6-10^7$ cm$^{-3}$ on a radial scale of $\approx 10^{14}-10^{15}$ cm from the white dwarf.  The emission from the nova ejecta colliding with a progenitor wind of mass loss rate $\dot{M}\lesssim 10^{-9} M_\odot$ yr$^{-1}$ is not sufficient to explain the observed radio peaks ($\S$\ref{sec:wind}).

The thin photo-ionized layer ahead of the forward shock plays a key role in our model, as its free-free optical depth determines the time, intensity and spectral indices near the peak of the emission (see $\S$ \ref{sec:radiopk}).  One-dimensional models robustly predict that higher frequency emission peaks at earlier times, imprinting a distinct evolution on the spectral index (Figure \ref{fig:tbs}) which is independent of whether the emission mechanism is thermal or non-thermal.  V1324 Sco and V5589 Sgr do exhibit this behavior, but in V1723 Aql the light curves at all frequencies peak at nearly the same time.  Possible explanations include the nova outflow colliding with a shell of much lower density than the mean density of the nova ejecta ($\S$\ref{sec:aql}).  Alternatively, the evolution may appear achromatic if the shock breaks out of the DES at different times across different parts of its surface, as would occur if the DES is not spherically symmetric (Fig.~\ref{fig:smeared}).

Because the light curve evolution is controlled by free-free opacity effects, one cannot readily distinguish thermal from non-thermal emission based on the spectral index evolution alone. Thermal emission models exhibit spectral index behaviour \citep{Metzger+14}  which is very similar to the non-thermal model described here (Figure \ref{fig:tbs}).  Non-thermal emission is best distinguished by the much higher brightness temperatures which can be achieved than for thermal emission.


The relativistic electrons (or positrons) responsible for the non-thermal synchrotron emission originate either from direct diffusive acceleration at the shock (leptonic scenario), or as secondary products from the decays of pions produced in proton-proton collisions (hadronic scenario).  In either case, the radio-emitting leptons are identical, or tightly related to, the particles which power the observed $\g$-ray emission.  For instance, if the observed $\gtrsim 100$ MeV gamma-rays are produced by relativistic bremsstrahlung emission (as favored in leptonic scenario by \citealt{Metzger+15}), then the energy of the radiating electrons (also $\gtrsim 100$ MeV) are very close to those which determine the peak of the radio synchrotron emission at later times (eq.~[\ref{eq:gammapk}]).  Likewise, gamma-rays from $\pi^{0}$ decay in hadronic scenarios are accompanied by electron/positron pairs with energies $\gtrsim 100$ MeV in the same range, as determined by the pion rest mass.  Once produced, relativistic leptons evolve approximately adiabatically downstream of the shock for conditions which characterize the radio maximum.  The only possible exception is Coulomb cooling in cases when free-free emission dominates lines in cooling the post-shock thermal gas (Fig.~\ref{fig:tbacc}).

In hadronic scenarios, only a small fraction of relativistic protons produce pions in the downstream before being advected into the central cold shell (from which radio emission is heavily attenuated by free-free absorption).  Relativistic protons therefore provide considerable pressure support in the post-shock cooling layer, preventing the gas from compressing by more than an order of magnitude ($\S$\ref{sec:pev}).  The possible role of clumping due to thermal instability of the radiative shock on this non-thermal pressure support deserves further attention, as the central shell may provide a shielded environment which is conducive to dust and molecule formation in novae (Derdzinski et al., in prep).   

The small number of pions produced in the rapidly-cooling post-shock layer implies that of the $\lesssim 10\%$ of the total shock power placed into relativistic protons, only a tiny fraction $\eps_e \sim 10^{-4}$ (eq.~[\ref{eq:eehadron}]) goes into $e^{\pm}$ pairs capable of producing detectable radio emission.  For physical values of $\eps_B\lesssim 0.1$, the critical product $\eps_e\eps_B^{3/4} \lesssim 10^{-5}$ which controls the brightness temperature is a few orders of magnitude below that required by data of $\sim 10^{-4}-10^{-3}$ (eq. [\ref{eq:firsteeconstr}-\ref{eq:lasteeconstr}]).  We therefore {\it disfavor the hadronic scenario for radio-producing leptons}.  This does not rule out a hadronic origin for the $\g$-ray emission, but it does suggest that the direct acceleration of electrons is occurring at the gamma-ray producing shocks.  

Within leptonic scenarios, the values of $\eps_e \eps_B^{3/4}$ required by the radio data are typically higher than those measured from PIC simulations (\citealt{Caprioli&Spitkovsky14a}) and by modeling Galactic SN remnants (\citealt{Morlino&Caprioli12}).  However, values of $\eps_e \eps_B^{3/4}\sim 10^{-4}-10^{-3}$ do appear similar to those inferred by modeling young radio supernovae (e.g.~\citealt{Chandra+12}).

 We confirm the finding of \citet{Metzger+14} that when CIE line cooling is included, thermal emission from radiative shocks cannot explain the early high brightness temperature radio peak.  Perhaps the best evidence for non-thermal radio emission comes from cases like V1324 Sco, where no X-ray emission is detected.  An X-ray non-detection implies a low velocity, radiative shock, which would be especially challenged to produce a significant radio flux in the early peak without a non-thermal contribution.  We can also rule out thermal emission from an adiabatic shock, at least under the assumption that only the first scaleheight behind the shock dominates that contributing to the observed emission ($\S$\ref{sec:fluxadiab}).  Future radiation hydrodynamical simulations of adiabatic shocks and their radio emission in the case of a steep density gradient are needed to determine whether thermal adiabatic shocks can be completely ruled out.

We have derived analytic expressions for the peak brightness temperature of non-thermal synchroton emission (eq.~[\ref{eq:Tnthestpk2}]).  Our model can reproduce the observed fluxes of V5589 Sgr, V1723 Aql, V1324 Sco and, in the case of V1324 Sco, details of the radio lightcurve. Non-detections of early radio peak from V959 Mon and V339 Del place upper limits on $\eps_e,\eps_B$.  The upper limits for V339 Del and V959 Mon are broadly consistent with the inferred values of $\eps_e, \eps_B$ values for novae with detected early radio peak. 

In summary, radio observations of novae provide a important tool for studying particle acceleration and magnetic field amplification in shocks which is complementary to $\g$-rays observations.  They also inform our understanding of the structure of the nova ejecta and internal shocks, which may vary considerably with time and as a function of polar angle relative to the binary axis.  

\section*{Acknowledgements}
We thank Laura Chomiuk, Tom Finzell, Yuri Levin, Koji Mukai, Michael Rupen, Jeno Sokoloski, and Jennifer Weston for helpful conversations.  We thank Jeno Sokoloski for providing us with the numerical tools to fit the thermal radio emission.  ADV and BDM gratefully acknowledge support from NASA grants NNX15AU77G (Fermi), NNX15AR47G (Swift), and NNX16AB30G (ATP), and NSF grant AST-1410950, and
the Alfred P. Sloan Foundation.

\appendix

\section{Non-Thermal Electron Cooling in the Post-Shock Cooling Layer}

\label{sec:timescales}

In calculating the radio synchrotron emission from nova shocks, we assume that relativistic electrons and $e^{\pm}$ pairs evolve adiabatically in the post-shock cooling layer.  This assumption is justified here by comparing various sources of cooling (Coulomb, bremsstrahlung, inverse Compton) to that of the background thermal plasma, $t_{\rm cool}$ (eq.~[\ref{eq:tcool}]).  We again focus on electrons or positrons with Lorentz factors of $\gamma_{\rm pk} \sim 200$ (eq.~[\ref{eq:gammapk}]), as these determine the synchrotron flux near the light curve peak, where $n_3 = 4n_4 = 4n_{\rm pk}$.

First note that thermal electrons and protons are well coupled behind the shock by Coulomb collisions and thus share a common temperature. 
At radio maximum ($n_4 = n_{\rm pk,\Delta}$) the Coulomb equilibration timescale for the thermal plasma,
$t_{\rm e-p} = 3.5 T_{\rm 3}^{3/2}/n_{\rm pk}$ s (\citealt{Huba07}), is short compared to the thermal cooling timescale $t_{\rm cool}$ (eq.~[\ref{eq:tcool}]),
\be
\frac{t_{\rm e-p}}{t_{\rm cool}} \approx
0.03 \,
\left(\frac{T}{T_3} \right)^{-0.2} \tvfe^{-0.4}.
\ee
This ratio evolves only weakly as gas cools behind the shock.

Relativistic electrons experience Coulomb losses on thermal background electrons on a timescale which is given by
\be
\frac{t_{e-e}}{t_{\rm cool}} =
\frac{2\gamma_{\rm pk}}{3c\sigma_{\rm T} n_3 \ln\Lambda \, t_{\rm cool}}=
10 \,
\tvfe^{-3.4} \, \left( \frac{\gamma_{\rm pk}}{200} \right) \, \left(\frac{\ln \Lambda}{20} \right)^{-1}.
\label{eq:tcoul}
\ee
Coulomb losses can thus be potentially important behind the shock where $n \gg n_{\rm pk}$.  

The cooling rate due to relativistic bremsstrahlung emission from interaction with thermal protons and electrons is well approximated by $\dot{\gamma}_{\rm rb} \approx (5/3) c\sigma_{\rm T} \alpha_{\rm fs} n_3 \gamma^{1.2}$, an expression which is accurate to $\sim 10-20\%$ between $\gamma\sim 10-10^3$, where $\alpha_{\rm fs} \simeq 1/137$ is the fine structure constant.  This yields
\be
\frac{t_{\rm rb}}{t_{\rm cool}} = \frac{\gamma_{\rm pk}}{\dot{\gamma}_{\rm rb} \, t_{\rm cool}} \simeq
\frac{3}{5c\sigma_{\rm T} \alpha_{\rm fs} n_3 \gamma_{\rm pk}^{0.2} \, t_{\rm cool}} \approx
45 \,
\tvfe^{-3.4} \, \left( \frac{\gamma_{\rm pk}}{200} \right)^{-0.2}.
\ee

The ratio of the synchrotron cooling timescale $t_{\rm syn} = 6\pi m_e c/(\sigma_T B_{\rm sh}^{2}\gamma)$
for electrons with $\gamma = \gamma_{\rm pk}$
to the {thermal} cooling time is given by
\be
\frac{t_{\rm syn}}{t_{\rm cool}} \approx 
100 \,
\eps_{B,-2}^{-1} \tvfe^{-5.4} \left( \frac{\gamma_{\rm pk}}{200} \right)^{-1}.
\ee

Finally, the characteristic timescale for inverse Compton cooling on the shock-produced thermal X-rays is $t_{\rm IC} = 3m_e c/( 4 \sigma_T U_{\rm rad}\gamma_{\rm pk})$, such that
\be
\frac{t_{\rm IC}}{t_{\rm cool}} &=&
\frac{3 m_e c }{4 \sigma_T U_{\rm rad}\gamma_{\rm pk} t_{\rm cool}} \approx
 1.7 \times 10^3 \,
\tvfe^{-6.4} \, \left( \frac{\gamma_{\rm pk}}{200} \right)^{-1},
\ee 
where  $U_{\rm rad} =(3/8)(\tvf/c)n_3 kT_3/\mu$
is the X-ray energy density.\footnote{Soft X-rays are in the Thomson regime for interacting with $\gamma \sim 100$ electrons.}

In summary, relativistic radio-emitting leptons cool in the post-shock gas on a timescale which is generally much longer than that of the thermal background plasma for the range of velocities $v_8 \lesssim 1$ of radiative shocks.  This justifies evolving their energies adiabatically behind the shock, with the possible exception of Coulomb losses, which are unimportant immediately behind the shock but may become so as gas compresses to higher densities $n \gg n_3$.


\begin{thebibliography}{}

\bibitem[\protect\citeauthoryear{{Abdo}, {Ackermann}, {Ajello}, {Atwood},
  {Baldini}, {Ballet}, {Barbiellini}, {Bastieri}, {Bechtol}, {Bellazzini} \& et
  al.}{{Abdo} et~al.}{2010}]{Abdo+10}
{Abdo} A.~A.,  {Ackermann} M.,  {Ajello} M.,  {Atwood} W.~B.,  {Baldini} L.,
  {Ballet} J.,  {Barbiellini} G.,  {Bastieri} D.,  {Bechtol} K.,  {Bellazzini}
  R.,    et al. 2010, Science, 329, 817

\bibitem[\protect\citeauthoryear{{Ackermann} et~al.,}{{Ackermann}
  et~al.}{2014}]{Ackermann+14}
{Ackermann} M.,  et~al., 2014, Science, 345, 554

\bibitem[\protect\citeauthoryear{{Bath} \& {Shaviv}}{{Bath} \&
  {Shaviv}}{1976}]{Bath&Shaviv76}
{Bath} G.~T.,  {Shaviv} G.,  1976, \mnras, 175, 305

\bibitem[\protect\citeauthoryear{{Bell}}{{Bell}}{2004}]{Bell04}
{Bell} A.~R.,  2004, \mnras, 353, 550

\bibitem[\protect\citeauthoryear{{Blandford} \& {Ostriker}}{{Blandford} \&
  {Ostriker}}{1978}]{Blandford&Ostriker78}
{Blandford} R.~D.,  {Ostriker} J.~P.,  1978, \apjl, 221, L29

\bibitem[\protect\citeauthoryear{Bode \& Evans}{Bode \&
  Evans}{2008}]{BodeEvans08}
Bode M.,  Evans A.,  2008, Classical Novae.
Cambridge Astrophysics, Cambridge University Press

\bibitem[\protect\citeauthoryear{{Caprioli} \& {Spitkovsky}}{{Caprioli} \&
  {Spitkovsky}}{2014a}]{Caprioli&Spitkovsky14a}
{Caprioli} D.,  {Spitkovsky} A.,  2014a, \apj, 783, 91

\bibitem[\protect\citeauthoryear{{Caprioli} \& {Spitkovsky}}{{Caprioli} \&
  {Spitkovsky}}{2014b}]{Caprioli&Spitkovsky14b}
{Caprioli} D.,  {Spitkovsky} A.,  2014b, \apj, 794, 46

\bibitem[\protect\citeauthoryear{{Caprioli} \& {Spitkovsky}}{{Caprioli} \&
  {Spitkovsky}}{2014c}]{Caprioli&Spitkovsky14c}
{Caprioli} D.,  {Spitkovsky} A.,  2014c, \apj, 794, 47

\bibitem[\protect\citeauthoryear{{Casanova}, {Jos{\'e}}, {Garc{\'{\i}}a-Berro},
  {Shore} \& {Calder}}{{Casanova} et~al.}{2011}]{Casanova+11}
{Casanova} J.,  {Jos{\'e}} J.,  {Garc{\'{\i}}a-Berro} E.,  {Shore} S.~N.,
  {Calder} A.~C.,  2011, \nat, 478, 490

\bibitem[\protect\citeauthoryear{{Chandra}, {Chevalier}, {Chugai}, {Fransson},
  {Irwin}, {Soderberg}, {Chakraborti} \& {Immler}}{{Chandra}
  et~al.}{2012}]{Chandra+12}
{Chandra} P.,  {Chevalier} R.~A.,  {Chugai} N.,  {Fransson} C.,  {Irwin} C.~M.,
   {Soderberg} A.~M.,  {Chakraborti} S.,    {Immler} S.,  2012, \apj, 755, 110

\bibitem[\protect\citeauthoryear{{Chevalier}}{{Chevalier}}{1982}]{Chevalier82}
{Chevalier} R.~A.,  1982, \apj, 259, 302

\bibitem[\protect\citeauthoryear{{Chevalier} \& {Imamura}}{{Chevalier} \&
  {Imamura}}{1982}]{Chevalier&Imamura82}
{Chevalier} R.~A.,  {Imamura} J.~N.,  1982, \apj, 261, 543

\bibitem[\protect\citeauthoryear{{Chomiuk}, {Linford}, {Yang}, {O'Brien},
  {Paragi}, {Mioduszewski}, {Beswick}, {Cheung}, {Mukai}, {Nelson}, {Ribeiro},
  {Rupen}, {Sokoloski}, {Weston}, {Zheng}, {Bode}, {Eyres}, {Roy} \&
  {Taylor}}{{Chomiuk} et~al.}{2014}]{Chomiuk+14b}
{Chomiuk} L.,  {Linford} J.~D.,  {Yang} J.,  {O'Brien} T.~J.,  {Paragi} Z.,
  {Mioduszewski} A.~J.,  {Beswick} R.~J.,  {Cheung} C.~C.,  {Mukai} K.,
  {Nelson} T.,  {Ribeiro} V.~A.~R.~M.,  {Rupen} M.~P.,  {Sokoloski} J.~L.,
  {Weston} J.,  {Zheng} Y.,  {Bode} M.~F.,  {Eyres} S.,  {Roy} N.,    {Taylor}
  G.~B.,  2014, \nat, 514, 339

\bibitem[\protect\citeauthoryear{{Chomiuk}, {Nelson}, {Mukai}, {Sokoloski},
  {Rupen}, {Page}, {Osborne}, {Kuulkers}, {Mioduszewski}, {Roy}, {Weston} \&
  {Krauss}}{{Chomiuk} et~al.}{2014}]{Chomiuk+14a}
{Chomiuk} L.,  {Nelson} T.,  {Mukai} K.,  {Sokoloski} J.~L.,  {Rupen} M.~P.,
  {Page} K.~L.,  {Osborne} J.~P.,  {Kuulkers} E.,  {Mioduszewski} A.~J.,  {Roy}
  N.,  {Weston} J.,    {Krauss} M.~I.,  2014, \apj, 788, 130

\bibitem[\protect\citeauthoryear{{Finzell}, {Chomiuk}, {Munari} \&
  {Walter}}{{Finzell} et~al.}{2015}]{Finzell+15}
{Finzell} T.,  {Chomiuk} L.,  {Munari} U.,    {Walter} F.~M.,  2015, \apj, 809,
  160

\bibitem[\protect\citeauthoryear{{Friedjung} \& {Duerbeck}}{{Friedjung} \&
  {Duerbeck}}{1993}]{Friedjung&Duerbeck93}
{Friedjung} M.,  {Duerbeck} H.~W.,  1993, NASA Special Publication, 507, 371

\bibitem[\protect\citeauthoryear{{Gallagher} \& {Starrfield}}{{Gallagher} \&
  {Starrfield}}{1978}]{Gallagher&Starrfield78}
{Gallagher} J.~S.,  {Starrfield} S.,  1978, \araa, 16, 171

\bibitem[\protect\citeauthoryear{{Huba}}{{Huba}}{2007}]{Huba07}
{Huba} J.~D.,  2007, {Nrl Plasma Formulary}.
{Wexford College Press}

\bibitem[\protect\citeauthoryear{{Kamae}, {Karlsson}, {Mizuno}, {Abe} \&
  {Koi}}{{Kamae} et~al.}{2006}]{Kamae+06}
{Kamae} T.,  {Karlsson} N.,  {Mizuno} T.,  {Abe} T.,    {Koi} T.,  2006, \apj,
  647, 692

\bibitem[\protect\citeauthoryear{{Kato}}{{Kato}}{2014}]{Kato14}
{Kato} T.~N.,  2014, ArXiv e-prints

\bibitem[\protect\citeauthoryear{{Kee}, {Owocki} \& {ud-Doula}}{{Kee}
  et~al.}{2014}]{Kee+14}
{Kee} N.~D.,  {Owocki} S.,    {ud-Doula} A.,  2014, \mnras, 438, 3557

\bibitem[\protect\citeauthoryear{{Krauss}, {Chomiuk}, {Rupen}, {Roy},
  {Mioduszewski}, {Sokoloski}, {Nelson}, {Mukai}, {Bode}, {Eyres} \&
  {O'Brien}}{{Krauss} et~al.}{2011}]{Krauss+11}
{Krauss} M.~I.,  {Chomiuk} L.,  {Rupen} M.,  {Roy} N.,  {Mioduszewski} A.~J.,
  {Sokoloski} J.~L.,  {Nelson} T.,  {Mukai} K.,  {Bode} M.~F.,  {Eyres}
  S.~P.~S.,    {O'Brien} T.~J.,  2011, \apjl, 739, L6

\bibitem[\protect\citeauthoryear{{Linford}, {Ribeiro}, {Chomiuk}, {Nelson},
  {Sokoloski}, {Rupen}, {Mukai}, {O'Brien}, {Mioduszewski} \&
  {Weston}}{{Linford} et~al.}{2015}]{Linford+15}
{Linford} J.~D.,  {Ribeiro} V.~A.~R.~M.,  {Chomiuk} L.,  {Nelson} T.,
  {Sokoloski} J.~L.,  {Rupen} M.~P.,  {Mukai} K.,  {O'Brien} T.~J.,
  {Mioduszewski} A.~J.,    {Weston} J.,  2015, \apj, 805, 136

\bibitem[\protect\citeauthoryear{{Livio}, {Shankar}, {Burkert} \&
  {Truran}}{{Livio} et~al.}{1990}]{Livio+90}
{Livio} M.,  {Shankar} A.,  {Burkert} A.,    {Truran} J.~W.,  1990, \apj, 356,
  250

\bibitem[\protect\citeauthoryear{{Lloyd}, {O'Brien} \& {Bode}}{{Lloyd}
  et~al.}{1996}]{Lloyd+96}
{Lloyd} H.~M.,  {O'Brien} T.~J.,    {Bode} M.~F.,  1996, in {Taylor} A.~R.,
  {Paredes} J.~M.,  eds, Radio Emission from the Stars and the Sun Vol.~93 of
  Astronomical Society of the Pacific Conference Series, {Models for the Radio
  Emission from Classical Novae}.
p.~200

\bibitem[\protect\citeauthoryear{{Lloyd}, {O'Brien} \& {Bode}}{{Lloyd}
  et~al.}{1997}]{Lloyd+97}
{Lloyd} H.~M.,  {O'Brien} T.~J.,    {Bode} M.~F.,  1997, \mnras, 284, 137

\bibitem[\protect\citeauthoryear{{Lloyd}, {O'Brien}, {Bode}, {Predehl},
  {Schmitt}, {Truemper}, {Watson} \& {Pounds}}{{Lloyd} et~al.}{1992}]{Lloyd+92}
{Lloyd} H.~M.,  {O'Brien} T.~J.,  {Bode} M.~F.,  {Predehl} P.,  {Schmitt}
  J.~H.~M.~M.,  {Truemper} J.,  {Watson} M.~G.,    {Pounds} K.~A.,  1992, \nat,
  356, 222

\bibitem[\protect\citeauthoryear{{Martin} \& {Dubus}}{{Martin} \&
  {Dubus}}{2013}]{Martin&Dubus13}
{Martin} P.,  {Dubus} G.,  2013, \aap, 551, A37

\bibitem[\protect\citeauthoryear{{Metzger}, {Caprioli}, {Vurm}, {Beloborodov},
  {Bartos} \& {Vlasov}}{{Metzger} et~al.}{2016}]{Metzger+16}
{Metzger} B.~D.,  {Caprioli} D.,  {Vurm} I.,  {Beloborodov} A.~M.,  {Bartos}
  I.,    {Vlasov} A.,  2016, \mnras, 457, 1786

\bibitem[\protect\citeauthoryear{{Metzger}, {Finzell}, {Vurm}, {Hasco{\"e}t},
  {Beloborodov} \& {Chomiuk}}{{Metzger} et~al.}{2015}]{Metzger+15}
{Metzger} B.~D.,  {Finzell} T.,  {Vurm} I.,  {Hasco{\"e}t} R.,  {Beloborodov}
  A.~M.,    {Chomiuk} L.,  2015, \mnras, 450, 2739

\bibitem[\protect\citeauthoryear{{Metzger}, {Hasco{\"e}t}, {Vurm},
  {Beloborodov}, {Chomiuk}, {Sokoloski} \& {Nelson}}{{Metzger}
  et~al.}{2014}]{Metzger+14}
{Metzger} B.~D.,  {Hasco{\"e}t} R.,  {Vurm} I.,  {Beloborodov} A.~M.,
  {Chomiuk} L.,  {Sokoloski} J.~L.,    {Nelson} T.,  2014, \mnras, 442, 713

\bibitem[\protect\citeauthoryear{{Morlino} \& {Caprioli}}{{Morlino} \&
  {Caprioli}}{2012}]{Morlino&Caprioli12}
{Morlino} G.,  {Caprioli} D.,  2012, \aap, 538, A81

\bibitem[\protect\citeauthoryear{{Mukai} \& {Ishida}}{{Mukai} \&
  {Ishida}}{2001}]{Mukai&Ishida01}
{Mukai} K.,  {Ishida} M.,  2001, \apj, 551, 1024

\bibitem[\protect\citeauthoryear{{Mukai}, {Orio} \& {Della Valle}}{{Mukai}
  et~al.}{2008}]{Mukai+08}
{Mukai} K.,  {Orio} M.,    {Della Valle} M.,  2008, \apj, 677, 1248

\bibitem[\protect\citeauthoryear{{Munari}, {Maitan}, {Moretti} \&
  {Tomaselli}}{{Munari} et~al.}{2015}]{Munari+15}
{Munari} U.,  {Maitan} A.,  {Moretti} S.,    {Tomaselli} S.,  2015, \na, 40, 28

\bibitem[\protect\citeauthoryear{{Nelson}, {Mukai}, {Chomiuk}, {Sokoloski},
  {Weston}, {Rupen}, {Mioduszewski} \& {Roy}}{{Nelson}
  et~al.}{2012}]{Nelson+12a}
{Nelson} T.,  {Mukai} K.,  {Chomiuk} L.,  {Sokoloski} J.,  {Weston} J.,
  {Rupen} M.,  {Mioduszewski} A.,    {Roy} N.,  2012, The Astronomer's
  Telegram, 4321

\bibitem[\protect\citeauthoryear{{Ness}, {Schwarz}, {Retter}, {Starrfield},
  {Schmitt}, {Gehrels}, {Burrows} \& {Osborne}}{{Ness} et~al.}{2007}]{Ness+07}
{Ness} J.-U.,  {Schwarz} G.~J.,  {Retter} A.,  {Starrfield} S.,  {Schmitt}
  J.~H.~M.~M.,  {Gehrels} N.,  {Burrows} D.,    {Osborne} J.~P.,  2007, \apj,
  663, 505

\bibitem[\protect\citeauthoryear{{O'Brien}, {Lloyd} \& {Bode}}{{O'Brien}
  et~al.}{1994}]{OBrien+94}
{O'Brien} T.~J.,  {Lloyd} H.~M.,    {Bode} M.~F.,  1994, \mnras, 271, 155

\bibitem[\protect\citeauthoryear{{Orio}}{{Orio}}{2004}]{Orio04}
{Orio} M.,  2004, in {Tovmassian} G.,  {Sion} E.,  eds, Revista Mexicana de
  Astronomia y Astrofisica Conference Series Vol.~20 of Revista Mexicana de
  Astronomia y Astrofisica Conference Series, {X-Ray Observations of Classical
  and recurrent Novae in Outburst}.
pp 182--186

\bibitem[\protect\citeauthoryear{{Osborne}}{{Osborne}}{2015}]{Osborne15}
{Osborne} J.~P.,  2015, Journal of High Energy Astrophysics, 7, 117

\bibitem[\protect\citeauthoryear{{Page} \& {Beardmore}}{{Page} \&
  {Beardmore}}{2013}]{Page&Beardmore13}
{Page} K.~L.,  {Beardmore} A.~P.,  2013, The Astronomer's Telegram, 5429

\bibitem[\protect\citeauthoryear{{Page}, {Osborne}, {Schwarz} \&
  {Walter}}{{Page} et~al.}{2012}]{Page+12}
{Page} K.~L.,  {Osborne} J.~P.,  {Schwarz} G.~J.,    {Walter} F.~M.,  2012, The
  Astronomer's Telegram, 4287, 1

\bibitem[\protect\citeauthoryear{{Park}, {Caprioli} \& {Spitkovsky}}{{Park}
  et~al.}{2014}]{Park+14}
{Park} J.,  {Caprioli} D.,    {Spitkovsky} A.,  2014, ArXiv e-prints

\bibitem[\protect\citeauthoryear{{Ribeiro}, {Munari} \& {Valisa}}{{Ribeiro}
  et~al.}{2013}]{Ribeiro+13}
{Ribeiro} V.~A.~R.~M.,  {Munari} U.,    {Valisa} P.,  2013, \apj, 768, 49

\bibitem[\protect\citeauthoryear{{Riquelme} \& {Spitkovsky}}{{Riquelme} \&
  {Spitkovsky}}{2011}]{Riquelme&Spitkovsky11}
{Riquelme} M.~A.,  {Spitkovsky} A.,  2011, \apj, 733, 63

\bibitem[\protect\citeauthoryear{{Rybicki} \& {Lightman}}{{Rybicki} \&
  {Lightman}}{1979}]{Rybicki&Lightman79}
{Rybicki} G.~B.,  {Lightman} A.~P.,  1979, {Radiative processes in
  astrophysics}

\bibitem[\protect\citeauthoryear{{Schaefer} et~al.,}{{Schaefer}
  et~al.}{2014}]{Schaefer+14}
{Schaefer} G.~H.,  et~al., 2014, \nat, 515, 234

\bibitem[\protect\citeauthoryear{{Schure}, {Kosenko}, {Kaastra}, {Keppens} \&
  {Vink}}{{Schure} et~al.}{2009}]{Schure+09}
{Schure} K.~M.,  {Kosenko} D.,  {Kaastra} J.~S.,  {Keppens} R.,    {Vink} J.,
  2009, \aap, 508, 751

\bibitem[\protect\citeauthoryear{{Schwarz}, {Ness}, {Osborne}, {Page}, {Evans},
  {Beardmore}, {Walter}, {Helton}, {Woodward}, {Bode}, {Starrfield} \&
  {Drake}}{{Schwarz} et~al.}{2011}]{Schwarz+11}
{Schwarz} G.~J.,  {Ness} J.-U.,  {Osborne} J.~P.,  {Page} K.~L.,  {Evans}
  P.~A.,  {Beardmore} A.~P.,  {Walter} F.~M.,  {Helton} L.~A.,  {Woodward}
  C.~E.,  {Bode} M.,  {Starrfield} S.,    {Drake} J.~J.,  2011, \apjs, 197, 31

\bibitem[\protect\citeauthoryear{{Schwarz}, {Shore}, {Starrfield} \&
  {Vanlandingham}}{{Schwarz} et~al.}{2007}]{Schwarz+07}
{Schwarz} G.~J.,  {Shore} S.~N.,  {Starrfield} S.,    {Vanlandingham} K.~M.,
  2007, \apj, 657, 453

\bibitem[\protect\citeauthoryear{{Seaquist} \& {Bode}}{{Seaquist} \&
  {Bode}}{2008}]{Seaquist&Bode08}
{Seaquist} E.~R.,  {Bode} M.~F.~.,  2008, {in Classical Novae}

\bibitem[\protect\citeauthoryear{{Seaquist}, {Duric}, {Israel}, {Spoelstra},
  {Ulich} \& {Gregory}}{{Seaquist} et~al.}{1980}]{Seaquist+80}
{Seaquist} E.~R.,  {Duric} N.,  {Israel} F.~P.,  {Spoelstra} T.~A.~T.,  {Ulich}
  B.~L.,    {Gregory} P.~C.,  1980, \aj, 85, 283

\bibitem[\protect\citeauthoryear{{Shore}}{{Shore}}{2012}]{Shore12}
{Shore} S.~N.,  2012, Bulletin of the Astronomical Society of India, 40, 185

\bibitem[\protect\citeauthoryear{{Shore}, {De Gennaro Aquino}, {Schwarz},
  {Augusteijn}, {Cheung}, {Walter} \& {Starrfield}}{{Shore}
  et~al.}{2013}]{Shore+13}
{Shore} S.~N.,  {De Gennaro Aquino} I.,  {Schwarz} G.~J.,  {Augusteijn} T.,
  {Cheung} C.~C.,  {Walter} F.~M.,    {Starrfield} S.,  2013, \aap, 553, A123

\bibitem[\protect\citeauthoryear{{Sokoloski}, {Luna}, {Mukai} \&
  {Kenyon}}{{Sokoloski} et~al.}{2006}]{Sokoloski+06}
{Sokoloski} J.~L.,  {Luna} G.~J.~M.,  {Mukai} K.,    {Kenyon} S.~J.,  2006,
  \nat, 442, 276

\bibitem[\protect\citeauthoryear{{Starrfield}, {Sparks}, {Truran} \&
  {Wiescher}}{{Starrfield} et~al.}{2000}]{Starrfield+00}
{Starrfield} S.,  {Sparks} W.~M.,  {Truran} J.~W.,    {Wiescher} M.~C.,  2000,
  \apjs, 127, 485

\bibitem[\protect\citeauthoryear{{Strong}, {Porter}, {Digel},
  {J{\'o}hannesson}, {Martin}, {Moskalenko}, {Murphy} \& {Orlando}}{{Strong}
  et~al.}{2010}]{Strong+10}
{Strong} A.~W.,  {Porter} T.~A.,  {Digel} S.~W.,  {J{\'o}hannesson} G.,
  {Martin} P.,  {Moskalenko} I.~V.,  {Murphy} E.~J.,    {Orlando} E.,  2010,
  \apjl, 722, L58

\bibitem[\protect\citeauthoryear{{Taylor}, {Pottasch}, {Seaquist} \&
  {Hollis}}{{Taylor} et~al.}{1987}]{Taylor+87}
{Taylor} A.~R.,  {Pottasch} S.~R.,  {Seaquist} E.~R.,    {Hollis} J.~M.,  1987,
  \aap, 183, 38

\bibitem[\protect\citeauthoryear{{Townsley} \& {Bildsten}}{{Townsley} \&
  {Bildsten}}{2005}]{Townsley&Bildsten05}
{Townsley} D.~M.,  {Bildsten} L.,  2005, \apj, 628, 395

\bibitem[\protect\citeauthoryear{{Vaytet}, {O'Brien}, {Page}, {Bode}, {Lloyd}
  \& {Beardmore}}{{Vaytet} et~al.}{2011}]{Vaytet+11}
{Vaytet} N.~M.~H.,  {O'Brien} T.~J.,  {Page} K.~L.,  {Bode} M.~F.,  {Lloyd} M.,
     {Beardmore} A.~P.,  2011, \apj, 740, 5

\bibitem[\protect\citeauthoryear{{Verner}, {Ferland}, {Korista} \&
  {Yakovlev}}{{Verner} et~al.}{1996}]{Verner+96}
{Verner} D.~A.,  {Ferland} G.~J.,  {Korista} K.~T.,    {Yakovlev} D.~G.,  1996,
  \apj, 465, 487

\bibitem[\protect\citeauthoryear{{Weston}, {Sokoloski}, {Chomiuk}, {Linford},
  {Nelson}, {Mukai}, {Finzell}, {Mioduszewski}, {Rupen} \& {Walter}}{{Weston}
  et~al.}{2015}]{Weston+15b}
{Weston} J.~H.~S.,  {Sokoloski} J.~L.,  {Chomiuk} L.,  {Linford} J.~D.,
  {Nelson} T.,  {Mukai} K.,  {Finzell} T.,  {Mioduszewski} A.,  {Rupen} M.~P.,
    {Walter} F.~M.,  2015, ArXiv e-prints

\bibitem[\protect\citeauthoryear{{Weston}, {Sokoloski}, {Metzger}, {Zheng},
  {Chomiuk}, {Krauss}, {Linford}, {Nelson}, {Mioduszewski}, {Rupen}, {Finzell}
  \& {Mukai}}{{Weston} et~al.}{2015}]{Weston+15a}
{Weston} J.~H.~S.,  {Sokoloski} J.~L.,  {Metzger} B.~D.,  {Zheng} Y.,
  {Chomiuk} L.,  {Krauss} M.~I.,  {Linford} J.,  {Nelson} T.,  {Mioduszewski}
  A.,  {Rupen} M.~P.,  {Finzell} T.,    {Mukai} K.,  2015, ArXiv e-prints

\bibitem[\protect\citeauthoryear{{Weston}, {Sokoloski}, {Zheng}, {Chomiuk},
  {Mioduszewski}, {Mukai}, {Rupen}, {Krauss}, {Roy} \& {Nelson}}{{Weston}
  et~al.}{2013}]{Weston+13}
{Weston} J.~H.~S.,  {Sokoloski} J.~L.,  {Zheng} Y.,  {Chomiuk} L.,
  {Mioduszewski} A.,  {Mukai} K.,  {Rupen} M.~P.,  {Krauss} M.~I.,  {Roy} N.,
   {Nelson} T.,  2013, ArXiv e-prints

\bibitem[\protect\citeauthoryear{{Williams} \& {Mason}}{{Williams} \&
  {Mason}}{2010}]{Williams&Mason10}
{Williams} R.,  {Mason} E.,  2010, \apss, 327, 207

\bibitem[\protect\citeauthoryear{{Williams}, {Mason}, {Della Valle} \&
  {Ederoclite}}{{Williams} et~al.}{2008}]{Williams+08}
{Williams} R.,  {Mason} E.,  {Della Valle} M.,    {Ederoclite} A.,  2008, \apj,
  685, 451

\bibitem[\protect\citeauthoryear{{Wolf}, {Bildsten}, {Brooks} \&
  {Paxton}}{{Wolf} et~al.}{2013}]{Wolf+13}
{Wolf} W.~M.,  {Bildsten} L.,  {Brooks} J.,    {Paxton} B.,  2013, \apj, 777,
  136

\bibitem[\protect\citeauthoryear{{Wolff} \& {Tautz}}{{Wolff} \&
  {Tautz}}{2015}]{Wolff&Tautz15}
{Wolff} M.,  {Tautz} R.~C.,  2015, ArXiv e-prints

\bibitem[\protect\citeauthoryear{{Yaron}, {Prialnik}, {Shara} \&
  {Kovetz}}{{Yaron} et~al.}{2005}]{Yaron+05}
{Yaron} O.,  {Prialnik} D.,  {Shara} M.~M.,    {Kovetz} A.,  2005, \apj, 623,
  398

\end{thebibliography}


\end{document}